\documentclass[paper,11pt]{article}
\pdfoutput=1

\usepackage[T1]{fontenc}
\usepackage{mdframed}
\usepackage{comment}
\usepackage{mathtools}
\usepackage{upgreek}
\usepackage{graphicx}
\usepackage{amsmath}
\usepackage{amsfonts}
\usepackage{amssymb}
\usepackage{bm}
\usepackage{amsthm}
\usepackage{slashed}
\usepackage{enumerate,enumitem}
\usepackage{amsbsy}
\usepackage{bbm}
\usepackage{mathrsfs}
\usepackage{url}
\usepackage{bbm}
\usepackage{color}
\usepackage{framed}
\usepackage{subfig}
\usepackage{caption}
\usepackage{float}
\usepackage{cancel}
\usepackage{epstopdf}
\usepackage{jheppub}
\usepackage{hyperref}

\def\({\left(}
\def\){\right)}
\def\[{\left[}
\def\]{\right]}
\newcommand{\rr}{R}

\newcommand{\bra}{\langle}
\newcommand{\ket}{\rangle}

\def\one{{\rm 1\kern -.9mm l}}              

\def\beq{\begin{equation}}
\def\eeq{\end{equation}}
\def\beqa{\begin{eqnarray}}
\def\eeqa{\end{eqnarray}}
\newcommand{\eqa}{\begin{eqnarray}}
\newcommand{\ena}{\end{eqnarray}}

\newcommand{\Z}{\mathbb{Z}}

\newcommand{\cR}{\mathcal{R}}

\DeclareMathAlphabet{\mathpzc}{OT1}{pzc}{m}{it}

\newcommand{\Bs}{{\bf{s}}}
\newcommand{\Bk}{{\bf{k}}}
\newcommand\blank[1]{}

\newcommand{\fract}[2]{{\textstyle\frac{#1}{#2}}}

\newcommand\ZZ{{\mathbb Z}}
\newcommand\RR{{\mathbb R}}
\newcommand\NN{{\mathbb N}}

\newcommand{\balpha}{\alpha\kern -6.7pt\alpha}
\newcommand{\bbalpha}{\alpha\kern-4.95pt\alpha}

\newcommand\en{\end{equation}}
\newcommand\bea{\begin{eqnarray}}
\newcommand\eea{\end{eqnarray}}

\newcommand{\One}{{\hbox{{\rm 1{\hbox to 1.5pt{\hss\rm1}}}}}}
\renewcommand{\One}{{\mathbb 1}}
\renewcommand{\One}{{\rm 1\!\!1}}

\newcommand{\ba}{\begin{eqnarray}}
\newcommand{\ea}{\end{eqnarray}}

\newcommand{\be}{\begin{equation}}
\newcommand{\ee}{\end{equation}}

\renewcommand{\log}{\ln}

\newcommand{\arcsinh}{\text{arcsinh}}

\newcommand{\kersg}{ \mathcal{ K } }
\newcommand{\TbT}{\text{T}\bar{\text{T}}}
\newcommand{\TbTs}{\text{T}\bar{\text{T}}_{\bf s}}
\newcommand{\JbT}{\text{J}\bar{\text{T}}}

\newcommand{\TbJ}{\text{T}\bar{\text{J}}}

\newcommand{\ta}{\tau}
\newcommand{\del}{\delta}
\renewcommand{\ell}{{\mathcal L}}

\newcommand{\T}{{\rm T}}
\newcommand{\Th}{\Theta}
\newcommand{\cI}{\mathcal{I}}
\newcommand{\fI}{\mathfrak{I}}
\setlist[itemize]{leftmargin=*}

\definecolor{purple_nice}{rgb}{0.4,0.2,0.7}
\definecolor{fuel_blue}{RGB}{42,162,185}
\definecolor{YInMn_blue}{RGB}{46, 80, 144}
\definecolor{ultramarine}{RGB}{63, 0, 255}
\definecolor{KLEIN_blue}{rgb}{0, 0.18, 0.65}

\def\XXint#1#2#3{{\setbox0=\hbox{$#1{#2#3}{\int}$ }
\vcenter{\hbox{$#2#3$ }}\kern-.6\wd0}}

\title{\boldmath Conserved currents and $\TbT_s$ irrelevant deformations of 2D   integrable field theories}

\author[1]{Riccardo Conti,}
\author[2]{Stefano Negro,}
\author[1]{Roberto Tateo}

\affiliation[1]{Dipartimento\ di Fisica and Arnold-Regge Center, Universit\`a di Torino, and INFN, Sezione di Torino, Via P. Giuria 1, I-10125 Torino, Italy.}
\affiliation[2]{ C.N. Yang Institute for Theoretical Physics, State University of New York at Stony Brook, NY 11794-3840. U.S.A.}

\emailAdd{riccardo.conti@to.infn.it}
\emailAdd{stefano.negro@stonybrook.edu}
\emailAdd{tateo@to.infn.it}
\abstract{
It has been recently discovered that the $\TbT$ deformation is closely-related to Jackiw-Teitelboim gravity. At classical level, the introduction of this perturbation induces an interaction between the stress-energy tensor and space-time and the deformed EoMs can be mapped, through a field-dependent change of coordinates, onto the corresponding undeformed ones. The effect of this perturbation on the quantum spectrum is non-perturbatively described by an inhomogeneous Burgers equation. In this paper, we point out that there exist infinite families of models where the geometry couples instead to generic combinations of  local conserved currents labelled by the Lorentz spin.
In spirit,  these generalisations are  similar to the $\JbT$ model as the resulting theories and the corresponding scattering phase factors are not Lorentz invariant. The link with the $\JbT$ model is discussed in detail.
While the classical setup described here is very general, we shall use the  sine-Gordon model and its CFT limit as explanatory quantum examples.  Most  of the final equations and considerations are, however, of broader validity or easily generalisable to more complicated systems.    
}
\begin{document}
\maketitle
\flushbottom

\section{Introduction}
The deformation of 2D quantum field theories  \cite{Smirnov:2016lqw, Cavaglia:2016oda} by the Zamolodchikov's $\TbT$ operator  \cite{Zamolodchikov:2004ce},  has recently attracted the attention of theoretical physicists due to the many important links with effective
string theory \cite{Dubovsky:2012wk, dubovsky2012effective,Caselle:2013dra, Chen:2018keo}  and the AdS/CFT correspondence \cite{McGough:2016lol,Turiaci:2017zwd, Giveon:2017nie, Giveon:2017myj,Asrat:2017tzd,Giribet:2017imm,Kraus:2018xrn, Cottrell:2018skz,  Baggio:2018gct,Babaro:2018cmq}. A remarkable  property of this perturbation, discovered  in \cite{Smirnov:2016lqw, Cavaglia:2016oda}, concerns the evolution of the quantum spectrum at finite volume $R$, with periodic boundary conditions,  in terms of the  $\TbT$ coupling constant $\tau$. The spectrum is governed by the hydrodynamic-type equation
\beq
\label{eq:Burgers}
\partial_{\ta} E(R,\ta) = \frac{1}{2} \partial_R\left( E^2(R,\ta) - P^2(R) \right) \;,
\eeq
where $E(R,\ta)$ and $P(R)$ are the eigenvalues of the energy and momentum operator, respectively, on a generic eigenstate $|n \ket$.  Important for the current purposes is that, under the perturbation, the evolution of the spectrum is equivalently encoded in the following Lorentz-type transformation
\beq
\label{eq:Lorentz}
 \left(\begin{array}{c}
E(R,\ta) \\
P(R)
\end{array}\right) =
\left(\begin{array}{cc}
\cosh{(\theta_0)} & \sinh{(\theta_0)} \\
\sinh{(\theta_0)} & \cosh{(\theta_0)} 
\end{array}\right) 
\left(\begin{array}{c}
E(\cR_0) \\
P(\cR_0)
\end{array}\right) \;,
\eeq
with $\cR_0$ and $\theta_0$ defined through
\beq
\sinh{\theta_0} = \frac{\ta\,P(R)}{\cR_0} = \frac{\ta\,P(\cR_0)}{R} \;,\quad \cosh{\theta_0} = \frac{R + \ta\,E(R,\ta)}{\cR_0} = \frac{\cR_0 - \ta\,E(\cR_0)}{R} \;.
\label{eq:Lorentz1}
\eeq
From (\ref{eq:Lorentz}), it follows that the solution to (\ref{eq:Burgers}) can be written, in implicit form, as
\beq
E^2(R,\ta) - P^2(R) = E^2(\cR_0) - P^2(\cR_0) \;,
\label{eq:Lorentz3}
\eeq
with the additional constraint
\beq
\partial_{\ta} R = -E(R,\ta)\;,
\label{eq:partialR}
\eeq
at fixed $\cR_0$, obtainable directly from (\ref{eq:Lorentz1}) (cf.  with the $s=1$ case of (\ref{eq:sol})).\\
As extensively discussed in \cite{Conti:2018tca} (see also \cite{Dubovsky:2017cnj}), the solutions to the classical EoMs associated to the $\TbT$-deformed Lagrangians \cite{Cavaglia:2016oda, Bonelli:2018kik, Conti:2018jho, Conti:2018tca} are obtained from the $\ta=0$ ones by a field-dependent coordinate transformation
\beq
dx^\mu= \left(\delta^{\mu}_{\;\;\nu} + \ta\,\widetilde{\bm T}^{\mu}_{\;\;\nu}(\mathbf y) \right) dy^\nu \;, \quad \mathbf y=(y^1,y^2)=(x',t')\;,
\label{eq:map1}
\eeq
with  
\beq
\widetilde{\bm T}^{\mu}_{\;\;\nu}(\mathbf y) = -g^{\mu\delta} \epsilon_{\delta\rho}\,\epsilon_{\sigma\nu}\,{\bm T}^{\rho\sigma}(\mathbf y)
\;,\quad \epsilon_{\mu\nu} = 
\(\begin{array}{cc}
0 & 1 \\
-1 & 0
\end{array}\)_{\mu\nu} \;,
\label{eq:Ttilde}
\eeq
where ${\bm T}^{\mu\nu}(\mathbf y)$ is the Hilbert stress-energy tensor associated to the undeformed theory, canonically defined as
\beq  
{\bm T}^{\mu\nu}(\mathbf y) = \frac{-2}{\sqrt g}\frac{\partial\mathcal{L}_g(\mathbf y)}{\partial g_{\mu\nu}} \;,\quad \sqrt{g} = \sqrt{\det{(g_{\mu\nu})}} \;,
\eeq
and $\mathcal{L}_g(\mathbf y)$ is the undeformed Lagrangian in the set of cartesian coordinates $\mathbf y$, minimally coupled to gravity through the metric $g_{\mu\nu}$. As shown in \cite{Conti:2018tca}, equation (\ref{eq:map1}) can be inverted as
\beqa
dy^\mu &=& \left(\delta^{\mu}_{\;\;\nu} - \ta\,\widetilde{\bm T}^{\mu}_{\;\;\nu}(\mathbf x,\ta) \right) dx^\nu \;, \quad \mathbf x=(x^1,x^2)=(x,t) \;,
\label{eq:map2}
\eeqa
with\footnote{In (\ref{eq:map2}), we corrected a sign typo made in \cite{Conti:2018tca}.}
\beq 
\widetilde{\bm T}^{\mu}_{\;\;\nu}(\mathbf x,\ta) = -g^{\mu\delta} \epsilon_{\delta\rho}\,\epsilon_{\sigma\nu}\,{\bm T}^{\rho\sigma}(\mathbf x,\ta) \;,
\eeq
where ${\bm T}^{\mu\nu}(\mathbf x,\ta) $ is now the Hilbert stress-energy tensor associated to the $\TbT$-deformed theory
\beq  
T^{\mu\nu}(\mathbf x,\ta) = \frac{-2}{\sqrt g}\frac{\partial\mathcal{L}_g(\mathbf x,\ta)}{\partial g_{\mu\nu}} \;,
\eeq
and $\mathcal{L}_g(\mathbf x,\ta)$ is the deformed Lagrangian in the set of cartesian coordinates $\mathbf x$. Following the convention of \cite{Conti:2018tca}, we will switch from cartesian to complex coordinates according to 
\beq 
\begin{cases}
z = x^1 + i\,x^2 \\
\bar z = x^1 - i\,x^2
\end{cases} \;,\quad 
\begin{cases}
w = y^1 + i\,y^2 \\
\bar w = y^1 - i\,y^2 
\end{cases} \;,
\label{eq:EuclLC}
\eeq
and we shall denote with $\mathbf z=(z,\bar z)$ and $\mathbf w=(w,\bar w)$ the two set of complex coordinates.\\
An important link with topological gravity was noticed and studied in  \cite{Dubovsky:2017cnj}, where it was shown
that JT gravity coupled to matter leads to a scattering phase matching that associated to the $\TbT$ perturbation \cite{Dubovsky:2012wk, dubovsky2012effective, Caselle:2013dra, Smirnov:2016lqw, Cavaglia:2016oda}. Equations (\ref{eq:map1}) and (\ref{eq:map2}) were also obtained in \cite{Conti:2018tca} starting from the deformed EoMs following, therefore, a completely independent line of thoughts compared to \cite{Dubovsky:2017cnj}. However, the final results turn out to be fully consistent with the proposal of \cite{Dubovsky:2017cnj}. 
In this paper, we shall argue that the approach of \cite{Conti:2018tca} admits natural generalisations corresponding to infinite families of geometric-type deformations of classical and quantum field theories.\\
Our analysis starts from (\ref{eq:map1}) and the observation  that the equality between the second mixed derivatives implies
\beq
\frac{\partial^2 x^{\mu}}{\partial y^{\rho} \partial y^{\sigma}} =
\frac{\partial^2 x^{\mu}}{\partial y^{\sigma} \partial y^{\rho}} \quad \Longleftrightarrow \quad
\partial_\mu {\bm T}^{\mu\nu} = 0\;.
\label{eq:mixed}
\eeq
Equation (\ref{eq:mixed}) suggests that a consistent and natural  generalisation of (\ref{eq:map1}) can be obtained by replacing the stress-energy tensor with an arbitrary (rank-two)  conserved current. Thus,  the main objective of this work is to study the generalisations of the change of coordinates (\ref{eq:map1}) obtained by replacing ${\bm T}^{\mu\nu}$ with a matrix built using the  higher-spin conserved currents,  typically present in free  or integrable theories. In complex coordinates $\mathbf z$, the spin $\Bs$ conserved currents ${\bf T}_{\Bs+1}$, ${\bf\bar T}_{\Bs+1}$, ${\bf\Theta}_{\Bs-1}$ and ${\bf\bar\Theta}_{\Bs-1}$ are related through the continuity equations
\beq
\bar\partial {\bf T}_{\Bs+1} = \partial {\bf\Theta}_{\Bs-1}\;,\quad  \partial {\bf\bar T}_{\Bs+1} = \bar \partial {\bf\bar\Theta}_{\Bs-1}\;,\quad  (\Bs  \in \NN) \;,
\label{eq:conserved0}
\eeq
where $\partial=\partial_z$ and $\bar\partial = \partial_{\bar z}$, the subscripts $\Bs+1$ and  $\Bs-1$ are the Lorentz spins of the corresponding field and the $\Bs=1$ case of (\ref{eq:conserved0}) corresponds to the energy and momentum conservation law. The replacement we shall perform in (\ref{eq:map1})--(\ref{eq:Ttilde}) is 
\beq
\label{eq:NoLorentz}
{\bm T} 
\quad \longrightarrow \quad
{\bm T}_\Bs \;,
\eeq
where the cartesian and complex components of ${\bm T}_\Bs$ are related through
\beqa  
({\bm T}_\Bs)_{11} = -\frac{1}{2\pi}\( {\bf\bar T}_{\Bs+1}+{\bf T}_{\Bs+1}-{\bf\bar{\Theta}}_{\Bs-1}-{\bf\Theta}_{\Bs-1} \)&,& ({\bm T}_\Bs)_{12} = \frac{i}{2\pi}\( {\bf\bar T}_{\Bs+1} - {\bf T}_{\Bs+1} + {\bf\bar\Theta}_{\Bs-1} - {\bf\Theta}_{\Bs-1} \), \notag \\
({\bm T}_\Bs)_{21} = \frac{i}{2\pi}\( {\bf\bar T}_{\Bs+1} - {\bf T}_{\Bs+1} - {\bf\bar\Theta}_{\Bs-1} + {\bf\Theta}_{\Bs-1} \) &,& ({\bm T}_\Bs)_{22} = \frac{1}{2\pi}\( {\bf\bar T}_{\Bs+1}+{\bf T}_{\Bs+1}+{\bf\bar\Theta}_{\Bs-1}+{\bf\Theta}_{\Bs-1} \). \notag \\
\eeqa
For simplicity, it will be useful to define the re-scaled quantities\footnote{Notice that this convention differs by a factor $2$ compared to \cite{Cavaglia:2016oda,Conti:2018tca}. This choice allows for a more transparent match between the classical and the quantum results of section \ref{sec:quantum}.}
\beq
{\rm T}_{\Bs+1}  = \frac{{\bf T}_{\Bs+1}}{2\pi} \;,\quad \Theta_{\Bs-1} = \frac{{\bf\Theta}_{\Bs-1}}{2\pi} \;,\quad \bar{\rm T}_{\Bs+1} = \frac{{\bf \bar T}_{\Bs+1}}{2\pi} \;,\quad \bar\Theta_{\Bs-1} = \frac{{\bf\bar\Theta}_{\Bs-1}}{2\pi} \;.
\eeq
One of the main consequences of the replacement (\ref{eq:NoLorentz}) is that  the resulting  generalised deformations break explicitly the (eventual) Lorentz symmetry of the original models. See appendix \ref{sec:deflogsol} for an explicit example where the difference between the $\TbT$ and the $\Bs\neq 1$ cases clearly emerge. The corresponding gravity-like systems have therefore many features in common with the $\JbT$ ($\TbJ$) models recently studied in \cite{Guica:2017lia, Bzowski:2018pcy,
Apolo:2018qpq, Aharony:2018ics, Chakraborty:2018vja, Nakayama:2018ujt, Araujo:2018rho}. As we shall see in section \ref{sec:TTspsformal}, in the reference frame $\mathbf w$, the deformed Hamiltonian -- formally integrated over the space-time in the $\mathbf z$ variable -- can be written as the integrated bare Hamiltonian plus a perturbing field:
\beqa
\int \mathcal{H}^{(\Bs)}(\mathbf z,\ta)\,dz\wedge d\bar z =
\int \mathcal{H}(\mathbf w)\,dw\wedge d\bar w + \ta \int {\bf \Phi}_{1,\Bs}(\mathbf w) \,dw\wedge d\bar w \;,
\eeqa
where the perturbing operator ${\bf \Phi}_{1,\Bs}(\mathbf w)$ reduces to
\beq
{\bf \Phi}_{1,\Bs}(\mathbf w)  = \T(w)\,\bar\T_{\Bs+1}(\bar w)+ \bar\T(\bar w)\,\T_{\Bs+1}(w)\,,
\eeq
when the original theory is a CFT. For this reason we shall denote this  newly-introduced class of systems as {\it geometrically} $\TbTs$-deformed theories.\footnote{It turns out that these perturbations are, in general, different from the ones recently studied in \cite{LeFloch:2019wlf}. Currently, we do not know if it exists a link between the class of deformations considered in \cite{LeFloch:2019wlf} and ours. In particular, it would be nice to understand whether or not the models of \cite{LeFloch:2019wlf} possess a geometric interpretation in terms of specific space-time coordinate transformations on the plane.} A precise connection between the $\Bs=0$ case and the $\JbT$ models will be established in section \ref{sec:TJ}.
As in \cite{Cavaglia:2016oda,Conti:2018jho,Conti:2018tca}, starting from  section \ref{sec:generalisedBurgers}, we shall
use the  sine-Gordon model as a specific quantum example. However, we would like to stress  that the techniques adopted   and some of the results, {\it i.e.} the generalised Burgers equations (\ref{eq:sol})--(\ref{eq:energyBu}), are at least formally, of much wider validity. 
\section{$\TbT$-deformed higher spin conserved currents}
\label{sec:TTbarHC}
The aim of this section is to  introduce an efficient method, based on the field-dependent coordinate transformation derived in \cite{Conti:2018tca}, to reconstruct the local Integrals of Motion (IMs) associated to the $\TbT$ deformation of a generic  integrable field theory.  The application of these ideas to the family of classical {\it geometrically} $\TbT_\Bs$ deformed models will be described in section \ref{sec:TTsHC}. 
\subsection{A strategy to reconstruct the $\TbT$-deformed higher conserved currents}
Let us consider the following pair of conjugated 1-forms
\beq
\label{eq:1forms}
\fI_{\Bk} = \T_{\Bk+1}(\mathbf z,\ta)\,dz + \Th_{\Bk-1}(\mathbf z,\ta)\,d\bar{z} \;,\quad \bar{\fI}_{\Bk} = \bar{\T}_{\Bk+1}(\mathbf z,\ta)\,d\bar{z} + \bar{\Th}_{\Bk-1}(\mathbf z,\ta)\,dz \;,\quad (\Bk\in\mathbb N) \;,
\eeq
where the components $\T_{\Bk+1}(\mathbf z,\ta)$, $\Th_{\Bk-1}(\mathbf z,\ta)$ and their complex conjugates are the higher conserved currents of the $\TbT$-deformed integrable theory which fulfil the continuity equations
\beq
\label{eq:conteq}
\bar\partial \T_{\Bk+1}(\mathbf z,\ta) = \partial\Th_{\Bk-1}(\mathbf z,\ta) \;,\quad \partial\bar\T_{\Bk+1}(\mathbf z,\ta) = \bar\partial \bar\Th_{\Bk-1}(\mathbf z,\ta) \;,
\eeq
and $\Bk$ denotes the Lorentz spin.
Using (\ref{eq:conteq}), it is easy to check that (\ref{eq:1forms}) are closed forms\footnote{Thus they are locally exact by the Poincar\'e lemma.}
\beq
d\fI_\Bk = \( \partial\Th_{\Bk-1} - \bar\partial\T_{\Bk+1} \)dz\wedge d\bar z = 0 \;,\quad d\bar{\fI}_\Bk = \( \partial\bar\T_{\Bk+1} - \bar\partial\bar\Th_{\Bk-1} \)dz\wedge d\bar z = 0 \;,
\eeq
therefore, for any given integration contour $C$, the following integrals
\beq
\label{eq:clint}
\int_C \fI_\Bk \;,\quad \int_C \bar{\fI}_\Bk \;,
\eeq
do not depend on deformations of $C$, at fixed end-points. Expressions (\ref{eq:clint}) can be used to define local IMs. From their very definition, differential forms are the right objects to be integrated over manifolds since they are independent of coordinates. In fact, since $1$-forms remain closed also under field-dependent coordinate transformations, we can construct the $\TbT$-deformed local IMs from the undeformed ones, using the change of coordinates introduced in \cite{Dubovsky:2017cnj,Conti:2018tca}. The strategy is the following:
\begin{itemize}
    \item Start from the $1$-forms (\ref{eq:1forms}) expressed in $\mathbf w$ coordinates
    \beqa
    \fI_{\Bk} = \T_{\Bk+1}(\mathbf w)\,dw + \Th_{\Bk-1}(\mathbf w)\,d\bar{w} \;,\quad \bar{\fI}_{\Bk} = \bar{\T}_{\Bk+1}(\mathbf w)\,d\bar{w} + \bar{\Th}_{\Bk-1}(\mathbf w)\,dw \;; \label{eq:undef1forms} 
    \eeqa
    where $\T_{\Bk+1}(\mathbf w)$ and $\Th_{\Bk-1}(\mathbf w)$ and their complex conjugates are the higher conserved currents of the undeformed theory which fulfil the continuity equations
    \beq
    \label{eq:contequndef}
    \partial_{\bar w} \T_{\Bk+1}(\mathbf w) = \partial_w\Th_{\Bk-1}(\mathbf w) \;,\quad \partial_w\bar\T_{\Bk+1}(\mathbf w) = \partial_{\bar w} \bar\Th_{\Bk-1}(\mathbf w) \;.
    \eeq
    \item Consider the change of coordinates $\mathbf w = \mathbf w(\mathbf z)$ (see \cite{Conti:2018tca}), which at differential level acts as follows
    \begin{gather}
    \label{eq:transfTT}
    \(\begin{array}{c}
    dw \\
    d\bar w
    \end{array}\) =
    \mathcal{J}^T
    \(\begin{array}{c}
    dz \\
    d\bar z
    \end{array}\) \;,\quad
    \(\begin{array}{c}
    \partial_w f \\
    \partial_{\bar w} f
    \end{array}\) =
    \mathcal{J}^{-1}
    \(\begin{array}{c}
    \partial f \\
    \bar\partial f
    \end{array}\) \;,\quad (\forall f:\mathbb{R}^2\rightarrow\mathbb{R}) \;,
    \end{gather}
    where the Jacobian and its inverse are
    \begin{gather}
    \label{eq:jacTT}
    \mathcal{J} = \( \begin{array}{cc}
    \partial w & \partial\bar w \\
    \bar\partial w & \bar\partial \bar w
    \end{array}\) = \frac{1}{\Delta(\mathbf w)}
    \( \begin{array}{cc}
    1+2\ta\,\Th_0(\mathbf w) & -2\ta\,\T_2(\mathbf w) \\
    -2\ta\,\bar\T_2(\mathbf w) & 1+2\ta\,\bar\Th_0(\mathbf w)
    \end{array} \) \;,  \\
    \label{eq:invjacTT}
    \mathcal{J}^{-1} = \( \begin{array}{cc}
    \partial_w z & \partial_w \bar z \\
    \partial_{\bar w}z & \partial_{\bar w} \bar z
    \end{array}\) = 
    \( \begin{array}{cc}
    1+2\ta\,\bar\Th_0(\mathbf w) & 2\ta\,\T_2(\mathbf w) \\
    2\ta\,\bar\T_2(\mathbf w) & 1+2\ta\,\Th_0(\mathbf w)
    \end{array} \) \;,
    \end{gather}
    and $$\Delta(\mathbf w)=\bigl(1+2\ta\,\Th_{0}(\mathbf w)\bigr)\bigl(1+2\ta\,\bar\Th_{0}(\mathbf w)\bigr)-4\ta^2\,\T_{2}(\mathbf w)\,\bar\T_{2}(\mathbf w) \;.$$
    \item Use the explicit expressions of $\T_2(\mathbf w)$, $\Th_0(\mathbf w)$ and their complex conjugates in terms of the fundamental fields of the theory to explicitly invert the map $\mathbf w(\mathbf z)$ at differential level. 
    Then, use the first expression of (\ref{eq:transfTT}) in (\ref{eq:undef1forms})
    \begin{multline}
    \label{eq:def1forms}
    \fI_\Bk = \frac{ \T_{\Bk+1}(\mathbf w(\mathbf z)) + 2\ta\bigl(\T_{\Bk+1}(\mathbf w(\mathbf z))\,\Th_0(\mathbf w(\mathbf z)) -\Th_{\Bk-1}(\mathbf w(\mathbf z))\,\T_2(\mathbf w(\mathbf z))\bigr) }{\Delta(\mathbf w(\mathbf z))}\,dz \\ +\frac{\Th_{\Bk-1}(\mathbf w(\mathbf z)) + 2\ta\bigl(\Th_{\Bk-1}(\mathbf w(\mathbf z))\,\bar\Th_0(\mathbf w(\mathbf z)) -\T_{\Bk+1}(\mathbf w(\mathbf z))\,\bar\T_2(\mathbf w(\mathbf z))\bigr)}{\Delta(\mathbf w(\mathbf z))}\,d\bar z \;,
    \end{multline}
    where $F(\mathbf w(\mathbf z))$ indicates that the fundamental fields in $\mathbf w$ coordinates involved in an arbitrary function $F$, have been replaced with fundamental fields in $\mathbf z$ coordinates according to the map $\mathbf w(\mathbf z)$.
    \item Read the $\TbT$-deformed higher conserved currents as components of (\ref{eq:def1forms}) in $\mathbf z$ coordinates:
    \beqa
    \label{eq:defcompTT}
    \T_{\Bk+1}(\mathbf z,\ta) &=& \frac{ \T_{\Bk+1}(\mathbf w(\mathbf z)) + 2\ta\bigl(\T_{\Bk+1}(\mathbf w(\mathbf z))\,\Th_0(\mathbf w(\mathbf z)) -\Th_{\Bk-1}(\mathbf w(\mathbf z))\,\T_2(\mathbf w(\mathbf z))\bigr) }{\Delta(\mathbf w(\mathbf z))} \;, \notag \\
    \Th_{\Bk-1}(\mathbf z,\tau) &=& \frac{\Th_{\Bk-1}(\mathbf w(\mathbf z)) + 2\ta\bigl(\Th_{\Bk-1}(\mathbf w(\mathbf z))\,\bar\Th_0(\mathbf w(\mathbf z)) -\T_{\Bk+1}(\mathbf w(\mathbf z))\,\bar\T_2(\mathbf w(\mathbf z))\bigr)}{\Delta(\mathbf w(\mathbf z))} \;. \notag \\
    \eeqa
\end{itemize}
By definition, the integrals (\ref{eq:clint}) are invariant under coordinate transformations, provided the integration contour $C$ is mapped into $C'$ accordingly 
\beq
\int_{C} \T_{\Bk+1}(\mathbf z,\ta)\,dz + \Th_{\Bk-1}(\mathbf z,\ta)\,d\bar{z} = \int_{C'} \T_{\Bk+1}(\mathbf w)\,dw + \Th_{\Bk-1}(\mathbf w)\,d\bar{w} \;.
\eeq
The conserved charges are obtained by integrating over the whole volume at constant time, in the $\mathbf x$ reference frame. However, as clearly emerges from (\ref{eq:map1}) and (\ref{eq:map2}), the equal time structure in the $\mathbf x-$plane gets distorted in the $\mathbf y-$plane and viceversa, which causes the emerging of a $\ta-$dependent flow in the charges. In the current setup, 
the resulting integrals of motion coincide  with $\TbT$-deformed ones. The application of this scheme to the more general change of variables defined through the replacement (\ref{eq:NoLorentz}), will  lead instead to different, but  equally interesting, families of deformed classical Hamiltonians (see, section \ref{sec:TTsHC}). \\
In order to make the above strategy more concrete, in the following section, we will explicitly discuss the construction of the $\TbT$-deformed higher currents for the massless free boson theory and comment on more general cases.
\subsection{The massless free boson}
\label{sec:TTbarHCfreebos}
Consider the Lagrangian of a single massless boson field $\phi$ in  complex coordinates $\bf w$
\beq  
\mathcal{L}(\mathbf w) = \partial_{w}\phi\,\partial_{\bar w}\phi \;.
\eeq
The EoMs are
\beq  
\label{eq:undefEoMs}
\partial_w\partial_{\bar w}\phi = 0 \;,
\eeq
therefore, without further external constraints, there exists an infinite number of options for the choice of the basis of conserved currents. For example, both 
\beqa
\label{eq:NG1formcomp1}
\T_{\Bk+1}^{(\text{POW})}(\mathbf w) = -\frac{1}{2}\(\partial_w\phi\)^{\Bk+1
} \;&,&\quad \Th_{\Bk-1}^{(\text{POW})}(\mathbf w) = 0 \;, \quad  (\Bk\in\mathbb N) \;,
\eeqa
and\footnote{The set of currents (\ref{eq:NG1formcomp2}) can be obtained as the massless limit of the Klein-Gordon hierarchy: 
$$\T_{\Bk+1}^{(\text{KG})}(\mathbf w) = -\frac{1}{2}\bigl(\partial_w^{\frac{1+\Bk}{2}}\phi\bigr)^2 \;,\quad \Th_{\Bk-1}^{(\text{KG})}(\mathbf w) = -\frac{m^2}{2}\bigl( \partial_w^{\frac{\Bk-1}{2}}\phi \bigr)^2 \;,\quad (\Bk\in 2\NN+1) \;,$$
with Lagrangian $\mathcal{L}^{(\text{KG})}(\mathbf w) = \partial_w\phi\,\partial_{\bar w}\phi + m^2\phi^2$.}
\beqa
\label{eq:NG1formcomp2}
\T_{\Bk+1}^{(\text{KG})}(\mathbf w) = -\frac{1}{2}\Bigl(\partial_w^{\frac{1+\Bk}{2}}\phi\Bigr)^2 \;&,&\quad \Th_{\Bk-1}^{(\text{KG})}(\mathbf w) = 0 \;,\quad (\Bk \in 2\NN+1) \;,
\eeqa
are possible  sets of higher conserved currents  since they fulfil (\ref{eq:contequndef}) on-shell. In general, any linear combination of the form
\beq
\label{eq:genericcurrent}
\T_{\Bk+1}^{(\text{GEN})}(\mathbf w) = \sum_{j=0}^{\Bk+1} c^{(\Bk)}_j\,(\partial_w\phi)^{\Bk+1-j}\,\partial_w^{j}\phi \;,\quad \Th_{\Bk-1}^{(\text{GEN})}(\mathbf w) = \sum_{j=0}^{\Bk-1} \bar c^{(\Bk)}_j\,(\partial_{\bar w}\phi)^{\Bk-1-j}\,\partial_{\bar w}^{j}\phi \;,
\eeq
automatically defines a conserved current with spin \Bk. Moreover, since the change of variables (\ref{eq:map1}) is non-linear, different choices of the current in (\ref{eq:NoLorentz}) should, at least in principle, give rise to different classical deformations of the original theory. \\
For simplicity, we will consider the sets (\ref{eq:NG1formcomp1})--(\ref{eq:NG1formcomp2}) separately. Following the strategy described in the previous section, we shall first derive the differential map, which means to express $(\partial_w f,\partial_{\bar w}f)^{\rm T}$ in terms of $(\partial f,\bar\partial f)^{\rm T}$, $\forall f:\mathbb{R}^2\to\mathbb{R}$. Setting $f=\phi$ in the second expression of (\ref{eq:transfTT}), we first write $(\partial_w\phi,\partial_{\bar w}\phi)^{\rm T}$ as a function of $(\partial \phi,\bar\partial\phi)^{\rm T}$ by solving the set of algebraic equations
\beq
\label{eq:systemTT}
\(\begin{array}{c}
\partial\phi \\
\bar\partial\phi
\end{array}\) = 
\mathcal{J}
\(\begin{array}{c}
\partial_w\phi \\
\partial_{\bar w}\phi
\end{array}\) \quad\longleftrightarrow\quad 
\begin{cases}
\partial\phi = \frac{\partial_w\phi}{1-\ta\(\partial_w\phi\)^2} \\
\bar\partial\phi = \frac{\partial_{\bar w}\phi}{1-\ta\(\partial_{\bar w}\phi\)^2}
\end{cases} \;,
\eeq
where we used
\beq  
\T_2(\mathbf w) = -\frac{1}{2}\(\partial_w\phi\)^2 \;,\quad \Th_0(\mathbf w) = 0 \;.
\eeq
The solution to (\ref{eq:systemTT}) is
\beq 
\label{eq:solsystemTT}
\partial_w \phi= \partial \phi - \frac{1}{4\ta}\(\frac{-1+\mathcal{S}}{\bar\partial\phi}\)^2 \bar\partial \phi \;,\quad \partial_{\bar w}\phi = - \frac{1}{4\ta}\(\frac{-1+\mathcal{S}}{\partial\phi}\)^2 \partial \phi + \bar\partial \phi \;,
\eeq
with
\beq 
\mathcal{S}=\sqrt{1+4\ta\,\partial\phi\,\bar\partial\phi} \;.
\eeq 
Now, plugging (\ref{eq:solsystemTT}) into the second expression of (\ref{eq:transfTT}), we find the differential map
\beq  
\label{eq:diffchangeTT}
\partial_w f= \partial f - \frac{1}{4\ta}\(\frac{-1+\mathcal{S}}{\bar\partial\phi}\)^2 \bar\partial f \;,\quad \partial_{\bar w}f = - \frac{1}{4\ta}\(\frac{-1+\mathcal{S}}{\partial\phi}\)^2 \partial f + \bar\partial f \;,\quad( \forall f:\mathbb{R}^2\rightarrow\mathbb{R}) \;.
\eeq
Using (\ref{eq:diffchangeTT}), we can easily derive the $\TbT$-deformed version of the original EoMs as follows
\beq 
\partial_w\(\partial_{\bar w}\phi\) = 0 \quad\longrightarrow\quad \(\partial-\frac{1}{4\ta}\(\frac{-1+\mathcal{S}}{\bar\partial\phi}\)^2\bar\partial\)\(- \frac{1}{4\ta}\(\frac{-1+\mathcal{S}}{\partial\phi}\)^2 \partial \phi + \bar\partial \phi\) = 0 \;,
\eeq
which, after some manipulations, can be recast into
\beq 
\label{eq:TTEoMs}
\partial\bar\partial\phi = \ta\,\frac{\partial^2\phi\,(\bar\partial\phi)^2 + \bar\partial^2\phi\,(\partial\phi)^2}{1+2\ta\,\partial\phi\,\bar\partial\phi} \;.
\eeq
The $\TbT$-deformed currents are obtained using (\ref{eq:diffchangeTT}) in the general expressions (\ref{eq:defcompTT}). Considering the set of currents (\ref{eq:NG1formcomp1}), namely setting $\T_{\Bk+1} = \T_{\Bk+1}^{(\text{POW})}$, $\Th_{\Bk-1} = \Th_{\Bk-1}^{(\text{POW})}$ and the same for their complex conjugates in (\ref{eq:defcompTT}), one finds
\beq
\label{eq:NG1formcomp1def}
\T_{\Bk+1}^{(\text{POW})}(\mathbf z,\ta) = -\frac{(\partial\phi)^{\Bk+1}}{2\mathcal{S}}\( \frac{2}{1+\mathcal{S}} \)^{\Bk-1} \;,\quad
\Th_{\Bk-1}^{(\text{POW})}(\mathbf z,\ta) = -\tau\, \frac{(\partial\phi)^{\Bk+1}(\bar\partial\phi)^2}{2\mathcal{S}} \( \frac{2}{1+\mathcal{S}} \)^{\Bk+1} \;,
\eeq
which coincides with the result first obtained in \cite{Cavaglia:2016oda} through perturbative computations. Observe that, using (\ref{eq:diffchangeTT}) in (\ref{eq:jacTT}), the Jacobian can be rewritten as
\beq
\label{eq:jacTTdef}
\mathcal{J} =
\( \begin{array}{cc}
1-2\ta\,\bar\Th_0(\mathbf z,\ta) & -2\ta\,\T_2(\mathbf z,\ta) \\
-2\ta\,\bar\T_2(\mathbf z,\ta) & 1-2\ta\,\Th_0(\mathbf z,\ta)
\end{array} \) \;,
\eeq
where $\T_2(\mathbf z,\ta)$, $\Th_0(\mathbf z,\ta)$ and their complex conjugates are the components of the $\TbT$-deformed stress-energy tensor, which can be read from (\ref{eq:NG1formcomp1def}), setting $\Bk=1$. Switching from complex $\mathbf z$ to cartesian $\mathbf x$ coordinates, it is easy to realize that expression (\ref{eq:jacTTdef}) leads to the inverse map (\ref{eq:map2}). The latter result can be generalised to the case of $N$-boson fields with generic potential (see \cite{Conti:2018tca}).\\
For the set of currents (\ref{eq:NG1formcomp2}), we can again derive exactly the associated $\TbT$-deformed currents, however their analytic expressions are more and more involved as $\Bk$ increases and we were unable to find a compact  formula valid for arbitrary spin $\Bk \in \NN$. We report here, as an example, the level $\Bk=3$ deformed current of the hierarchy (\ref{eq:NG1formcomp2}):
\beqa
\T_4(\mathbf z,\ta) &=& -\frac{(\partial\phi)^2}{2\mathcal{S}}\(\frac{(\mathcal{S}-1)^4 \bar\partial^2\phi-16 \ta ^2 (\bar\partial\phi)^4\,\partial^2\phi}{4\ta (\mathcal{S}-1)\left(\mathcal{S}^2+1\right) (\bar\partial\phi)^3}\)^2 \;, \notag \\
\Th_2(\mathbf z,\ta) &=& -\frac{(\mathcal{S}-1)^2}{8\ta\,\mathcal{S}}\(\frac{(\mathcal{S}-1)^4 \bar\partial^2\phi-16 \ta ^2 (\bar\partial\phi)^4\,\partial^2\phi}{4\ta (\mathcal{S}-1)\left(\mathcal{S}^2+1\right) (\bar\partial\phi)^3}\)^2 \;.
\eeqa
Finally, it is important to stress that the method presented in this section is completely general and can be applied to a generic integrable model, provided the stress-energy tensor and the conserved currents are known in terms of fundamental fields. We have explicitly computed the $\TbT$-deformed conserved currents with  $\Bk=3,5$ for the sine-Gordon model using the generalised change of variables described in  \cite{Conti:2018tca}. 
Again the resulting expressions are extremely complicated and we will not present them here. 
\section{Deformations induced by conserved currents with higher Lorentz spin}
\label{sec:TTsHC}
In the following sections we shall  generalise the change of variables (\ref{eq:transfTT}) to deformations built from conserved currents with generic positive and negative integer spins. Conventionally,  we will denote   with $s =|\Bs| \ge 0$ ($k =|\Bk| \ge 0$) the absolute value of the spin, and  set  $s' = - s \le  0 $ ($k' = - k \le  0 $).   
We will discuss separately perturbations induced by conserved currents  with spin $s= \Bs >0$, $s'=\Bs < 0$ and $\Bs \rightarrow 0$, providing explicit examples. 
\subsection{Deformations related to charges with positive spin $\Bs >0$}
\label{sec:TTspsformal}
\noindent
The natural generalisation of (\ref{eq:transfTT}), which ensures the equality of mixed partial derivatives is
\begin{gather}
\label{eq:JacobianTTs}
\(\begin{array}{c}
dw \\
d\bar w
\end{array}\) =
\(\mathcal{J}^{(s)}\)^T
\(\begin{array}{c}
dz \\
d\bar z
\end{array}\) \;,\quad
\(\begin{array}{c}
\partial_w f \\
\partial_{\bar w} f
\end{array}\) =
\(\mathcal{J}^{(s)}\)^{-1}
\(\begin{array}{c}
\partial f \\
\bar\partial f
\end{array}\) \;,\quad ( \forall f:\mathbb{R}^2\rightarrow\mathbb{R} )\;,
\end{gather}
where
\beqa
\label{eq:generalJacobianHC1}
\mathcal{J}^{(s)} &=& \(\begin{array}{cc}
\partial w & \partial\bar w \\
\bar\partial  w & \bar\partial\bar w
\end{array}\) = 
\frac{1}{\Delta^{(s)}(\mathbf w)}\(\begin{array}{cc}
1+2\ta\,\Th_{s-1}(\mathbf w) & -2\ta\,\T_{s+1}(\mathbf w) \\
-2\ta\,\bar\T_{s+1}(\mathbf w) & 1+2\ta\,\bar\Th_{s-1}(\mathbf w)
\end{array}\) \;,  \\
\label{eq:generalJacobianHC2}
\(\mathcal{J}^{(s)}\)^{-1} &=& \(\begin{array}{cc}
\partial_ w z & \partial_ w \bar{z} \\
\partial_{\bar w} z & \partial_{\bar w} \bar z
\end{array}\) =
\(\begin{array}{cc}
1+2\ta\,\bar\Th_{s-1}(\mathbf w) & 2\ta\,\T_{s+1}(\mathbf w) \\
2\ta\,\bar\T_{s+1}(\mathbf w) & 1+2\ta\,\Th_{s-1}(\mathbf w)
\end{array}\) \;,
\eeqa
with $s>0$, 
\beqa
\Delta^{(s)}(\mathbf w)=\bigl(1+2\ta\,\Th_{s-1}(\mathbf w)\bigr)\bigl(1+2\ta\,\bar\Th_{s-1}(\mathbf w)\bigr)-4\ta^2\,\T_{s+1}(\mathbf w)\,\bar\T_{s+1}(\mathbf w) \;,
\eeqa
and $s=1$ corresponds to the $\TbT$ deformation, $\mathcal{J}^{(1)} = \mathcal{J}$. In fact, using the continuity equations (\ref{eq:contequndef}) in (\ref{eq:generalJacobianHC2}), one finds that the second mixed partial derivatives are identical
\beq
\label{eq:mixder}
\partial_{\bar w}(\partial_w z) = \partial_w(\partial_{\bar w}z) \;,\quad \partial_{\bar w}(\partial_w \bar z) = \partial_w(\partial_{\bar w}\bar z) \;.
\eeq
Consider now the 1-forms
\beq
\label{eq:1formsTTs}
\fI_{k} = \T_{k+1}^{(s)}(\mathbf z,\ta)\,dz + \Th_{k-1}^{(s)}(\mathbf z,\ta)\,d\bar{z} \;,\quad \bar{\fI}_{k} = \bar{\T}_{k+1}^{(s)}(\mathbf z,\ta)\,d\bar{z} + \bar{\Th}_{k-1}^{(s)}(\mathbf z,\ta)\,dz \;,
\eeq
where the components $\T_{k+1}^{(s)}(\mathbf z,\ta)$, $\Th_{k-1}^{(s)}(\mathbf z,\ta)$ and their complex conjugates are the level-$k$ conserved currents of the integrable theory, deformed according to the generalised change of variables (\ref{eq:JacobianTTs}).  They fulfil the continuity equations \beq
\label{eq:conteqTTs}
\bar\partial \T_{k+1}^{(s)}(\mathbf z,\ta) = \partial\Th_{k-1}^{(s)}(\mathbf z,\ta) \;,\quad \partial\bar\T_{k+1}^{(s)}(\mathbf z,\ta) = \bar\partial \bar\Th_{k-1}^{(s)}(\mathbf z,\ta) \;.
\eeq
Using the strategy described in section \ref{sec:TTbarHC}, we perform the change of variables (\ref{eq:JacobianTTs}) in (\ref{eq:undef1forms}) and obtain
\beqa
\fI_k &=& \frac{ \T_{k+1}(\mathbf w(\mathbf z)) + 2\ta\bigl(\T_{k+1}(\mathbf w(\mathbf z))\,\Th_{s-1}(\mathbf w(\mathbf z)) -\Th_{k-1}(\mathbf w(\mathbf z))\,\T_{s+1}(\mathbf w(\mathbf z))\bigr) }{\Delta^{(s)}(\mathbf w(\mathbf z))}\,dz  \notag \\ &+& \frac{\Th_{k-1}(\mathbf w(\mathbf z)) + 2\ta\bigl(\Th_{k-1}(\mathbf w(\mathbf z))\,\bar\Th_{s-1}(\mathbf w(\mathbf z)) -\T_{k+1}(\mathbf w(\mathbf z))\,\bar\T_{s+1}(\mathbf w(\mathbf z))\bigr)}{\Delta^{(s)}(\mathbf w(\mathbf z))}\,d\bar z \;, \notag \\
\label{eq:def1formsTTs}
\eeqa
from which the components of the deformed currents can be extracted
\beqa
\label{eq:defcompTTs}
\T_{k+1}^{(s)}(\mathbf z,\ta) &=& \frac{ \T_{k+1}(\mathbf w(\mathbf z)) + 2\ta\bigl(\T_{k+1}(\mathbf w(\mathbf z))\,\Th_{s-1}(\mathbf w(\mathbf z)) -\Th_{k-1}(\mathbf w(\mathbf z))\,\T_{s+1}(\mathbf w(\mathbf z))\bigr) }{\Delta^{(s)}(\mathbf w(\mathbf z))} \;, \notag \\
\Th_{k-1}^{(s)}(\mathbf z,\ta) &=& \frac{\Th_{k-1}(\mathbf w(\mathbf z)) + 2\ta\bigl(\Th_{k-1}(\mathbf w(\mathbf z))\,\bar\Th_{s-1}(\mathbf w(\mathbf z)) -\T_{k+1}(\mathbf w(\mathbf z))\,\bar\T_{s+1}(\mathbf w(\mathbf z))\bigr)}{\Delta^{(s)}(\mathbf w(\mathbf z))} \;. \notag \\
\eeqa
Let us consider the following combinations of the components of level-$k$ conserved currents
\beqa
\label{eq:calI}
\cI_k(\mathbf w) = -\bigl(\T_{k+1}(\mathbf w) + \Th_{k-1}(\mathbf w)\bigr) \;&,&\quad \bar\cI_k(\mathbf w) = -\bigl(\bar\T_{k+1}(\mathbf w) + \bar\Th_{k-1}(\mathbf w)\bigr) \;, \\
\cI_k^{(s)}(\mathbf z,\ta) = -\bigl(\T_{k+1}^{(s)}(\mathbf z,\ta) + \Th_{k-1}^{(s)}(\mathbf z,\ta)\bigr) \;&,&\quad \bar\cI_k^{(s)}(\mathbf z,\ta) = -\bigl(\bar\T_{k+1}^{(s)}(\mathbf z,\ta) + \bar\Th_{k-1}^{(s)}(\mathbf z,\ta)\bigr) \;. \notag \\
\label{eq:calIdef}
\eeqa
Then, following the standard convention, the level-$k$ Hamiltonian and momentum density are\footnote{In the following, ``c.c.'' denotes the replacement $\lbrace\T(\mathbf z,\ta),\Th(\mathbf z,\ta) \rbrace \to \lbrace\bar\T(\mathbf z,\ta),\bar\Th(\mathbf z,\ta)\rbrace$.}
\beqa
\mathcal{H}_k^{(s)}(\mathbf z,\ta) &=& \cI_k^{(s)}(\mathbf z,\ta) + \bar\cI_k^{(s)}(\mathbf z,\ta)
\notag \\ &=&
\frac{1}{\Delta^{(s)}(\mathbf w(\mathbf z))} \biggl[ \mathcal{H}_k(\mathbf w(\mathbf z)) + 2\ta \bigl( \mathcal{T}_{+,k}(\mathbf w(\mathbf z))\, \bar{\mathcal{T}}_{-,s}(\mathbf w(\mathbf z)) + \text{c.c.} \bigr) \biggr] \;,
\label{eq:Hks}
\eeqa
\beqa
\mathcal{P}_k^{(s)}(\mathbf z,\ta) &=& \cI_k^{(s)}(\mathbf z,\ta) - \bar\cI_k^{(s)}(\mathbf z,\ta)
\notag \\ &=&
\frac{1}{\Delta^{(s)}(\mathbf w(\mathbf z))} \biggl[ \mathcal{P}_k(\mathbf w(\mathbf z)) +2\ta \bigl( \mathcal{T}_{-,k}(\mathbf w(\mathbf z))\, \bar{\mathcal{T}}_{-,s}(\mathbf w(\mathbf z)) - \text{c.c.} \bigr) \biggr] \;,
\label{eq:Pks}
\eeqa
where
\beqa
\mathcal{H}_k(\mathbf w) = \cI_k(\mathbf w) + \bar\cI_k(\mathbf w)
\;,\quad
\mathcal{P}_k(\mathbf w) = \cI_k(\mathbf w) - \bar\cI_k(\mathbf w)
\;,
\eeqa
are the undeformed Hamiltonian and momentum density, and the quantities $\mathcal{T}_{\pm,n}$ and $\bar{\mathcal{T}}_{\pm,n}$ correspond to the combinations
\beq
\label{eq:defcalT}
\mathcal{T}_{\pm,n}(\mathbf w) = \T_{n+1}(\mathbf w) \pm \bar\Th_{n-1}(\mathbf w) \;,\quad \bar{\mathcal{T}}_{\pm,n}(\mathbf w) = \bar\T_{n+1}(\mathbf w) \pm \Th_{n-1}(\mathbf w) \;,
\eeq
Integrating (\ref{eq:Hks}) and (\ref{eq:Pks}) we find
\beqa
\int \mathcal{H}_k^{( s)}(\mathbf z,\ta)\,dz\wedge d\bar z &=& \int \biggl[ \mathcal{H}_k(\mathbf w) + 2\ta \bigl( \mathcal{T}_{+,k}(\mathbf w)\,\bar{\mathcal{T}}_{-,s}(\mathbf w) + \text{c.c.} \bigr) \biggr]\,dw\wedge d\bar w  \notag \\ &=&
\int \mathcal{H}_k(\mathbf w)\,dw\wedge d\bar w - 2\ta \int (\fI_k + \bar{\fI}_k) \wedge (\fI_s - \bar{\fI}_s) \;,\label{eq:dressingH}
\eeqa
\beqa
\int \mathcal{P}_k^{(s)}(\mathbf z,\ta)\,dz\wedge d\bar z &=& \int \biggl[ \mathcal{P}_k(\mathbf w) + 2\ta \bigl( \mathcal{T}_{-,k}(\mathbf w)\,\bar{\mathcal{T}}_{-,s}(\mathbf w) - \text{c.c.} \bigr) \biggr]\,dw\wedge d\bar w \notag \\ &=&
\int \mathcal{P}_k(\mathbf w)\,dw\wedge d\bar w - 2\ta \int (\fI_k - \bar{\fI}_k) \wedge (\fI_s - \bar{\fI}_s) \;. \label{eq:dressingP}
\eeqa
We now interpret the result (\ref{eq:dressingH}) as follows: $\mathcal{H}_k^{( s)}(\mathbf z,\ta)\,dz\wedge d\bar z$ coincides with the corresponding bare quantity $\mathcal{H}_k(\mathbf w)\,dw\wedge d\bar w$ deformed by the operator 
\beq
\label{eq:pertoperform}
{\bf \Phi}_{k,s}(\mathbf w)\,dw\wedge d\bar w = -2\,(\fI_k + \bar{\fI}_k) \wedge (\fI_s - \bar{\fI}_s) \;,
\eeq
\beq 
\label{eq:pertoper}
{\bf \Phi}_{k,s}(\mathbf w) = 2\(\mathcal{T}_{+,k}(\mathbf w)\,\bar{\mathcal{T}}_{-,s}(\mathbf w) + \bar{\mathcal{T}}_{+,k}(\mathbf w)\,\mathcal{T}_{-,s}(\mathbf w)\) \;,
\eeq
together with a non-trivial dressing given by the change of variables (\ref{eq:JacobianTTs}). In the $s=1$ case, {\it i.e.} the $\TbT$ example, the operator (\ref{eq:pertoper}) associated to the $k=1$ Hamiltonian becomes
\beq
{\bf \Phi}_{1,1}(\mathbf w) = \T_2(\mathbf w)\,\bar\T_2(\mathbf w) - \Th_0(\mathbf w)\,\bar\Th_0(\mathbf w) \;,
\eeq
which, but for the change of coordinates, coincides with the bare $\TbT$ operator. In analogy with the $\TbT$ result \cite{Dubovsky:2017cnj,Conti:2018tca}, one may be tempted to interpret (\ref{eq:pertoper}) as the perturbing operator of the level-$k$ Hamiltonian.  However, the coordinate transformation (\ref{eq:JacobianTTs}) also introduces $O(\tau)$ corrections which can, in principle, completely spoil this naive picture. In addition, even when the initial theory is a CFT and the bare operator (\ref{eq:pertoper}) is completely symmetric in $s$ and $k$
\beq
{\bf \Phi}_{k,s}(\mathbf w)  = \T_{k+1}(w)\,\bar\T_{s+1}(\bar w)+ \bar\T_{k+1}(\bar w)\,\T_{s+1}(w)\;, \label{eq:TTs}
\eeq
the change of variables spoils the $s \leftrightarrow k$ symmetry, since it involves only the level-$s$ currents.
Examples of this phenomena  will neatly emerge from the study of the $\Bs \le 0$ deformations of the free massless boson theory.\\
From the explicit  examples  discussed in section \ref{sec:classBurgers}, it will clearly emerge that the $\ta$ deformations of the original momentum and Hamiltonian, obtained using the change of variables, do not always coincide with the generators of space and time translations, in the cartesian $\mathbf x=(x^1,x^2)=(x,t)$ directions. Although all the deformed charges
\beq
\label{eq:clint2}
I^{(\Bs)}_\Bk(R,\ta) = \int_0^R \cI^{(\Bs)}_\Bk({\bf x},\tau)\,dx \;,\quad \bar I^{(\Bs)}_\Bk(R,\ta) = \int_0^R \bar{\cI}^{(\Bs)}_\Bk({\bf x},\ta)\,dx \;,
\eeq
are conserved in $t$, they evolve the system along ``generalised space-time'' directions, which differ from the original undeformed ones for $\Bk\neq\Bs$.
Due to this fact, the Lagrangians associated to the deformed EoMs do not, in general, correspond to the deformed Hamiltonians. Therefore, these theories are more complicated compared to the $\TbT$ and $\JbT$ examples studied in the previous literature.\\
One of the main objectives of the following sections will be the identification of the additional scattering phase factors needed for the  characterisation of the finite-size quantum spectrum of the deformed charges (\ref{eq:clint2}).
\subsection{Deformations related to charges with negative spin $\Bs<0$}
\label{sec:TTsmsformal}
\noindent
To gain precise information about the spectrum, it turns out to be particularly convenient to first extend the current setup to the $\Bs<0$ cases. In order to define perturbations induced by higher conserved currents with negative spin $\Bs$, we replace $s\rightarrow s'$ in the definition of the generalised Jacobian (\ref{eq:generalJacobianHC1}) and (\ref{eq:generalJacobianHC2}), obtaining
\beqa
\label{eq:generalJacobianHC3}
\mathcal{J}^{(s')} &=& 
\frac{1}{\Delta^{(s')}(\mathbf w)}\(\begin{array}{cc}
1+2\ta\,\bar\T_{s+1}(\mathbf w) & -2\ta\,\bar\Th_{s-1}(\mathbf w) \\
-2\ta\,\Th_{s-1}(\mathbf w) & 1+2\ta\,\T_{s+1}(\mathbf w)
\end{array}\) \;, \\
\bigl(\mathcal{J}^{(s')}\bigr)^{-1} &=& 
\(\begin{array}{cc}
1+2\ta\,\T_{s+1}(\mathbf w) & 2\ta\,\bar\Th_{s-1}(\mathbf w) \\
2\ta\,\Th_{s-1}(\mathbf w) & 1+2\ta\,\bar\T_{s+1}(\mathbf w)
\end{array}\)  \;,
\label{eq:generalJacobianHC4}
\eeqa
with $s'=\Bs<0$ , $s=|\Bs|$ and
\beqa
\Delta^{(s')}(\mathbf w)=\bigl(1+2\ta\,\T_{s+1}(\mathbf w)\bigr)\bigl(1+2\ta\,\bar\T_{s+1}(\mathbf w)\bigr)-4\ta^2\,\Th_{s-1}(\mathbf w)\,\bar\Th_{s-1}(\mathbf w) \;.
\eeqa
We arrived  to (\ref{eq:generalJacobianHC3})--(\ref{eq:generalJacobianHC4}) by implementing  the  spin-flip  symmetry (see, for example, \cite{Smirnov:2016lqw})
\beq
\label{eq:negsId}
\Th_{s'-1} = \bar\T_{s+1} \;,\quad \T_{s'+1} = \bar\Th_{s-1} \;,
\eeq
which corresponds to the following reflection property at the level of the $1$-forms
\beq  
\label{eq:classrefl1form}
\fI_{s'} = \bar{\fI}_s \;,\quad \bar{\fI}_{s'} = \fI_s \;,
\eeq
or, equivalently
\beq
\label{eq:classrefl}
\cI_{s'} = \bar{\cI}_s \;,\quad \bar{\cI}_{s'} = \cI_s \;.
\eeq
Using the continuity equations (\ref{eq:contequndef}), it is easy to verify that (\ref{eq:generalJacobianHC4}) fulfils again the conditions (\ref{eq:mixder}), therefore it defines a consistent field-dependent change of variables. Repeating the computations (\ref{eq:def1formsTTs})--(\ref{eq:dressingP}) using (\ref{eq:generalJacobianHC3})--(\ref{eq:generalJacobianHC4}) one finds that  (\ref{eq:defcompTTs}) become
\beqa 
\label{eq:defcompTTnegs}
\T_{k+1}^{(s')}(\mathbf z,\ta) &=& \frac{\T_{k+1}(\mathbf w(\mathbf z)) +  2\ta\(\T_{k+1}(\mathbf w(\mathbf z))\,\bar\T_{s+1}(\mathbf w(\mathbf z)) - \Th_{k-1}(\mathbf w(\mathbf z))\,\bar\Th_{s-1}(\mathbf w(\mathbf z))\)}{\Delta^{(s')}(\mathbf w(\mathbf z))} \;, \notag \\
\Th_{k-1}^{(s')}(\mathbf z,\ta) &=& \frac{\Th_{k-1}(\mathbf w(\mathbf z)) +  2\ta\(\Th_{k-1}(\mathbf w(\mathbf z))\,\T_{s+1}(\mathbf w(\mathbf z)) - \T_{k+1}(\mathbf w(\mathbf z))\,\Th_{s-1}(\mathbf w(\mathbf z))\)}{\Delta^{(s')}(\mathbf w(\mathbf z))} \;, \notag \\
\eeqa
while (\ref{eq:dressingH})--(\ref{eq:dressingP}) become
\beqa
\int \mathcal{H}_k^{(s')}(\mathbf z,\ta)\,dz\wedge d\bar z &=& \int \biggl[ \mathcal{H}_k(\mathbf w) -
2\ta \bigl( \mathcal{T}_{+,k}(\mathbf w)\, \bar{\mathcal{T}}_{-,s}(\mathbf w) + \text{c.c.} \bigr) \biggr]\,dw\wedge d\bar w  \nonumber \\
&=&\int \mathcal{H}_k(\mathbf w)\,dw\wedge d\bar w + 2\ta \int (\fI_k + \bar{\fI}_k) \wedge (\fI_s - \bar{\fI}_s) \;,
\label{eq:dressingHnegs}
\eeqa
\beqa
\int \mathcal{P}_k^{(s')}(\mathbf z,\ta)\,dz\wedge d\bar z &=& \int \biggl[ \mathcal{P}_k(\mathbf w) -
2\ta \bigl( \mathcal{T}_{-,k}(\mathbf w)\, \bar{\mathcal{T}}_{-,s}(\mathbf w) - \text{c.c.} \bigr) \biggr]\,dw\wedge d\bar w  \nonumber \\
&=&\int \mathcal{P}_k(\mathbf w)\,dw\wedge d\bar w + 2\ta \int (\fI_k - \bar{\fI}_k) \wedge (\fI_s - \bar{\fI}_s) \;,
\label{eq:dressingPnegs}
\eeqa
which are formally equal to (\ref{eq:dressingH}) and (\ref{eq:dressingP}), except for the sign of $\ta$.\\
However, the positive and negative spin sectors are deeply different, especially in what concerns the non-zero momentum states. They are not simply related by a change of sign in the coupling constant $\tau$ as the comparison between (\ref{eq:dressingH})--(\ref{eq:dressingP}) and (\ref{eq:dressingHnegs})--(\ref{eq:dressingPnegs}) would naively suggests. In fact, the change of variables and the corresponding perturbing operators are different. By studying in detail the $\Bs<0$ perturbations of the massless free boson model (see  section \ref{sec:classBurgers} below), the difference with respect to the $\Bs=1$ perturbation, {\it i.e.} the $\TbT$, clearly emerges.
\subsection{The classical Burgers-type equations}
\label{sec:classBurgers}
In this section, we consider deformations of the massless free boson theory induced by the coordinate transformations (\ref{eq:generalJacobianHC3})--(\ref{eq:generalJacobianHC4}), and we derive the higher conserved currents of the deformed models. As already discussed in section \ref{sec:TTbarHC}, the most general level-$\Bk$ current of the hierarchy can be expressed in the form (\ref{eq:genericcurrent}). While the structure of the deformed currents does not emerge clearly by working with the general combination (\ref{eq:genericcurrent}), we observed that the subset (\ref{eq:NG1formcomp1}) is analytically much easier to treat since it does not mix with the others. This property allows to obtain compact expressions for the deformed currents which are formally identical to the exact quantum results of section \ref{sec:generalisedBurgers}.\\
Using (\ref{eq:NG1formcomp1}) in (\ref{eq:generalJacobianHC3})--(\ref{eq:generalJacobianHC4}), the coordinate transformations read explicitly
\beqa
\mathcal{J}^{(s')} &=& 
\label{eq:Jacobiannegs}
\(\begin{array}{cc}
\frac{1}{1-\ta\,(\partial_{w}\phi)^{s+1}} & 0 \\
0 & \frac{1}{1-\ta\,(\partial_{\bar w}\phi)^{s+1}}
\end{array}\) \;, \\
\label{eq:invJacobiannegs}
\bigl(\mathcal{J}^{(s')}\bigr)^{-1} &=& 
\(\begin{array}{cc}
1-\ta\,(\partial_{w}\phi)^{s+1} & 0 \\
0 & 1-\ta\,(\partial_{\bar w}\phi)^{s+1}
\end{array}\) \;.
\eeqa
Repeating the same computation performed in section \ref{sec:TTbarHCfreebos}, we first express $(\partial_w\phi,\partial_{\bar w}\phi)^{\rm T}$ in terms of $(\partial\phi,\bar\partial\phi)^{\rm T}$, by solving the set of equations
\beq
\label{eq:systemTTs}
\(\begin{array}{c}
\partial\phi \\
\bar\partial\phi
\end{array}\) = 
\mathcal{J}^{(s')}
\(\begin{array}{c}
\partial_w\phi \\
\partial_{\bar w}\phi
\end{array}\) \quad\longleftrightarrow\quad 
\begin{cases}
\partial\phi = \frac{\partial_w\phi}{1-\ta\(\partial_w\phi\)^{s+1}} \\
\bar\partial\phi = \frac{\partial_{\bar w}\phi}{1-\ta\(\partial_{\bar w}\phi\)^{s+1}}
\end{cases} \;.
\eeq
The solutions to (\ref{eq:systemTTs}) can be written in terms of generalized hypergeometric functions for any value of $s'=-s$ as
\beq
\label{eq:solsystemTTs}
\partial_w\phi = \tilde{F}_s\(-2\ta\frac{(s+1)^{s+1}}{s^s}\frac{(\partial\phi)^{s+1}}{2}\)\partial\phi \;,\quad \partial_{\bar w}\phi = \tilde{F}_s\(-2\ta\frac{(s+1)^{s+1}}{s^s}\frac{(\bar\partial\phi)^{s+1}}{2}\)\bar\partial\phi \;, 
\eeq
with
\beq  
\label{eq:hypergeo1}
\tilde{F}_n(x) = {}_{n}F_{n-1}\(\frac{1}{n+1},\dots,\frac{n}{n+1}; \frac{2}{n},\dots,\frac{n-1}{n},\frac{n+1}{n};x\) \;,\quad (n\in\mathbb{N}-\left\lbrace 0 \right\rbrace) \;.
\eeq
Plugging (\ref{eq:solsystemTTs}) into the second expression of (\ref{eq:JacobianTTs}) (with the replacement $s\to s'=-s$), one finds the differential map
\begin{gather}
\partial_w f = \tilde{F}_s\(-2\ta\frac{(s+1)^{s+1}}{s^s}\frac{(\partial\phi)^{s+1}}{2}\)\partial f\;,\quad \partial_{\bar w}f = \tilde{F}_s\(-2\ta\frac{(s+1)^{s+1}}{s^s}\frac{(\bar\partial\phi)^{s+1}}{2}\)\bar\partial f\;,
\label{eq:diffchangeTTs}
\end{gather}
$\forall f:\mathbb{R}^2\rightarrow \mathbb{R}$. From (\ref{eq:diffchangeTTs}), it follows immediately that the deformed EoMs are
\beq
\label{eq:TTsEoMs}
\partial\bar\partial\phi = 0 \;,
\eeq
which reflects the fact that the $\Bs<0$ perturbations of CFT's do not mix the holomorphic and anti-holomorphic derivatives, as already emerged from (\ref{eq:systemTTs}). Using the technique described in section \ref{sec:TTbarHC}, we can now derive the deformed currents. Plugging the differential map (\ref{eq:diffchangeTTs}) into (\ref{eq:defcompTTnegs}), we obtain
\beqa
\label{eq:compnegs}
\T_{s+1}^{(s')}(\mathbf z,\ta) &=& -\frac{s}{2\ta\,(s+1)}\[ -1 + F_s\(-2\ta\frac{(s+1)^{s+1}}{s^s}\frac{(\partial\phi)^{s+1}}{2}\)\] \;,\quad \Th_{s-1}^{(s')}(\mathbf z,\ta) = 0 \;, \notag \\
\bar\T_{s+1}^{(s')}(\mathbf z,\ta) &=& -\frac{s}{2\ta\,(s+1)}\[ -1 + F_s\(-2\ta\frac{(s+1)^{s+1}}{s^s}\frac{(\bar\partial\phi)^{s+1}}{2}\)\] \;,\quad \bar\Th_{s-1}^{(s')}(\mathbf z,\ta) = 0 \;, \notag \\
\eeqa
with 
\beq
\label{eq:hypergeo2}
F_n(x)= {}_{n}F_{n-1}\( -\frac{1}{n+1},\frac{1}{n+1},\dots,\frac{n-1}{n+1};\frac{1}{n},\frac{2}{n},\dots,\frac{n-1}{n};x \) \;,\quad (n\in\mathbb{N}-\lbrace 0 \rbrace) \;,
\eeq
where we used the following relation between generalised hypergeometric functions\footnote{Relation (\ref{eq:hyperrel}) can be easily checked expanding at every perturbative order around $x=0$.}
\beq  
\label{eq:hyperrel}
F_n(x) = \frac{1}{n}\(-1 + \frac{(n+1)^{n+2}}{(n+1)^{n+1}+n^n\,x\(\tilde{F}_n(x)\)^{n+1}} \) \;,
\eeq
to trade $\tilde{F}_n(x)$ with $F_n(x)$.
From (\ref{eq:TTsEoMs}), it follows that $\phi(\mathbf z,\ta) = \varphi(z,\ta) + \bar\varphi(\bar z,\ta)$ and therefore $\T_{s+1}^{(s')}(\mathbf z,\ta)$ and $\bar\T_{s+1}^{(s')}(\mathbf z,\ta)$ depend only on $z$ and $\bar z$, respectively. Again we observe that, using (\ref{eq:diffchangeTTs}) in (\ref{eq:Jacobiannegs}), the Jacobian can be rewritten in terms of the deformed components (\ref{eq:compnegs}) as
\beq 
\mathcal{J}^{(s')} = 
\(\begin{array}{cc}
1-2\ta\,\T_{s+1}^{(s')}(\mathbf z,\ta) & -2\ta\,\bar\Th_{s-1}^{(s')}(\mathbf z,\ta) \\
-2\ta\,\Th_{s-1}^{(s')}(\mathbf z,\ta) & 1-2\ta\,\bar\T_{s+1}^{(s')}(\mathbf z,\ta)
\end{array}\) \;,
\eeq
which confirms that the coordinate transformation is invertible.\\
In terms of the quantities (\ref{eq:calI}) and (\ref{eq:calIdef}), (\ref{eq:compnegs}) can be more transparently written as
\beqa
\label{eq:cIdens}
\cI_s^{(s')}(\mathbf z,\ta) &=& \frac{s}{2\ta\,(s+1)}\[ -1 + F_s\(-2\ta\frac{(s+1)^{s+1}}{s^s}\,\cI_s(\mathbf z)\)\] \;, \notag \\
\bar\cI_s^{(s')}(\mathbf z,\ta) &=& \frac{s}{2\ta\,(s+1)}\[ -1 + F_s\(-2\ta\frac{(s+1)^{s+1}}{s^s}\,\bar\cI_s(\mathbf z)\)\] \;.
\eeqa
Quite remarkably, the latter expressions are solutions to simple algebraic equations of the form
\beq  
\label{eq:algeqs}
\cI_s^{(s')}(\mathbf z,\ta) = \frac{\cI_s(\mathbf z)}{\(1+2\ta\,\cI_s^{(s')}(\mathbf z,\ta)\)^s} \;,\quad \bar\cI_s^{(s')}(\mathbf z,\ta) = \frac{\bar\cI_s(\mathbf z)}{\(1+2\ta\,\bar\cI_s^{(s')}(\mathbf z,\ta)\)^s} \;.
\eeq
Moreover, one can show that the combinations $\cI_k^{(s')}(\mathbf z,\ta)$ and $\bar\cI_k^{(s')}(\mathbf z,\ta)$ of generic level-$k$ deformed currents are
\beqa 
\label{eq:cIdenslev1}
\cI_k^{(s')}(\mathbf z,\ta) = \cI_k(\mathbf z)\, \left\lbrace 1+\frac{s}{s+1}\[-1+F_s\(-2\ta\frac{(s+1)^{s+1}}{s^s}\,\cI_s(\mathbf z)\)\]\right\rbrace^{-k} \;, \notag \\
\bar\cI_k^{(s')}(\mathbf z,\ta) = \bar\cI_k(\mathbf z)\, \left\lbrace 1+\frac{s}{s+1}\[-1+F_s\(-2\ta\frac{(s+1)^{s+1}}{s^s}\,\bar\cI_s(\mathbf z)\)\]\right\rbrace^{-k} \;,
\eeqa
and fulfil the following equations
\beq
\label{eq:Iksdens}
\cI_k^{(s')}(\mathbf z,\ta) = \frac{\cI_k(\mathbf z)}{\(1+2\ta\,\cI_s^{(s')}(\mathbf z,\ta)\)^k} \;,\quad \bar\cI_k^{(s')}(\mathbf z,\ta) = \frac{\bar\cI_k(\mathbf z)}{\(1+2\ta\,\bar\cI_s^{(s')}(\mathbf z,\ta)\)^k} \;,
\eeq
which generalise (\ref{eq:algeqs}).\\
Before moving to the next section, let us make a few important remarks:
\begin{enumerate}
    \item The $\TbT$ and the $\JbT$ examples discussed in \cite{Guica:2017lia,Kraus:2018xrn}, taught us that, at least formally, the evolution equations for the quantized spectra already emerge at classical level after replacing the classical densities with their average value over the volume $R$:
    \beqa 
    \cI_k(\mathbf z) \longrightarrow \frac{I_k(R)}{R} = \frac{I_k^{(+)}(R)}{R} \;&,&\quad \bar\cI_k(\mathbf z) \longrightarrow \frac{\bar I_k(R)}{R} = \frac{ I_k^{(-)}(R)}{R} \;, \notag \\
    \cI_k^{(\Bs)}(\mathbf z,\ta) \longrightarrow \frac{I_k^{(\Bs)}(R,\ta)}{R} = \frac{I_k^{(\Bs,+)}(R,\ta)}{R} \;&,&\quad \bar\cI_k^{(\Bs)}(\mathbf z,\ta) \longrightarrow \frac{ \bar I_k^{(\Bs)}(R,\ta)}{R} = \frac{ I_k^{(\Bs,-)}(R,\ta)}{R} \;. \notag \\
    \label{eq:densTocharge}
    \eeqa
    In (\ref{eq:densTocharge}), the labels $(+)$ and $(-)$ stand for the right and left light-cone component of the conserved charges, respectively (cf. section \ref{sec:quantum}). 
    
    Implementing (\ref{eq:densTocharge}) in (\ref{eq:Iksdens}) gives
    \beq
    \label{eq:algnegs}
    I_k^{(s',\pm)}(R,\ta) = \frac{R^k\,I_k^{(\pm)}(R)}{\(R+2\ta\,I_s^{(s',\pm)}(R,\ta)\)^k} \;,
    \eeq
    which coincides with the CFT quantum result (\ref{eq:EsCFTdef}) of section \ref{sec:CFTlimit}.
    \item Although (\ref{eq:algnegs}) were derived  for the $\Bs<0$ case, it is natural to conjecture  that they can be extended also to $\Bs \ge 0$. From the  reflection property (\ref{eq:classrefl}), we find:
    \beq
    \label{eq:algposs}
    I_k^{(s,\pm)}(R,\ta) = \frac{R^k\,I_k^{(\pm)}(R)}{\(R+2\ta\,\bar I_s^{(s,\mp)}(R,\ta)\)^k} \;,
    \eeq
    which again match the $\Bs>0$ CFT quantum result quoted in (\ref{eq:spositive}).
    \item It is straightforward to check that (\ref{eq:algnegs}) is a solution to the generalised Burgers-type equations (\ref{eq:BurgersIpm})--(\ref{eq:sol}) for the deformed quantum spectrum, which hold also for massive models. In fact, as shown in Appendix \ref{Appendix:burgers}, in the CFT case equations (\ref{eq:BurgersIpm})--(\ref{eq:sol}) can be recast into the simpler form
    \beqa
    \label{eq:simplBurgers}
    \partial_\ta I_k^{(s,\pm)}(R,\ta) &=& 2 I_s^{(s,\mp)}(R,\ta)\,\partial_R I_k^{(s,\pm)}(R,\ta) \;,\quad (s>0) \;, \notag \\
    \partial_\ta I_{k'}^{(s',\pm)}(R,\ta) &=& 2 I_{s'}^{(s',\pm)}(R,\ta)\,\partial_R I_{k'}^{(s',\pm)}(R,\ta) \;,\quad (s'=-s<0) \;.
    \eeqa
    \item
    From the point of view of the infinite tower of  conservation laws, the map (\ref{eq:Jacobiannegs})-(\ref{eq:invJacobiannegs}) corresponds to a non trivial $\tau$-dependent mixing of the charges. Every conserved quantity, including the deformations of the Hamiltonian and momentum densities
    \beqa
    \mathcal{H}^{(s')}(\mathbf z,\ta)=\cI_1^{(s')}(\mathbf z,\ta) + \bar\cI_1^{(s')}(\mathbf z,\ta)\,, \\
     \mathcal{P}^{(s')}(\mathbf z,\ta)=\cI_1^{(s')}(\mathbf z,\ta) - \bar\cI_1^{(s')}(\mathbf z,\ta)\,, 
    \eeqa
    are smoothly deformed, leading to a trajectory in the space of models which share the same set of conserved charges. As already mentioned at the end of section \ref{sec:TTspsformal}, a direct consequence of the mixing among the conserved quantities, and of the associated {\it generalised} time variables,  is that the interpretation of 
    \beq
    \label{eq:EPclass}
    E^{(s')}(R,\ta) = \int_0^R \mathcal{H}^{(s')}(\mathbf x,\ta)\, dx\;,\quad P^{(s')}(R,\ta) = \int_0^R \mathcal{P}^{(s')}(\mathbf x,\ta)\, dx \;,
    \eeq
    as the generators of translations in  $t$ and $x$ is no longer valid at $\tau \ne 0$. Formally, the generators of translations in $\mathbf x$  correspond to the unperturbed ($\tau=0$) Hamiltonian and momentum.
    \item The result (\ref{eq:TTsEoMs}) shows that the coordinate transformation (\ref{eq:Jacobiannegs})-(\ref{eq:invJacobiannegs}) is, on the plane, an automorphism of the space of classical solutions of the free boson theory. In addition, the undeformed action is mapped into itself by the change of variables (\ref{eq:Jacobiannegs}):
    \beq
    \int \mathcal{L}(\mathbf w)\,dw\wedge d\bar w = \int \mathcal{L}(\mathbf z)\,dz\wedge d\bar z \;,\quad \mathcal{L}(\mathbf z) = \partial\phi\,\bar\partial\phi \;.
    \eeq
\end{enumerate}
\subsection{Deformations related to charges with spin $\Bs \rightarrow 0$ and the $\JbT$-type models}
\label{sec:s0def}
In this section we shall consider perturbations of the massless free boson model induced by the $U(1)_L\times U(1)_R$ currents defined through
\beqa
J_+(\mathbf w) = -2i\,\T_1(\mathbf w) = i\,\partial_w\phi \;&,&\quad J_-(\mathbf w) = -2i\,\Th_{-1}(\mathbf w) = 0 \;, \notag \\
\bar J_+(\mathbf w) = 2i\,\bar\T_1(\mathbf w) = -i\,\partial_{\bar w}\phi \;&,&\quad \bar J_-(\mathbf w) = 2i\,\bar\Th_{-1}(\mathbf w) = 0 \;,
\label{eq:U1currents}
\eeqa
where $\T_1(\mathbf w)$, $\Th_{-1}(\mathbf w)$ and their complex conjugates corresponds to the case $\Bk=0$ in (\ref{eq:NG1formcomp1}), while the additional factor $i$ in (\ref{eq:U1currents}) is used to write the holomorphic and anti-holomorphic components of the stress-energy tensor in Sugawara form (see \cite{Chakraborty:2018vja})
\beq  
\T_2(\mathbf w) = \frac{1}{2}\(J_+(\mathbf w)\)^2 \;,\quad \bar\T_2(\mathbf w) = \frac{1}{2}\(\bar J_+(\mathbf w)\)^2 \;.
\eeq
Since also the currents (\ref{eq:U1currents}) are components of closed $1$-forms
\beq  
\label{eq:U1form}
\fI_0 = J_+(\mathbf w)\,dw + J_-(\mathbf w)\,d\bar w \;,\quad \bar{\fI}_0 = \bar J_+(\mathbf w)\,d\bar w + \bar J_-(\mathbf w)\,dw \;,
\eeq
we shall use the strategy described in section \ref{sec:TTbarHC} to derive the corresponding deformations. Following the same spirit of the $\Bs>0$ and $\Bs<0$ deformations, the most general change of coordinates built out of the $U(1)$ currents (\ref{eq:U1currents}) is of the form
\beqa
\label{eq:generalJacobianHC5}
\mathcal{J}^{(0)} &=& \frac{1}{\Delta^{(0)}(\mathbf w)}
\(\begin{array}{cc}
1+\ta^{(4)}\,\partial_{\bar w}\phi & -\ta^{(2)}\,\partial_w\phi \\
-\ta^{(3)}\,\partial_{\bar w}\phi & 1+\ta^{(1)}\,\partial_w\phi
\end{array}\) \;, \\
\bigl(\mathcal{J}^{(0)}\bigr)^{-1} &=& 
\(\begin{array}{cc}
1+\ta^{(1)}\,\partial_w\phi & \ta^{(2)}\,\partial_w\phi \\
\ta^{(3)}\,\partial_{\bar w}\phi & 1+\ta^{(4)}\,\partial_{\bar w}\phi
\end{array}\)  \;,
\label{eq:generalJacobianHC6}
\eeqa
where $\ta^{(i)} \;,\; (i=1,\dots 4)$, are four different deformation parameters and
\beqa
\Delta^{(0)}(\mathbf w)=1+\ta^{(1)}\,\partial_w\phi+\ta^{(4)}\,\partial_{\bar w}\phi + \bigl(\ta^{(1)}\ta^{(4)}-\ta^{(2)}\ta^{(3)}\bigr)\,\partial_w\phi\,\partial_{\bar w}\phi \;.
\eeqa
Since the general case is quite cumbersome, we will restrict our analysis to some particular limits. 
\vspace{0.5cm}\\
\fbox{\parbox{6.6cm}{
{\bf Case} $\ta^{(1)}=\ta^{(4)}=0 \;|\; \ta^{(2)}=\ta^{(3)}=-\ta$}}
\vspace{0.5cm}\\
With this choice, (\ref{eq:generalJacobianHC5})--(\ref{eq:generalJacobianHC6}) are reduced to
\begin{gather}
\label{eq:jacs0p}
\mathcal{J}^{(0)} = 
\frac{1}{1-\ta^2\,\partial_w\phi\,\partial_{\bar w}\phi}\( \begin{array}{cc}
1 & \ta\,\partial_w\phi \\
\ta\,\partial_{\bar w}\phi & 1
\end{array} \) \;,  \\
\label{eq:invjacs0p}
\(\mathcal{J}^{(0)}\)^{-1} = 
\( \begin{array}{cc}
1 & -\ta\,\partial_w\phi \\
-\ta\,\partial_{\bar w}\phi & 1
\end{array} \) \;,
\end{gather}
which correspond to the case $s=0$ in (\ref{eq:generalJacobianHC1})--(\ref{eq:generalJacobianHC2}).
Following the procedure described in detail in section \ref{sec:TTbarHCfreebos}, one can easily derive the differential map
\beq
\label{eq:diffchanges0p}
\partial_w f = \partial f - \frac{\ta\,\partial\phi}{1+\ta\,\bar\partial\phi}\bar\partial f \;,\quad \partial_{\bar w} f = \bar\partial f - \frac{\ta\,\bar\partial\phi}{1+\ta\,\partial\phi}\partial f \;,
\eeq
from which the deformed EoMs are
\beq
\partial\bar\partial\phi = \ta\,\frac{\partial\phi\,\bar\partial^2\phi\(1+\ta\,\partial\phi\)+\bar\partial\phi\,\partial^2\phi\(1+\ta\,\bar\partial\phi\)}{1+\ta\( \partial\phi+\bar\partial\phi \)+2\ta^2\,\partial\phi\,\bar\partial\phi} \;.
\eeq
Then, setting $s=0$ in (\ref{eq:defcompTTs}) and using (\ref{eq:diffchanges0p}), we get both the deformed level-$0$ currents
\beq
\label{eq:complev0s0p}
\T_1^{(0)}(\mathbf z,\ta) = -\frac{1}{2}\frac{\partial\phi\( 1 + \ta\,\partial\phi \)}{1+\ta\(\partial\phi+\bar\partial\phi\)} \;,\quad \Th_{-1}^{(0)}(\mathbf z,\ta) = -\frac{1}{2}\frac{\ta\,\partial\phi\,\bar\partial\phi}{1+\ta\(\partial\phi+\bar\partial\phi\)} \;,
\eeq
and the components of the deformed stress-energy tensor
\beq 
\label{eq:complev1s0p}
\T_2^{(0)}(\mathbf z,\ta) = -\frac{1}{2}\,\frac{(\partial\phi)^2\(1+\ta\,\partial\phi\)}{\(1+\ta\,\bar\partial\phi\)\(1+\ta\(\partial\phi+\bar\partial\phi\)\)} \;,\quad \Th_0^{(0)}(\mathbf z,\ta) = -\frac{1}{2}\,\frac{\ta\,\bar\partial\phi\,(\partial\phi)^2}{\(1+\ta\,\bar\partial\phi\)\(1+\ta\(\partial\phi+\bar\partial\phi\)\)} \;.
\eeq
Plugging (\ref{eq:diffchanges0p}) into (\ref{eq:jacs0p}), we notice that the Jacobian can be rewritten in terms of the deformed level-$0$ currents (\ref{eq:complev0s0p}) as
\beq  
\mathcal{J}^{(0)} = 
\(\begin{array}{cc}
1-2\ta\,\bar\Theta_{-1}^{(0)}(\mathbf z,\ta) & -2\ta\,\T_1^{(0)}(\mathbf z,\ta) \\
-2\ta\,\bar\T_1^{(0)}(\mathbf z,\ta) & 1-2\ta\,\Theta_{-1}^{(0)}(\mathbf z,\ta)
\end{array}\) \;,
\eeq
which again confirms that the coordinate transformation is invertible. In terms of the quantities (\ref{eq:calI}) and (\ref{eq:calIdef}), one finds the following relations
\beqa 
\cI_0^{(0)}(\mathbf z,\ta) = \frac{1}{2}\partial\phi = \cI_0(\mathbf z) \;&,&\quad \bar\cI_0^{(0)}(\mathbf z,\ta) = \frac{1}{2}\bar\partial\phi = \bar\cI_0(\mathbf z)\;, \notag \\
\cI_1^{(0)}(\mathbf z,\ta) = \frac{1}{2}\,\frac{(\partial\phi)^2}{1+\ta\,\bar\partial\phi} = \frac{\cI_1(\mathbf z)}{1+2\ta\,\bar\cI_0^{(0)}(\mathbf z,\ta)} \;&,&\quad \bar\cI_1^{(0)}(\mathbf z,\ta) = \frac{1}{2}\,\frac{(\bar\partial\phi)^2}{1+\ta\,\partial\phi} = \frac{\bar\cI_1(\mathbf z)}{1+2\ta\,\cI_0^{(0)}(\mathbf z,\ta)} \;. \notag \\
\eeqa
\fbox{\parbox{6.6cm}{
{\bf Case} $\ta^{(1)}=\ta^{(4)}=-\ta \;|\; 
\ta^{(2)}=\ta^{(3)}=0$}}
\vspace{0.5cm}\\
With this choice (\ref{eq:generalJacobianHC5})--(\ref{eq:generalJacobianHC6}) reduced to
\begin{gather}
\label{eq:jacs0m}
\mathcal{J}^{(0)}=
\frac{1}{\(1-\ta\,\partial_w\phi\)\(1-\ta\,\partial_{\bar w}\phi\)}\(\begin{array}{cc}
1-\ta\,\partial_{\bar w}\phi & 0 \\
0 & 1-\ta\,\partial_w\phi
\end{array}\) \;, \\
\label{eq:invjacs0m}
\(\mathcal{J}^{(0)}\)^{-1}=
\(\begin{array}{cc}
1-\ta\,\partial_w\phi & 0 \\
0 & 1-\ta\,\partial_{\bar w}\phi 
\end{array}\) \;,
\end{gather}
which corresponds to the case $s'=-s=0$ in (\ref{eq:generalJacobianHC3})--(\ref{eq:generalJacobianHC4}). The deformed level-$0$ currents and the components of the deformed stress energy tensor can be immediately read from (\ref{eq:compnegs}), setting $s'=-s = 0$:
\beqa
\label{eq:complev0s0m}
\T_1^{(0)}(\mathbf z,\ta) = -\frac{1}{2}\partial\phi \;&,&\quad \Th_{-1}^{(0)}(\mathbf z,\ta) = 0 \;, \\
\label{eq:complev1s0m}
\T_2^{(0)}(\mathbf z,\ta) = -\frac{1}{2}\frac{(\partial\phi)^2}{1+\ta\,\partial\phi} \;&,&\quad \Th_0^{(0)}(\mathbf z,\ta) = 0 \;,
\eeqa
and, in terms of the quantities (\ref{eq:calIdef}) and (\ref{eq:calI}), from (\ref{eq:cIdenslev1}) one has
\beqa
\cI_0^{(0)}(\mathbf z,\ta) = \frac{1}{2}\partial\phi = \cI_0(\mathbf z) \;&,&\quad \bar\cI_0^{(0)}(\mathbf z,\ta) = \frac{1}{2}\bar\partial\phi = \bar\cI_0(\mathbf z) \;, \notag \\
\cI_1^{(0)}(\mathbf z,\ta) = \frac{1}{2}\frac{(\partial\phi)^2}{1+\ta\,\partial\phi} = \frac{\cI_1(\mathbf z)}{1+2\ta\,\cI_0^{(0)}(\mathbf z,\ta)} \;&,&\quad \bar\cI_1^{(0)}(\mathbf z,\ta) = \frac{1}{2}\frac{(\bar\partial\phi)^2}{1+\ta\,\bar\partial\phi} = \frac{\bar\cI_1(\mathbf z)}{1+2\ta\,\bar\cI_0^{(0)}(\mathbf z,\ta)} \;. \notag \\
\eeqa
\fbox{\parbox{6.3cm}{
{\bf Case} $\ta^{(1)}=\ta^{(3)}=\del \;|\; \ta^{(2)}=\ta^{(4)}=\ta$}}
\vspace{0.5cm}\\
In this particular case (\ref{eq:generalJacobianHC5})--(\ref{eq:generalJacobianHC6}) reduce to
\begin{gather}
\mathcal{J}^{(0)}=
\(\begin{array}{cc}
1 - \del\,\partial\phi & -\ta\,\partial\phi \\
-\del\,\bar\partial\phi & 1-\ta\,\bar\partial\phi
\end{array}\) \;, \notag \\
\(\mathcal{J}^{(0)}\)^{-1}=
\(\begin{array}{cc}
1 + \del\,\partial_w\phi & \ta\,\partial_w\phi \\
\del\,\partial_{\bar w}\phi & 1+\ta\,\partial_{\bar w}\phi
\end{array}\) \;,
\label{eq:s0jacobian}
\end{gather}
which corresponds to an explicit change of variables $\mathbf w(\mathbf z)$ of the form
\beq 
\label{eq:s0change}
w = z - \del\phi \;,\quad \bar w = \bar z - \ta\phi \;.
\eeq
In \cite{Conti:2018tca}, it was observed that the $\TbT$-deformed solutions fulfil a non-linear evolution equation. Its extension to generic spin $\Bs$ is, in complex coordinates $\mathbf z$
\beq 
\label{eq:gensBurgersClass}
\partial_\ta \phi^{(s)}(\mathbf z,\ta) + (\partial_{\tau} z) \,\partial\phi^{(s)}(\mathbf z,\ta) + (\partial_{\tau}\bar z)\,\bar\partial\phi^{(s)}(\mathbf z,\ta) = 0 \;,
\eeq
with\footnote{Starting from (\ref{eq:solrel}), {\it i.e.} $\phi^{(s)}({\bf z},\tau) = \phi({\bf w}({\bf z}^{(s)}),0)$, we let ${\bf z}$ depend on $\tau$ such that $\frac{d}{d\tau}{\bf w}({\bf z}^{(s)}) = 0$: $$\frac{d}{d\tau}\phi^{(s)}({\bf z},\tau) = 0 = \partial_{\tau}\phi^{(s)}({\bf z},\tau) + (\partial_{\tau} z)\partial_{z}\phi^{(s)}({\bf z},\tau)+ (\partial_{\tau} \bar{z})\partial_{\bar z}\phi^{(s)}({\bf z},\tau)\;.$$ From the definition of the change of coordinates (\ref{eq:JacobianTTs}) $$z = w({\bf z}) + 2\tau \intop^{\bf w({\bf z})}\left(\bar{\Th}_{s-1}\left({\bf w}\right) dw + \bar{\T}_{s+1}\left({\bf w}\right) d\bar{w}\right) \;,\quad \bar z = \bar w({\bf z}) + 2\tau \intop^{\bf w({\bf z})}\left(\T_{s+1}\left({\bf w}\right) dw + \Th_{s-1}\left({\bf w}\right) d\bar{w}\right) \;,$$ using the definition (\ref{eq:undef1forms}) and the coordinate independence property of differential forms, we arrive to (\ref{eq:TTschange}).}
\beq 
\label{eq:TTschange}
\partial_\ta z = 2\int^{\mathbf z} \bar{\fI}_\Bs \;,\quad \partial_\ta \bar z = 2\int^{\mathbf z} \fI_\Bs \;.
\eeq
The multi-parameter variant of (\ref{eq:gensBurgersClass})  associated to the coordinate transformation (\ref{eq:generalJacobianHC5}) is, instead
\beq 
\label{eq:multiBurgersclass}
\partial_{\ta^{(i)}} \phi(\mathbf z,\vec\ta) + \bigl(\partial_{\ta^{(i)}} z\bigr)\,\partial\phi(\mathbf z,\vec\ta) + \bigl(\partial_{\ta^{(i)}} \bar z\bigr)\,\bar\partial\phi(\mathbf z,\vec\ta) = 0 \;,\quad \vec\ta = \lbrace \ta^{(i)} \rbrace \;.
\eeq
In general, equations (\ref{eq:gensBurgersClass})--(\ref{eq:multiBurgersclass}) cannot be explicitly integrated, however, in the case of the change of coordinates (\ref{eq:s0change}), equations (\ref{eq:multiBurgersclass}) become a set of inviscid Burgers equations for the function $\phi(z,\bar z,\ta,\del)$ in the variables $z$, $\bar z$, $\ta$ and $\del$
\beq
\partial_\ta\phi(z,\bar z,\ta,\del) + \frac{1}{2}\bar\partial\(\phi^2(z,\bar z,\ta,\del)\) = 0 \;,\quad \partial_\del\phi(z,\bar z,\ta,\del) + \frac{1}{2}\partial\(\phi^2(z,\bar z,\ta,\del)\) = 0 \;,
\eeq 
whose solution can be expressed, in implicit form, as
\beq  
\phi(z,\bar z,\ta,\del) = \phi(z-\del\phi,\bar z -\ta\phi,0,0) \;.
\eeq
Using the method discussed in sections \ref{sec:TTbarHC} and \ref{sec:TTsHC}, we can write down the deformed EoMs, {\it i.e.}
\beq  
\label{eq:s0EoMs}
\partial\bar\partial\phi = -\frac{\ta\,\partial\phi\,\bar\partial^2\phi\(1-\del\,\partial\phi\)+\del\,\bar\partial\phi\,\partial^2\phi\(1-\ta\,\bar\partial\phi\)}{1-\del\,\partial\phi-\ta\,\bar\partial\phi+2\ta\del\,\partial\phi\,\bar\partial\phi} \;,
\eeq
and the tower of deformed higher conserved currents from the undeformed ones. In particular, the components of the deformed $U(1)$ currents (\ref{eq:U1currents}) are
\beqa 
\label{eq:U1currentsdef}
J_+^{(0)}(\mathbf z,\ta,\del) = i\,\partial\phi\,\frac{1-\del\,\partial\phi}{1-\del\,\partial\phi-\ta\,\bar\partial\phi} \;&,&\quad J_-^{(0)}(\mathbf z,\ta,\del) =  -i\,\frac{\del\,\partial\phi\,\bar\partial\phi}{1-\del\,\partial\phi-\ta\,\bar\partial\phi} \;, \notag \\
\bar J_+^{(0)}(\mathbf z,\ta,\del) = -i\,\bar\partial\phi\,\frac{1-\ta\,\partial\phi}{1-\del\,\partial\phi-\ta\,\bar\partial\phi} \;&,&\quad \bar J_-^{(0)}(\mathbf z,\ta,\del) = i\,\frac{\ta\,\partial\phi\,\bar\partial\phi}{1-\del\,\partial\phi-\ta\,\bar\partial\phi} \;,
\eeqa
while the components of the deformed stress-energy tensor are
\beqa 
\label{eq:U1stressdef}
\T_2^{(0)}(\mathbf z,\ta,\del) = -\frac{1}{2}\,(\partial\phi)^2\,\frac{1-\del\,\partial\phi}{(1-\del\,\partial\phi-\ta\,\bar\partial\phi)^2} \;,\quad \Th_0^{(0)}(\mathbf z,\ta,\del) = \frac{1}{2}\,\frac{\del\,(\partial\phi)^2\,\bar\partial\phi}{(1-\del\,\partial\phi-\ta\,\bar\partial\phi)^2} \;, \notag \\
\bar\T_2^{(0)}(\mathbf z,\ta,\del) = -\frac{1}{2}\,(\bar\partial\phi)^2\,\frac{1-\ta\,\bar\partial\phi}{(1-\del\,\partial\phi-\ta\,\bar\partial\phi)^2} \;,\quad \bar\Th_0^{(0)}(\mathbf z,\ta,\del) = \frac{1}{2}\,\frac{\ta\,(\bar\partial\phi)^2\,\partial\phi}{(1-\del\,\partial\phi-\ta\,\bar\partial\phi)^2} \;. \notag \\
\eeqa
Therefore, the deformed Hamiltonian and momentum density are
\beqa
\label{eq:s0Ham}
\mathcal{H}^{(0)}(\mathbf z,\ta,\del) &=& 
\cI_1^{(0)}(\mathbf z,\ta,\del) + \bar\cI_1^{(0)}(\mathbf z,\ta,\del) \notag \\ &=& -\frac{\partial\phi\,\bar\partial\phi\( \del\,\partial\phi + \ta\,\bar\partial\phi \)-(\partial\phi)^2(1-\del\,\partial\phi)-(\bar\partial\phi)^2(1-\ta\,\bar\partial\phi)}{2\( 1-\del\,\partial\phi-\ta\,\bar\partial\phi \)^2} \;,
\eeqa
\beqa
\label{eq:s0calP}
\mathcal{P}^{(0)}(\mathbf z,\ta,\del) &=& 
\cI_1^{(0)}(\mathbf z,\ta,\del) - \bar\cI_1^{(0)}(\mathbf z,\ta,\del) \notag \\ &=& \frac{(\partial\phi + \bar\partial\phi)\(\partial\phi\,(1-\del\,\partial\phi) -\bar\partial\phi\,(1-\ta\,\bar\partial\phi)\)}{2\( 1-\del\,\partial\phi-\ta\,\bar\partial\phi \)^2} \;,
\eeqa
and the corresponding deformed Lagrangian\footnote{A straightforward way to obtain the Lagrangian (\ref{eq:s0Lag}) from the Hamiltonian (\ref{eq:s0Ham}), is to start from a formal series expansion of $\mathcal{L}^{(0)}(\mathbf z,\ta,\del)$ around $\ta=\del=0$ and fix the unknown coefficients by matching the Legendre transformation of $\mathcal{L}^{(0)}(\mathbf z,\ta,\del)$ with (\ref{eq:s0Ham}).}
\beq  
\label{eq:s0Lag}
\mathcal{L}^{(0)}(\mathbf z,\ta,\del) = \frac{\partial\phi\,\bar\partial\phi}{1-\del\,\partial\phi-\ta\,\bar\partial\phi} \;,
\eeq
whose associated EoMs coincide with (\ref{eq:s0EoMs}). Notice that, while the deformed action is mapped exactly into the undeformed one under (\ref{eq:s0change}), the integral of (\ref{eq:s0Ham}) transforms with an additional term
\beq
\int \mathcal{H}^{(0)}(\mathbf z,\ta,\del)\,dz\wedge d\bar z = \int \[ \mathcal{H}(\mathbf w) + i\,(\del-\ta)\bigl( J_+(\mathbf w)\,\bar \T_2(\mathbf w) - \text{c.c.} \bigr)\]\,dw\wedge d\bar w \;. \notag \\
\eeq
In order to unambiguously identify the perturbing operator we must Legendre transform (\ref{eq:s0Ham}). First of all, we move from complex $\mathbf z$ coordinates to cartesian coordinates $\mathbf x = (x^1,x^2) = (x,t)$ according to the convention (\ref{eq:EuclLC}). Then, inverting the Legendre map one finds
{\small\begin{gather}
\Pi = i\,\frac{\partial\mathcal{L}^{(0)}(\mathbf x,\ta,\del)}{\partial\phi_t} \;, \notag \\
\phi_t= i\,\frac{\bigl(1+2\,\Pi\,(\ta-\del)\bigr)\bigl(-2+(\ta+\del)\,\phi_x\bigr)+2\sqrt{\bigl(1+2\,\Pi\,(\ta-\del)\bigr)\bigl(1-\ta\,\phi_x\bigr)\bigl(1-\del\, \phi_x\bigr)}}{(\ta-\del)\bigl(1+2\,\Pi\,(\ta-\del)\bigr)} \;.
\label{eq:s0Leg}
\end{gather}}
Plugging (\ref{eq:s0Leg}) in (\ref{eq:s0Ham})--(\ref{eq:s0calP}), or equivalently, performing the canonical Legendre transformation, one gets
{\small
\beqa  
\label{eq:s0HamLeg}
\mathcal{H}^{(0)}(\mathbf x,\ta,\del) &=& 
\cI_1^{(0)}(\mathbf x,\ta,\del) + \bar\cI_1^{(0)}(\mathbf x,\ta,\del) = i\,\Pi\,\phi_t + \mathcal{L}^{(0)}(\mathbf x,\ta,\del) \notag \\
&=& -\frac{\bigl(1+\Pi\,(\ta -\del)\bigr)\bigl(-2+(\ta+\del)\,\phi_x\bigr)+2\sqrt{\bigl(1+2\,\Pi\,(\ta-\del)\bigr)\bigl(1-\ta\,\phi_x\bigr)\bigl(1-\del\,\phi_x\bigr)}}{(\ta -\del)^2} \;, \notag \\
\eeqa}
and
\beq  
\label{eq:s0calPLeg}
\mathcal{P}^{(0)}(\mathbf x,\ta,\del) = 
\cI_1^{(0)}(\mathbf x,\ta,\del) - \bar\cI_1^{(0)}(\mathbf x,\ta,\del) = -\Pi\,\phi_x \;.
\eeq
From (\ref{eq:s0calPLeg}), we find that the momentum density is unaffected by the perturbation, since
\beq 
\mathcal{P}^{(0)}(\mathbf x,\ta,\del)  = -\Pi\,\phi_x = \mathcal{P}(\mathbf x) \;,
\eeq
where $\mathcal{P}(\mathbf x)$ is the unperturbed momentum density. Finally, expanding (\ref{eq:s0HamLeg}) at the first order in $\ta$ and $\del$, we can identify the perturbing operator at first order
\beqa
\label{eq:s0Hampert}
\mathcal{H}^{(0)}(\mathbf x,\ta,\del) \underset{\substack{\ta\to 0 \\ \del\to 0}}{\sim} \mathcal{H}(\mathbf x) + 2\,\bigl(\ta\,J_2(\mathbf x)\,\bar\T_2(\mathbf x) + \del\,\bar J_2(\mathbf x)\,\T_2(\mathbf x)\bigr) + \mathcal{O}(\ta\del) \;,
\eeqa  
where $\mathcal{H}(\mathbf x) = \cI_1(\mathbf x) + \bar\cI_1(\mathbf x) = \Pi^2 + \frac{1}{4}\,\phi_x^2$ is the undeformed Hamiltonian and
\beqa
\cI_1(\mathbf x) = -\T_2(\mathbf x) = \frac{1}{8}\(2\,\Pi - \phi_x\)^2 \;,\quad \bar\cI_1(\mathbf x) = -\bar\T_2(\mathbf x) = \frac{1}{8}\(2\,\Pi + \phi_x\)^2 \;.
\eeqa
are the holomorphic and anti-holomorphic components of the undeformed stress energy tensor, respectively. In (\ref{eq:s0Hampert}), we denoted as $J_\mu(\mathbf x)$ and $\bar J_\mu(\mathbf x)$ , $(\mu=1,2)$, the cartesian components of the undeformed $U(1)$ currents (\ref{eq:U1currents})
\beqa 
J_1(\mathbf x) &=& J_+(\mathbf x) - J_-(\mathbf x) =  \frac{i}{2}\(-2\Pi + \phi_x\) \;, \notag \\
J_2(\mathbf x) &=& i\,\bigl( J_+(\mathbf x) + J_-(\mathbf x)\bigr) = \frac{1}{2}\(2\Pi - \phi_x\) \;, \notag \\
\bar J_1(\mathbf x) &=& \bar J_+(\mathbf x) - \bar J_-(\mathbf x) =  \frac{i}{2}\(-2\Pi - \phi_x\) \;, \notag \\
\bar J_2(\mathbf x) &=& -i\,\bigl( \bar J_+(\mathbf x) + \bar J_-(\mathbf x)\bigr) = \frac{1}{2}\(-2\Pi - \phi_x\) \;,
\eeqa
which fulfil the continuity equations
\beq 
\partial_\mu J_\mu(\mathbf x) = 0 \;,\quad \partial_\mu \bar J_\mu(\mathbf x) = 0 \;.
\eeq
In a similar way, we define the cartesian components of the deformed $U(1)$ currents (\ref{eq:U1currentsdef}) as
\begin{gather}
J_1^{(0)}(\mathbf x,\ta,\del) = J_+^{(0)}(\mathbf x,\ta,\del) - J_-^{(0)}(\mathbf x,\ta,\del) \;,\quad  J_2^{(0)}(\mathbf x,\ta,\del) = i\,\bigl( J_+^{(0)}(\mathbf x,\ta,\del) + J_-^{(0)}(\mathbf x,\ta,\del)\bigr) \;, \notag \\
\bar J_1^{(0)}(\mathbf x,\ta,\del) = \bar J_+^{(0)}(\mathbf x,\ta,\del) - \bar J_-^{(0)}(\mathbf x,\ta,\del)  \;,\quad \bar J_2^{(0)}(\mathbf x,\ta,\del) = -i\,\bigl( \bar J_+^{(0)}(\mathbf x,\ta,\del) + \bar J_-^{(0)}(\mathbf x,\ta,\del)\bigr) \;, 
\end{gather}
which again fulfil the continuity equations
\beq 
\label{eq:s0conteqdef}
\partial_\mu J_\mu^{(0)}(\mathbf x,\ta,\del) = 0 \;,\quad \partial_\mu \bar J_\mu^{(0)}(\mathbf x,\ta,\del) = 0 \;.
\eeq
Therefore, from (\ref{eq:s0conteqdef}), one finds that the quantities
\beq 
Q(R,\ta,\del) = \int_0^R J_2^{(0)}(\mathbf x,\ta,\del)\,dx^1 \;,\quad \bar Q(R,\ta,\del) = \int_0^R \bar J_2^{(0)}(\mathbf x,\ta,\del)\,dx^1 \;.
\eeq
are the conserved charges associated to the $U(1)_L\times U(1)_R$ symmetry in the deformed theory.
\vspace{0.5cm}\\
\fbox{\parbox{6.3cm}{
{\bf Case} $\ta^{(1)}=\ta^{(3)}=0 \;|\; \ta^{(2)}=\ta^{(4)}=\ta$}}
\vspace{0.5cm}\\
This case can be easily retrieved from the previous one by sending $\del\to 0$. It corresponds to the change of coordinates associated to the $\JbT$ deformation (see \cite{Guica:2017lia,Chakraborty:2018vja}). First we notice that, setting $\delta=0$ in (\ref{eq:U1currentsdef}) and (\ref{eq:U1stressdef}), the deformation preserves the Sugawara construction for the holomorphic sector
\beq 
\T_2^{(\JbT)}(\mathbf z,\ta) = \T_2^{(0)}(\mathbf z,\ta,0) = -\frac{1}{2}\,\frac{(\partial\phi)^2}{(1-\ta\,\bar\partial\phi)^2} = \frac{1}{2}\(J_+^{(\JbT)}(\mathbf z,\ta)\)^2 \;,\quad J_+^{(\JbT)}(\mathbf z,\ta) = J_+^{(0)}(\mathbf z,\ta,0) \;,
\eeq 
but this is not true for the anti-holomorphic sector. Then, we observe that the Lagrangian (\ref{eq:s0Lag}) reduces to
\beq 
\mathcal{L}^{(\JbT)}(\mathbf z,\ta) = \mathcal{L}^{(0)}(\mathbf z,\ta,0) = \frac{\partial\phi\,\bar\partial\phi}{1-\ta\,\bar\partial\phi} \;,
\eeq
and the corresponding Legendre transformed Hamiltonian (\ref{eq:s0HamLeg}) becomes
\beq
\label{eq:JTHamLeg}
\mathcal{H}^{(\JbT)}(\mathbf x,\ta) = \mathcal{H}^{(0)}(\mathbf x,\ta,0) = \mathcal{P}(\mathbf x) + \frac{2}{\ta^2}\( \bigl(1+\ta\,J_2(\mathbf x)\bigr) - \mathcal{S}^{(\JbT)} \) \;,
\eeq
where 
\beq 
\mathcal{S}^{(\JbT)} = \sqrt{\bigl(1+\ta\,J_2(\mathbf x)\bigr)^2 - 2\ta^2\,\bar\cI_1(\mathbf x)} \;.
\eeq 
Writing (\ref{eq:JTHamLeg}) as $\mathcal{H}^{(\JbT)}(\mathbf x,\ta) = \cI^{(\JbT)}_1(\mathbf x,\ta) + \bar\cI^{(\JbT)}_1(\mathbf x,\ta)$ with
\beqa 
\cI^{(\JbT)}_1(\mathbf x,\ta) &=& -\(\T_2^{(\JbT)}(\mathbf x,\ta) +\Th_0^{(\JbT)}(\mathbf x,\ta)\) = \mathcal{P}(\mathbf x) + \frac{1}{\ta^2}\(\bigl(1+\ta\,J_2(\mathbf x)\bigr) - \mathcal{S}^{(\JbT)}\) \;, \notag \\
\bar\cI^{(\JbT)}_1(\mathbf x,\ta) &=& -\(\bar\T_2^{(\JbT)}(\mathbf x,\ta) + \bar\Th_0^{(\JbT)}(\mathbf x,\ta)\) = \frac{1}{\ta^2}\(\bigl(1+\ta\,J_2(\mathbf x)\bigr) - \mathcal{S}^{(\JbT)}\) \;,
\label{eq:JTIpImLeg}
\eeqa
we find that 
\beq
\label{eq:JTJ2}
J_2^{(\JbT)}(\mathbf x,\ta) = -\frac{1}{\ta} \( 1 + \mathcal{S}^{(\JbT)}\) = J_2(\mathbf x) - \ta\,\bar\cI^{(\JbT)}_1(\mathbf x,\ta) \;.
\eeq
As already discussed in \cite{Conti:2018jho} for the $\TbT$ deformation, we argue that also in the $\JbT$ case the energy density of the right and left movers (\ref{eq:JTIpImLeg}) has the same formal expression of the spectrum obtained in \cite{Bzowski:2018pcy} (cf. (\ref{eq:JTIpIm})), where the classical densities are replaced by the corresponding integrated quantities. In addition, also the deformation of the $U(1)$ current density (\ref{eq:JTJ2}) admits a straightforward generalisation at the quantum level (cf. (\ref{eq:JTQpdef})). Analogous considerations apply to the case $\ta^{(1)}=\ta^{(3)}=\ta$, $\ta^{(2)}=\ta^{(4)}=0$ , which corresponds to the $\TbJ$ deformation.
\vspace{0.5cm}\\
\fbox{\parbox{5.8cm}{
{\bf Case} $\ta^{(1)}=\ta^{(2)}=\ta^{(3)}=\ta^{(4)}=\ta$}}
\vspace{0.5cm}\\
All the equations (\ref{eq:s0jacobian})--(\ref{eq:s0HamLeg}) can be obtained setting $\delta=\ta$. In this case, the square root in (\ref{eq:s0HamLeg}) disappears, and the Hamiltonian takes the simple form
\beq 
\mathcal{H}^{(0)}(\mathbf x,\ta) = \mathcal{H}^{(0)}(\mathbf x,\ta,\ta) =
\frac{\mathcal{H}(\mathbf x) + \ta\,\mathcal{P}(\mathbf x)\( J_2(\mathbf x)+\bar J_2(\mathbf x) \) + \ta^2\,\mathcal{P}^2(\mathbf x)}{1+\ta\(J_2(\mathbf x) - \bar J_2(\mathbf x)\)} \;.
\eeq
Again, we split the Hamiltonian as $\mathcal{H}^{(0)}(\mathbf x,\ta) = \cI_1^{(0)}(\mathbf x,\ta) + \bar\cI_1^{(0)}(\mathbf x,\ta)$ with
\beqa  
\cI_1^{(0)}(\mathbf x,\ta) =
-\bigl(\T_2^{(0)}(\mathbf x,\ta) + \Th_0^{(0)}(\mathbf x,\ta)\bigr) = 
\frac{\cI_1(\mathbf x)+\ta\,J_2(\mathbf x)\,\mathcal{P}(\mathbf x)+\frac{\ta^2}{2}\mathcal{P}^2(\mathbf x)}{1+\ta\(J_2(\mathbf x) - \bar J_2(\mathbf x)\)} \;, \notag
\eeqa
\beqa
\label{eq:s0symIpm}
\bar\cI_1^{(0)}(\mathbf x,\ta) = -\bigl(\bar\T_2^{(0)}(\mathbf x,\ta) + \bar\Th_0^{(0)}(\mathbf x,\ta)\bigr) = 
\frac{\bar\cI_1(\mathbf x)+\ta\,\bar J_2(\mathbf x)\,\mathcal{P}(\mathbf x)+\frac{\ta^2}{2}\mathcal{P}^2(\mathbf x)}{1+\ta\(J_2(\mathbf x) - \bar J_2(\mathbf x)\)} \;.
\eeqa
Finally, the deformed $U(1)$ currents fulfil
\beqa 
\label{eq:s0symQpm}
J_2^{(0)}(\mathbf x,\ta) = J_2(\mathbf x) + \ta\,\mathcal{P}(\mathbf x) \;,\quad 
\bar J_2^{(0)}(\mathbf x,\ta) = \bar J_2(\mathbf x) + \ta\,\mathcal{P}(\mathbf x) \;.
\eeqa
We will see in section \ref{sec:TJ} that the quantum version of this perturbation (cf. (\ref{eq:s0IpIm})--(\ref{eq:s0QpQmdef})) can be obtained by introducing two different scattering factors in the NLIEs (\ref{eq:DDVCFT}).
\section{The quantum spectrum}
\label{sec:quantum}
This second part of the paper is devoted to the study of the quantum version of the perturbations of  classical field theories described in the preceding sections. As in \cite{Cavaglia:2016oda}, a NLIE will be the starting point for the derivation of the Burgers-type equations for the spectrum of the deformed conserved charges $I_\Bk^{(\Bs,\pm)}(R,\ta)$. Although we arrived to (\ref{eq:algnegs}) by considering the free boson model and only a very specific set of conserved currents, in the following we will assume the general validity of (\ref{eq:algnegs}) and of (\ref{eq:algposs}), obtained from (\ref{eq:algnegs}) using the reflection property (\ref{eq:classrefl}). In order to put our classical/quantum identification on a more solid foundation, it would be very important to extend the proof of (\ref{eq:algnegs}) to the set of conserved currents (\ref{eq:genericcurrent}) and also to massive theories.
\subsection{The scattering phase}
\label{sec:phasefactor}
We will now argue that the results of section \ref{sec:classBurgers} unambiguously suggest that the quantisation of these deformations is associated to a specific family of non-Lorentz invariant scattering phase factors.  Consider first the quantum version of (\ref{eq:algnegs}), where the holomorphic  and anti-holomorphic 
sectors are not coupled together by the interaction. Then,
the level-$k$ Hamiltonian and momentum operators factorise as 
\beqa
\hat E_k^{(s')}(R,\ta) &=&     \hat I_k^{(s',+)}(R,\ta)\otimes \mathbb{I} + \mathbb{I} \otimes  \hat I_k^{(s',-)}(R,\ta)\,, \\
\hat P_k^{(s')}(R,\ta) &=& \hat I_k^{(s',+)}(R,\ta)\otimes \mathbb{I}  - \mathbb{I} \otimes\ \hat I_k^{(s',-)}(R,\ta)\, ,
\eeqa
and their action on a generic multi-particle  state 
\beq
|N^{(+)},N^{(-)} \ket_{\tau} = |N^{(+)} \ket_{\tau}  \otimes |N^{(-)} \ket_{\tau}  =|\theta_1^{(+)},\theta_2^{(+)},\dots,\theta_{N^{(+)}} \ket_{\tau} \otimes |\theta_1^{(-)},\theta_2^{(-)},\dots, \theta_{N^{(-)}}^{(-)} \ket_{\tau}  \;,
\eeq
is, in the asymptotically in the  $R \rightarrow \infty $ limit, determined by \footnote{We have adopted here the convention of \cite{Zamolodchikov:1991vx}, where the single particle energy and momentum for right $(+)$ and left $(-)$ movers are parametrised as $\left( \frac{\hat{m}}{2} e^{\pm \theta},\pm \frac{\hat{m}}{2}e^{\pm  \theta} \right)$.}
\beq
\hat I_k^{(s',\pm)}(R,\tau) |N^{(\pm)} \ket_{\tau}  = \frac{\hat{\gamma}_k}{2} \(\sum_{i=1}^{N^{(\pm)}}   e^{\pm k \theta_i^{(\pm)}}\)|N^{(\pm)} \ket_{\tau} \,.
\label{eq:asy3}
\eeq
Notice that, in the deformed massless boson theory under consideration, there is only one species of elementary excitations and  the  set of rapidities $\{ \theta_i^{(\pm)} \}$ completely characterises an asymptotic quantum state.
In addition
\beqa
\hat I_k^{(\pm)}(R)|N^{(\pm)} \ket_0 &=& \( \frac{\hat{\gamma}_k}{2} \) \( \frac{2}{\hat{m}} \)^k \sum_{i=1}^{N^{(\pm)}} \left(\frac{2 \pi n_i^{(\pm)}}{R} \right)^k |N^{(\pm)} \ket_0\;, 
\label{eq:asy4}
\eeqa
where $n_i^{(\pm)} \in \Z^+$,
$\hat{m}= \hat{\gamma}_1$ and, we used the fact that the undeformed theory is free. 
Considering now equations  (\ref{eq:algnegs}), assuming that, at least
asymptotically in the  $R \rightarrow \infty$ limit:
\beq
[\hat I_k^{(s',\pm)}(R,\ta), \hat I_{k'}^{(s',\pm)}(R,\ta)] = 0\;,
\eeq
we have
\beq
\label{eq:algnegs1}
\prescript{}{0}{\bra} N^{(\pm)}|\hat I_k^{(s',\pm)}(R,\ta) \(R+2\ta\hat I_s^{(s',\pm)}(R,\ta)\)^k|N^{(\pm)}\ket_{\tau} = \,\prescript{}{0}{\bra} N^{(\pm)}| \hat I_k^{(\pm)}(R)|N^{(\pm)}\ket_{\tau}\, R^k,
\eeq
and using (\ref{eq:asy3}) and (\ref{eq:asy4}), we  find  
\beq
\(\sum_{i=1}^{N^{(\pm)}}  e^{\pm k \theta_i^{(\pm)}} \)\(R+2\ta\, \frac{\hat \gamma_{s}}{2} \sum_{j=1}^{N^{(\pm)}}  e^{\mp s' \theta_j^{(\pm)}}\right)^k = \(\frac{2}{\hat{m}} \)^{k}  \sum_{i=1}^{N^{(\pm)}} \left(2 \pi n_i^{(\pm)} \right)^k \;,\quad (\forall k \in \Z) \;,
\label{eq:BA0}
\eeq
with $s'=\Bs<0$.
The only consistent solutions  to  (\ref{eq:BA0}) are
\beq
\pm R\frac{\hat{m}}{2} e^{\pm \theta_i^{(\pm)}} \pm  \ta\, \frac{\hat{m}}{2}\hat  \gamma_{s'} \sum_{j=1}^{N^{(\pm)}}  e^{\pm (\theta_i^{(\pm)}-s' \theta_j^{(\pm)})} = \pm 2 \pi n_i^{(\pm)} \;,\quad ( i=1,2, \dots,N^{(\pm)})\;,
\label{eq:BA4}
\eeq
{\it i.e.} the asymptotic Bethe Ansatz (BA) equations for our models. The two body scattering amplitudes involving right  and left  movers are:
\beq
\delta^{(s')}_{(\pm,\mp)}( \theta, \theta') = 0 \;, \quad \delta^{(s')}_{(\pm,\pm)}( \theta, \theta')=\pm \tau \, \frac{\hat{m}}{2} \hat{\gamma}_{s'}\,e^{\pm ( \theta - s'\,\theta' )} \;,\quad (s'=\Bs<0)\;.
\label{eq:CDDs20}
\eeq
Similarly,  starting from (\ref{eq:algposs}) we find:
\beq
\delta^{(s)}_{(\pm,\mp)}( \theta, \theta') = \pm \tau \, \frac{\hat{m}}{2} {\hat \gamma}_s 
\, e^{\pm ( \theta - s\, \theta' )} 
\;, \quad \delta^{(s)}_{(\pm,\pm)}( \theta, 
\theta')=0 \;,\quad (s=\Bs > 0) \;.
\label{eq:CDDs7}
\eeq
The results presented in this section, strongly support the idea
that the classical theories introduced in this paper through a field dependent change of coordinates, can be  consistently quantised  within the exact S-matrix approach  through the introduction of specific Lorentz breaking phase factors.  The natural generalisation  of  (\ref{eq:CDDs20}) and (\ref{eq:CDDs7}) for a massive field theory is:
\beq
\delta^{(\Bs)}( \theta, \theta') = \tau \, m\gamma_\Bs \sinh( \theta - \Bs\, \theta' ) \;,
\label{eq:CDDs9}
\eeq
with asymptotic BA equations
\beq
R\,m  \sinh(\theta_i) +  \ta\, m \gamma_s \sum_{j=1}^{N} \sinh(\theta_i- \Bs \, \theta_j)  =  2 \pi n_i \;,\quad (n_i \in \ZZ\;,\; i=1,2, \dots,N)\;.
\label{eq:BA41}
\eeq
Summing over all the rapidities we see that, apart for $\mathbf{s}=1$, the kinetic total momentum is not quantised, i.e.
\beq
P(R,\ta) = \mathcal P_1^{(\Bs)}(R,\ta)=\sum_{i=1}^N m \sinh(\theta_i)  \ne  \frac{2 \pi}{R} n\,,\quad  n=\sum_{i=1}^N  n_i\;,\quad (\mathbf{s} \ne 1)\;, 
\eeq
since, at classical level, there is no trace of translational invariance  breaking, this result shows that the generator of space translations 
$\check{P}$ and the deformed momentum $P$ do not, in general, coincide. This fact is in agreement with the discussion of section \ref{sec:classBurgers}. From (\ref{eq:BA41}) it is easy to show that a natural definition of quantised momentum is, in the large $R$ limit:
\beq
{\check P}(R)  =  P(R,\tau) + \frac{\tau}{R} \left( P(R,\tau) \mathcal{E}^{(\Bs)}_\Bs(R,\tau) - E(R,\tau) \mathcal{P}^{(\Bs)}_\Bs(R,\tau)\right)=  \frac{2 \pi}{R}  n \;,\quad(n \in \ZZ)\;,
\label{eq:pcheck1}
\eeq
with 
\beq
\mathcal{E}^{(\Bs)}_\Bk= \sum_{i=1}^N \gamma_s \cosh(\Bk \,\theta_i)\;,\quad
\mathcal{P}^{(\Bs)}_\Bk= \sum_{i=1}^N \gamma_s \sinh(\Bk \, \theta_i) \;,\quad E=\mathcal{E}^{(\Bs)}_1 \;.
\eeq
Relation (\ref{eq:pcheck1}) can also be written as:
\beq
\check{P} = \check{P}^{(+)} - \check{P}^{(-)}\;,\quad \check{P}^{(\pm)}= I_{1}^{(\Bs,\pm)} + \frac{2 \tau}{R}  I_{1}^{(\Bs,\pm)} I_{\Bs}^{(\Bs,\mp)}\;. 
\label{eq:Psplit}
\eeq
Finally, notice that the quantisation of $P$ on a circle would be preserved by taking  symmetrised versions of the scattering phases (\ref{eq:CDDs20}), (\ref{eq:CDDs7}) or  (\ref{eq:CDDs9}). In the  massive case:
\beq
\delta'^{(\tilde \Bs,\Bs)}( \theta, \theta') = \frac{\tau}{2} \, m\gamma_\Bs \left( \sinh( \tilde \Bs\,  \theta - \Bs\, \theta' ) + \sinh( \Bs\, \theta -  \tilde \Bs\,\theta' ) \right)  \;.
\label{eq:CDDs10}
\eeq
with $\tilde \Bs=1$.
However, this is not the phase factor that our classical analysis suggests for the spectrum of the deformed charges (\ref{eq:clint2}). We have some concrete evidences \cite{Unpublished} that the phase factors (\ref{eq:CDDs10}), with arbitrary integers $
\tilde \Bs$ and $\Bs$, correspond instead to the 
\beq
T_{\Bs+1} \bar T_{\tilde \Bs+1} +\bar T_{\Bs+1} T_{\tilde \Bs+1} - \Theta_{\Bs-1} \bar \Theta_{\tilde \Bs-1} -\bar \Theta_{\Bs-1} \Theta_{\tilde \Bs-1}\;, 
\eeq
perturbations, recently discussed in \cite{LeFloch:2019rut}. Furthermore, the scattering phase
\beq
\tilde{\delta}^{(\Bs)}( \theta, \theta') = \tau \, m\gamma_\Bs \sinh( \Bs\, \theta -  \theta' ) \;,
\eeq
appears to be  related, instead, to the spectrum of the corresponding mirror deformed Hamiltonians  (See, for example,  \cite{Arutyunov:2007tc} for a rigorous definition of mirror theory in a similar, non relativistic invariant, Bethe Ansatz context.) of the models under consideration \cite{Unpublished}.
\subsection{Burgers-type  equations for the  spectrum}
\label{sec:generalisedBurgers}
The next goal is to try to build the quantum version of the classical models described in section \ref{sec:TTsHC}, by including the scattering phase factors (\ref{eq:CDDs9}) into the NLIE for the sine-Gordon model confined on a infinite cylinder of circumference $R$. The sine-Gordon NLIE is \cite{KlumperPearce0,KlumperPearce, DDV,FioravantiDDV, Feverati:1998dt}
\beqa
\label{eq:sGDDV}
f_\nu(\theta) =\nu(R,\alpha_0\,|\,\theta)- \int_{\mathcal{C}_1} d \theta' \, \kersg(\theta -\theta' ) \, \log\left( 1 + e^{-f_\nu(\theta')} \right) + \int_{\mathcal{C}_2} d\theta' \, \kersg(\theta - \theta') \, \log\left( 1 + e^{f_\nu(\theta')} \right) \;, \notag \\
\eeqa
where we have set
\beq
\label{eq:sGdriving}
\nu(R,\alpha_0\,|\,\theta)=i 2 \pi \alpha_0 -i\, m\rr \sinh(\theta) \;,
\eeq
to denote the {\it driving term}. In (\ref{eq:sGDDV})--(\ref{eq:sGdriving}), $m$ is the sine-Gordon soliton mass,  $\alpha_0$ is the quasi-momentum and $\kersg(\theta)$ is the kernel defined as
\beq
\kersg(\theta) = \frac{1}{2 \pi i } \partial_{\theta} \log S_{sG}(\theta) \;,
\eeq
where
\beq 
\log S_{sG}(\theta) = - i \int_{\RR^+} \frac{d p}{p} \sin( p \theta) \,\frac{ \sinh\left( \pi p (\zeta - 1)/2 \right) }{ \cosh\left( \pi p/2 \right) \, \sinh\left( \pi p \zeta/2 \right) }\;,\quad \zeta= \frac{\beta^2}{1-\beta^2} \;.
\eeq
The {\it quasi-momentum} or {\it vacuum parameter} $\alpha_0$ emerges \cite{Bazhanov:1994ft,Bazhanov:1996aq} by imposing  periodic  boundary conditions on the sine-Gordon field:  $\varphi(x+R ,t)= \varphi(x,t)$. 
Due to the periodicity of the potential in the sine-Gordon model
\beqa
\mathcal{L}_{sG} = \frac{1}{8 \pi} \left( \left(\partial_t \varphi \right)^2 -\left(\partial_x \varphi\)^2 \) + 2 \mu \cos(\sqrt{2} \beta \varphi)\;,\quad \(\mu \propto (m)^{2 -2 \beta^2}\)\;,
\label{eq:sglag}
\eeqa
the Hilbert space splits into orthogonal subspaces $\mathcal{H}_{\alpha_0}$, characterised by the quasi-momentum $\alpha_0$,
\beq
\varphi \rightarrow \varphi +\frac{2\pi}{\sqrt{2}\beta} 
\quad \longleftrightarrow \quad {|\Psi_{\alpha_0} \ket} \rightarrow e^{i 2\pi \alpha_0} {|\Psi_{\alpha_0} \ket} \;,\quad (\,{|\Psi_{\alpha_0} \ket} \in\mathcal{H}_{\alpha_0}\,) \;,
\label{eq:vsta}
\eeq
with $\alpha_0 \in [-1/2,1/2]$. Twisted boundary conditions of the form:
\beqa
\varphi(x+R ,t)= \varphi(x,t) + \frac{2\pi}{\sqrt{2}\beta} n\;,\quad (n \in \ZZ) \;,
\eeqa
are also natural in the sine-Gordon model since they correspond, in the infinite volume limit, to field configurations with non-trivial topological charge:
\beqa
Q_x= \frac{\sqrt{2} \beta}{2 \pi} \int_{0}^R \partial_x \varphi \, dx \;.
\eeqa
Energy levels  in the  twisted sectors are also described by the same NLIE at specific values of the quasi-momentum. Furthermore,  $\alpha_0$ can also be related to a background charge (cf. equation (\ref{eq:centralc})). Therefore, (\ref{eq:sGDDV}) also describes minimal models of the Virasoro algebra, $\mathcal{M}_{p,q}$ perturbed  by the operator $\phi_{13}$ \cite{Feverati:1999sr}.\\ 
The integration contours $\mathcal{C}_{1}$, $\mathcal{C}_{2}$ in (\ref{eq:sGDDV}) are state dependent. For the ground-state in an arbitrary subspace  $\mathcal{H}_{\alpha_0}$, one may take them to be straight lines slightly displaced from the real axis: $\mathcal{C}_{1} = \mathbb{R} + i 0^{+} $, $\mathcal{C}_{2} = \mathbb{R} - i 0^+$. Equations describing excited states have the same form \cite{Bazhanov:1996aq, DT, FioravantiDDV, Feverati:1998dt} but  with integration contours encircling a number of singularities $\{\theta_i\}$ with  $\left(1+ e^{f(\theta_i)}\right) =0$. However, in this paper we can ignore these subtleties, since the final evolution equations do not depend explicitly on the specific details of the integration contours. Setting
{\small\beq
\label{eq:EE0}
b_k^{(\pm)}(r) =  \int_{\mathcal{C}_1} \frac{d \theta}{2 \pi i}  \left(\pm \frac{r}{2} e^{\pm  \theta}\right)^k  \log\left( 1 + e^{-f_\nu(\theta)} \right)-\int_{\mathcal{C}_2} \frac{d \theta}{2 \pi i}  \left(\pm \frac{r}{2} e^{\pm  \theta}\right)^k  \log\left( 1 + e^{f_\nu(\theta)} \right) \;,\quad (k\in 2 \NN+1) \;,
\eeq}
with $r=m R$, and\footnote{The explicit dependence on some parameters, {\it e.g.} the mass $m$, will be sometime omitted not to weigh down the notation.}
\beq
I_{k}^{(\pm)}(R,m) = \left(\frac{2 \pi}{R}\right)^{k} \frac{b_{k}^{(\pm)}(r)}{C_k} \;,
\eeq
where 
\beq
\label{eq:cn}
C_k  = \frac{1}{2 k} \left(\frac{4 \pi}{\beta^2} \right)^{\fract{k+1}{2}} \frac{\Gamma\left(\fract{k}{2}(\zeta+1)\right)}{\Gamma\left(\fract{k +3}{2} \right) \, \Gamma\left(\fract{k}{2}\zeta\right)} \left( \frac{\Gamma(1 + \zeta/2)}{\Gamma(3/2 + \zeta/2)}\right)^{k} \;,
\eeq
then,  $\left\lbrace I_{k}^{(+)}, I_{k}^{(-)}\right\rbrace$ are the eigenvalues of the quantum operators associated to the classical conserved charges defined in (\ref{eq:densTocharge}), {\it i.e}
\beq 
I_{k}^{(+)} \longleftrightarrow \int_0^R  \cI_k(\mathbf x)\,dx\;,\quad I_{k}^{(-)} \longleftrightarrow \int_0^R  \bar\cI_k(\mathbf x)\,dx\;.
\eeq
It is also convenient to define the alternative set of conserved charges
\beq
\mathcal{E}_k(\rr) = I_{k}^{(+)}(\rr) + I_{k}^{(-)}(\rr) \,,\quad  \mathcal{P}_k(\rr) = I_{k}^{(+)}(\rr) - I_{k}^{(-)}(\rr) \,,\quad  
\eeq
where $E(\rr)=\mathcal{E}_1(\rr)$ and  $P(\rr)=\mathcal{P}_1(\rr)$ are the total energy and momentum of the state, respectively.
In addition, the following reflection property holds
\beq
I_{-k}^{(\pm)}(R)=  I_{k}^{(\mp)}(R) \;,\quad (k>0) \;,
\label{eq:ref}
\eeq
which is the quantum analog of (\ref{eq:classrefl}).\\
Motivated by the results  of section \ref{sec:phasefactor}, we conjecture that the  geometric-type deformations described in section \ref{sec:TTsHC}, associated to generic combinations of the higher spin conserved currents, correspond to the inclusion  of extra scattering phases of the form
\beq
\delta( \theta, \theta') = \sum_{\Bs} \tau^{(\Bs)} \, \gamma_1\gamma_{\Bs} \sinh( \theta -  \Bs \, \theta' ) \;, 
\label{eq:CDDs0}
\eeq
where the $\tau^{(\Bs)}$ are independent coupling parameters and
\beq
\gamma_{\Bs}= \frac{(2 \pi m)^{s}}{C_s}  \;,\quad \gamma_{1}=m\;, 
\eeq
with $s=|\Bs|$. In (\ref{eq:CDDs0}) the sum runs, in principle, over positive and negative odd integers.\\
However, since the whole analysis can be straightforwardly analytically extended to arbitrary values of $\Bs$, 
in the following we shall relax this constraint,  at least to include the  case $\Bs \rightarrow 0$ and the set of nonlocal conserved charges \cite{Lukyanov:2010rn}. 
The scattering phase factor  (\ref{eq:CDDs0}) leads to   multi-parameter perturbations of the spectrum of the original QFT which, due to the intrinsic non-linearity of the problem, can effectively be studied only on the case-by-case basis. A detailed analytic and numerical study of specific multi-parameter deformations of a massive QFT, such as the sine-Gordon model,  appears to be a very challenging long-term objective. Most of our checks have been performed considering conformal field theories  deformed (explicitly at leading order) by a single irrelevant composite field. Therefore, in this paper, we shall restrict the analysis to scattering phase factors with only a single non vanishing irrelevant coupling:
\beq
\delta^{(\Bs)}( \theta, \theta') = \tau \, \gamma_1 \gamma_\Bs \sinh( \theta - \Bs\, \theta' ) \;, 
\label{eq:CDDs}
\eeq
which modifies the kernel appearing in the NLIE (\ref{eq:sGDDV}) as
\beqa
\label{eq:modkernel}
\kersg(\theta-\theta') \rightarrow \kersg(\theta-\theta') - \frac{1}{2 \pi} \partial_{\theta} \delta^{(\Bs)}(\theta-\Bs\,\theta') = \kersg(\theta- \theta') - \tau\,m \frac{\gamma_\Bs}{2 \pi} \cosh(\theta-\Bs\,\theta') \;.
\eeqa
Inserting (\ref{eq:modkernel}) in (\ref{eq:sGDDV}), after simple manipulations, we find that the deformed version of $f_\nu(\theta)$ fulfils (\ref{eq:sGDDV}) with
\beq  
\label{eq:modsGDDV2}
\nu=\nu(\mathcal{R}_0 ,\alpha_0\,|\,\theta-\theta_0) \;,
\eeq
where $\mathcal{R}_0$ and $\theta_0$ are defined through
\beq
\label{eq:defineRtheta}
\mathcal{R}_0 \,\cosh\left( \theta_0 \right)  =  R + \tau\,\mathcal{E}^{(\Bs)}_\Bs(\rr , \tau) \;,\quad \mathcal{R}_0 \,  \sinh\left(  \theta_0 \right) = \tau\,\mathcal{P}^{(\Bs)}_\Bs(\rr,\tau) \;.
\eeq
Equations (\ref{eq:defineRtheta}) imply
\beq
(\mathcal{R}_0)^2 = \(R + \tau\,\mathcal{E}^{(\Bs)}_\Bs(\rr , \tau)\)^2 - \(\tau\,\mathcal{P}^{(\Bs)}_\Bs(\rr,\tau)\)^2 \;.
\eeq
The quantities $\mathcal{E}^{(\Bs)}_\Bk(\rr, \tau)$ and $\mathcal{P}^{(\Bs)}_\Bk(\rr, \tau)$ denote the $\Bk$-th higher conserved charges of the theory deformed with the $\Bs$-th perturbation
\beq
\label{eq:defDDVHC}
\mathcal{E}^{(\Bs)}_\Bk(\rr, \tau) = I_{\Bk}^{(\Bs,+)}(\rr,\ta) + I_{\Bk}^{(\Bs,-)}(\rr,\ta)\;,\quad  \mathcal{P}^{(\Bs)}_\Bk(\rr, \tau) = I_{\Bk}^{(\Bs,+)}(\rr,\ta) - I_{\Bk}^{(\Bs,-)}(\rr,\ta) \;,\quad  
\eeq
and $I_{\Bk}^{(\Bs,\pm)}$ are again defined through (\ref{eq:EE0})--(\ref{eq:cn}) but with the deformed driving term (\ref{eq:modsGDDV2}).\\
Formula (\ref{eq:modsGDDV2}) shows that the solutions of the deformed NLIE are modified simply by a redefinition of the length $R$ and by a rapidity shift.
One can easily show that the deformed charges are related to the undeformed ones through
\beq
\label{eq:Ikspm}
I_{\Bk}^{(\Bs,\pm)}(R,\tau) = 
e^{\pm \Bk \theta_0} I_{\Bk}^{(\pm)}(\cR_0) \;,
\eeq
which, together with (\ref{eq:defDDVHC}), leads to a generalisation of the Lorentz-type transformation (\ref{eq:Lorentz}) derived in \cite{Conti:2018jho}
\beq
\label{eq:genLorentz}
\(\begin{array}{c}
\mathcal{E}^{(\Bs)}_\Bk(R,\ta) \\
\mathcal{P}^{(\Bs)}_\Bk(R,\ta)
\end{array}\) =
\(\begin{array}{cc}
\cosh{(\Bk\,\theta_0)} & \sinh{(\Bk\,\theta_0)} \\
\sinh{(\Bk\,\theta_0)} & \cosh{(\Bk\,\theta_0)}
\end{array}\)
\(\begin{array}{c}
\mathcal{E}_\Bk(\mathcal{R}_0) \\
\mathcal{P}_\Bk(\mathcal{R}_0)
\end{array}\) \;.
\eeq
From (\ref{eq:genLorentz}) one has
\beq
\(\mathcal{E}^{(\Bs)}_\Bk(R,\ta)\)^2 - \(\mathcal{P}^{(\Bs)}_\Bk(R,\ta)\)^2 = \(\mathcal{E}_\Bk(\cR_0)\)^2 - \(\mathcal{P}_\Bk(\cR_0)\)^2 \;,
\label{eq:genLorentzEsPs}
\eeq
and, in particular, for $\Bk=1$
\beq
\label{eq:genLorentzEP}
E^2(R,\ta) - P^2(R,\ta) = E^2(\cR_0) - P^2(\cR_0) \;,
\eeq
where $E(R,\ta)= \mathcal{E}^{(\Bs)}_1(R,\ta)$ and $P(R,\ta)= \mathcal{P}^{(\Bs)}_1(R,\ta)$. To find the generalisations of the Burgers equation (\ref{eq:Burgers}) describing the evolution of the spectrum under the deformations with generic Lorentz spin $\Bs$, we differentiate both sides of (\ref{eq:Ikspm}) w.r.t. $\ta$ at fixed $\cR_0$, getting
\beq
\label{eq:BurgersIpm}
\partial_\ta I_{\Bk}^{(\Bs,\pm)}(R,\tau) + R' \partial_R  I_{\Bk}^{(\Bs,\pm)}(R,\tau)  =  \pm\Bk \, \theta'_0  I_{\Bk}^{(\Bs,\pm)}(R,\tau) \;.
\eeq
All that is left to do is compute $\partial_\ta R = R'$ and $\partial_\ta \theta_0 = \theta'_0$. We start by rewriting (\ref{eq:defineRtheta}) as
\beq
\label{eq:defineRtheta0}
\begin{cases}
\mathcal{R}_0 = R \,e^{-\theta_0}  + 2 \tau \, e^{(\Bs-1) \theta_0} I_{\Bs}^{(+)}(\cR_0) \\
\mathcal{R}_0 = R \,e^{\theta_0} + 2 \tau \, e^{- (\Bs-1) \theta_0} I_{\Bs}^{(-)}(\cR_0)
\end{cases} \;,
\eeq
then, differentiating both equations in (\ref{eq:defineRtheta0}) w.r.t. $\ta$ we obtain
\beq
\label{eq:defineRtheta1}
\begin{cases}
0 = R' \,e^{-\theta_0} - R \,e^{-\theta_0} \theta'_0 +2 e^{(\Bs-1) \theta_0} I_{\Bs}^{(+)}(\cR_0)  + 2 \tau  e^{(\Bs-1) \theta_0}(\Bs-1)\, \theta'_0\,I_{\Bs}^{(+)}(\cR_0) \\
0 = R' \,e^{\theta_0}+  R \,e^{\theta_0} \theta'_0  + 2  e^{- (\Bs-1) \theta_0} I_{\Bs}^{(-)}(\cR_0) -2  \tau  e^{- (\Bs-1) \theta_0} (\Bs-1)\, \theta'_0\,I_{\Bs}^{(-)}(\cR_0)
\end{cases} \;,
\eeq
whose solution is
\beqa
\label{eq:sol}
\partial_\ta R= R' &=& -\mathcal{E}^{(\Bs)}_\Bs(R,\ta) + \frac{(\Bs-1)\ta\(\mathcal{P}^{(\Bs)}_\Bs(R,\ta)\)^2 }{(\Bs-1) \tau\, \mathcal{E}^{(\Bs)}_\Bs(R,\ta)-R} \;, \notag \\
\partial_\ta \theta_0 =\theta'_0 &=& -\frac{\mathcal{P}^{(\Bs)}_\Bs(R,\ta)}{(\Bs-1) \tau\,\mathcal{E}^{(\Bs)}_\Bs(R,\ta)-R} \;.
\eeqa
The equations for the total energy and momentum are then
\beq
\begin{cases}
\label{eq:BurgersEP}
\partial_\ta E(R,\tau) + R' \partial_R  E(R,\tau)  =  \theta'_0  P(R,\tau) \\
\partial_\ta P(R,\tau) + R' \partial_R  P(R,\tau)  = \theta'_0  E(R,\tau)
\end{cases} \;,
\eeq
which are driven, through $R'$ and $\theta'_0$ by  the evolution equations for
$\mathcal{E}^{(\Bs)}_\Bs$ and $\mathcal{P}^{(\Bs)}_\Bs$
\beq
\label{eq:energyBu}
\begin{cases}
\partial_\ta \mathcal{E}^{(\Bs)}_\Bs(R,\tau) + R' \partial_R  \mathcal{E}^{(\Bs)}_\Bs(R,\tau) = \Bs\,\theta'_0\,\mathcal{P}^{(\Bs)}_\Bs(R,\tau) \\
\partial_\ta \mathcal{P}^{(\Bs)}_\Bs(R,\tau) + R' \partial_R  \mathcal{P}^{(\Bs)}_\Bs(R,\tau)  = \Bs\,\theta'_0 \,\mathcal{E}^{(\Bs)}_\Bs(R,\tau)
\end{cases} \;.
\eeq
Again it is worth to stress that the deformed momentum  $P$, defined through (\ref{eq:BurgersEP}) is not  always quantized for $\tau \ne 0$, as it flows according to the complicated non-linear equations (\ref{eq:BurgersEP})-(\ref{eq:energyBu}). One can argue (see for example \cite{Cavaglia:2010nm}) that a quantised object is
\beq
{\check P}(R) = \frac{1}{2 \pi R} \left(
\int_{\mathcal{C}_1} d \theta \,  p(\theta) \log\left( 1 + e^{-f_\nu(\theta)} \right)-\int_{\mathcal{C}_2} d \theta \, p(\theta)  \log\left( 1 + e^{f_\nu(\theta)} \right) \right)\, ,
\label{eq:quantised}
\eeq
with
\beq
p(\theta) =\partial_{\theta} \nu(\mathcal{R}_0 ,\alpha_0\,|\,\theta-\theta_0) = - i m \mathcal{R}_0 \cosh( \theta-\theta_0)\,.
\label{eq:smallp}
\eeq
Using (\ref{eq:defineRtheta}) and (\ref{eq:smallp})   in (\ref{eq:quantised}), we find 
\beq
{\check P}  = P(R) = P(R,\tau) + \frac{\tau}{R} \left( P(R,\tau) \mathcal{E}^{(\Bs)}_\Bs(R,\tau) - E(R,\tau) \mathcal{P}^{(\Bs)}_\Bs(R,\tau)\right)\,,
\eeq
which is the same formula obtained in section \ref{eq:pcheck1} from BA considerations.
\subsection{The CFT limit of the NLIE}
\label{sec:CFTlimit}
The CFT limit of the  sine-Gordon model is described by a pair of decoupled NLIEs corresponding to the right- $(+)$ and the left- $(-)$ mover sectors. The latter equations can be  obtained by sending $m$ to zero  as $m = \hat{m} \epsilon$ with $\epsilon \rightarrow 0^+$ and simultaneously $\theta$ to  $\pm \infty$ as $\theta \rightarrow  \theta \pm \log(\epsilon)$ such that 
$\hat{m}  e^{\pm \theta}$ remain finite. The resulting equations are identical to (\ref{eq:sGDDV}) but with the term $m \sinh(\theta)$ replaced by $\frac{\hat{m} }{2}e^\theta$  and  $-\frac{\hat{m} }{2} e^{-\theta}$, for the right- and left-mover sectors, respectively
\beqa
\label{eq:DDVCFT}
f_{\nu^{(\pm)}}(\theta) &=& \nu^{(\pm)}(R,\alpha^{(\pm)}_0\,|\,\theta) \notag \\
&-& \int_{\mathcal{C}_1} d \theta' \, \kersg(\theta -  \theta' ) \, \log\left( 1 + e^{- f_{\nu^{(\pm)}}( \theta' )} \right) +  \int_{\mathcal{C}_2} d  \theta'  \, \kersg(\theta - \theta' ) \, \log\left( 1 + e^{f_{\nu^{(\pm)}}( \theta' )} \right) \;, \notag \\
\eeqa
where
\beq 
\nu^{(\pm)}(R,\alpha^{(\pm)}_0\,|\,\theta)=i 2\pi \alpha_0^{(\pm)} \mp i \frac{\hat{m} \rr}{2} \,  e^{\pm \theta} \;,
\label{eq:twoche}
\eeq
and $\hat{m}$ sets the energy scale.
In (\ref{eq:twoche}), we have allowed for the possibility of two independent vacuum parameters $\alpha_0^{(\pm)}$ in the two chiral sectors. The resulting NLIEs are, in principle, suitable for the description of twisted boundary conditions and  more general states in the  $c=1$ CFT, compared to the set strictly emerging from the  $m\rightarrow 0$ sine-Gordon model. For example, they can accommodate the states with odd fermionic numbers of the  massless Thirring model, which require anti-periodic boundary conditions. The integration contours $\mathcal{C}_{1}$, $\mathcal{C}_{2}$ are still state dependent. In particular, they  should be deformed away from the initial ground-state configuration $\mathcal{C}_{1} = \mathbb{R} + i 0^{+} $, $\mathcal{C}_{2} = \mathbb{R} - i 0^+$ when the parameters  $\pm \alpha_0^{(\pm)}$ are analytically extended to large negative values. Setting
{\small \beq
\label{eq:EECFT}
\hat{b}_k^{(\pm)} =   \int_{\mathcal{C}_1} \frac{d \theta}{2 \pi i}  \left(\pm \frac{\hat r}{2} e^{\pm  \theta}\right)^k  \log\left( 1 + e^{-f_{\nu^{(\pm)}}(\theta)} \right)-\int_{\mathcal{C}_2} \frac{d \theta}{2 \pi i}  \left(\pm \frac{\hat r}{2} e^{\pm  \theta}\right)^k  \log\left( 1 + e^{f_{\nu^{(\pm)}}(\theta)} \right) \;,\; (k \in 2 \NN +1) \;,
\eeq}
with  $\hat r= \hat m R$, then
\beq
\label{eq:DDVcharges}
I_{k}^{(\pm)}(R)= \left(\frac{2 \pi}{R}\right)^{k} \frac{\hat{b}_{k}^{(\pm)}}{C_k} = \left(\frac{1}{R}\right)^{k} 2 \pi a_{k}^{(\pm)} \;,
\eeq
where the $a_k^{(\pm)}$ are state-dependent constants, {\it i.e.} $R$ and $\hat m$ independent. Using again the spin-flip symmetry $I_{-k}^{(\pm)}(R)=I_{k}^{(\mp)}(R)$, we can extend the discussion to $\Bk \in 2\ZZ+1$, with $k=|\Bk|$. Some of the state-dependent coefficients $I_{k}^{(\pm)}(2 \pi)$  can be found in  \cite{Bazhanov:1994ft}. In particular, the energy and momentum of a generic state are
\beqa
E(R) &=& I^{(+)}_1(R) + I^{(-)}_1(R)=  \frac{2 \pi }{R}\left(n^{(+)}  - \frac{c_0^{(+)}}{24} \right)+\frac{2 \pi }{R}\left(n^{(-)} - \frac{c_0^{(-)}}{24} \right) \;, \notag \\
P(R) &=& I^{(+)}_1(R) - I^{(-)}_1(R)=  \frac{2 \pi }{R}\left(h^{(+)} - h^{(-)}\right) \;,
\eeqa
with effective central charges
\beq
c_0^{(\pm)} =  1 - 24\beta^2\(\alpha_0^{(\pm)}\)^2 = 1 - 24 h_0^{(\pm)} \;,
\label{eq:centralc}
\eeq
where $h_0^{(\pm)}$ are the holomorphic and antiholomorphic highest weights and $h^{(\pm)}=h_0^{(\pm)} + n^{(\pm)}$, $(n^{(\pm)} \in \NN)$.

For the current purposes, it is convenient to think about the massless limit of the sine-Gordon model as a perturbation by a relevant field of the compactified free boson with Lagrangian given by (\ref{eq:sglag}) with $\mu=0$ (see \cite{Feverati:1998dt} for more details):
\beq
\varphi(x+R,t) = \varphi(x,t) + \frac{2\pi\,n}{\sqrt{2}\beta} \;,\quad (n \in\ZZ) \;.
\label{eq:shift2}
\eeq
In (\ref{eq:shift2}), $\mathbf{r}=(\sqrt{2}\beta)^{-1}$ is standard  compactification  radius of the bosonic field. 
Then, the highest weights are now labelled by a pair of integers $(n,\tilde{n})$, where $\tilde{n}\beta^2/\sqrt{4\pi}$ is the quantized charge associated to the total field momentum
\beq  
Q_t = \frac{\sqrt{2} }{2 \pi\beta}\int_0^R \partial_t\varphi\,dx = \frac{2\sqrt{2}}{\beta}\int_0^R \Pi\,dx\;,
\eeq
and $n$ is the winding number corresponding to the topological charge
\beq 
Q_x = \frac{\sqrt{2} \beta}{2 \pi }\int_0^R \partial_x\varphi\,dx \;.
\eeq
Then the combinations
\beq  
 Q_0^{(\pm)} = \pi \(Q_x \pm Q_t\)= \pi\( n \pm \tilde{n} \beta^2 \) \;,
\label{eq:topQ}
\eeq
are the two different charges, associated  to the $U(1)_R \times U(1)_L$ symmetry of the $c=1$ compactified boson. Notice that $Q_0^{(\pm)}$ differ from the the standard Kac-Moody $U(1)_R \times U(1)_L$ charges by a multiplicative factor $\beta$ which spoils explicitly the $\beta \rightarrow 1/\beta$ symmetry. We adopted this unconventional definition for the topological charges since, as a reminiscence of the sine-Gordon model, it emerges more  naturally from the current setup.  The anagolous of the Bloch wave states in (\ref{eq:vsta}) are now created by the action on the CFT vacuum state of the vertex operators
\beq 
\mathcal{V}_{(n,\tilde{n})}(\mathbf z) = \exp{\(\frac{\sqrt{2}}{2 \pi \beta} Q_0^{(+)}\,\phi(z) + \frac{\sqrt{2}}{ 2 \pi \beta} Q_0^{(-)}\,\bar{\phi}(\bar z)\)} \;,\quad \varphi(\mathbf z) = \phi(z) + \bar\phi(\bar z) \;,
\label{eq:vertex}
\eeq
with  left and right conformal dimensions given by
\beq
h^{(\pm)}_0 = \frac{1}{4 \pi^2 \beta^2} \( Q_0^{(\pm)} \)^2 =\frac{1}{4} \(\frac{n}{ \beta} \pm  \tilde{n}  \beta \)^2 \,.
\eeq
Considering (\ref{eq:centralc}), in section \ref{eq:furhterD} we will make the following  identification:
\beq 
Q_0^{(\pm)} = 2\pi\alpha_0^{(\pm)}\beta^2 \;.
\label{eq:ide}
\eeq
However, relation (\ref{eq:ide}) is valid only at formal level since $\alpha_0^{(\pm)}$ are continuous parameters which can also account, for example, for twisted boundary conditions while, at fixed $\beta$, the charges $Q_0^{(\pm)}$ can only assume the discrete set of values given in (\ref{eq:topQ}).
Using  (\ref{eq:ide})  in (\ref{eq:vertex}) we find
\beq 
\mathcal{V}_{(n,\tilde{n})}(\mathbf z) = \exp{\(\sqrt{2}\beta \alpha_0^{(+)}\,\phi(z) + \sqrt{2}\beta \alpha_0^{(-)}\,\bar\phi(\bar z)\)} \;,
\eeq
which, for  $\tilde{n}=0$ and  under the field-shift (\ref{eq:shift2}), display the  same quasi-periodicity  properties of the finite volume sine-Gordon Bloch states (\ref{eq:vsta}). 
In this limit, the $\ta$ dependent phase factor (\ref{eq:CDDs}) splits, for $\Bs>0$, into 
\beq
\delta^{(s)}_{(\pm,\mp)}( \theta, \theta') = \pm \tau \, \fract{\hat{m}}{2} \hat{ \gamma}_s\, e^{\pm ( \theta - s\, \theta' )} \;, \quad \delta^{(s)}_{(\pm,\pm)}( \theta, \theta')=0 \;,
\label{eq:CDDs1}
\eeq
with
\beq
\hat{\gamma}_{\Bs}= \frac{(2 \pi \hat{m})^{s}}{C_s}  \,,\quad \hat{\gamma}_{1}=\hat{m}\,,
\eeq
breaking conformal invariance by explicitly introducing  a coupling between the right and the left mover sectors.
Setting
\beq
\mathcal{R}_0^{(s,\pm)} = \rr + 2\ta\,I_{s}^{(s,\mp)}(R,\ta) \;,
\label{eq:R00}
\eeq
the resulting NLIE is identical to (\ref{eq:DDVCFT}) with driving term $\nu^{(\pm)}\bigl(\mathcal{R}_0^{(s,\pm)},\alpha_0\,|\,\theta\bigr)$. For $\Bs<0$, the two chiral sectors remain decoupled
\beq
\delta^{(s')}_{(\pm,\mp)}( \theta, \theta') = 0 \;, \quad \delta^{(s')}_{(\pm,\pm)}( \theta, \theta')=\pm \tau \, \frac{\hat{m}}{2} \hat{\gamma}_{s'}\,e^{\pm ( \theta - s'\,\theta' )} \;,\quad (s'=\Bs<0)\;,
\label{eq:CDDs2}
\eeq
and, due to the reflection property $I_s^{(s',\pm)}=I_{s'}^{(s',\mp)}$, the driving terms become $\nu^{(\pm)}\bigl(\mathcal{R}_0^{(s',\mp)},\alpha_0\,|\,\theta\bigr)$. In turn, the length redefinition (\ref{eq:R00}) implies
\beqa
\label{eq:IkfunIkspos}
I_k^{(s,\pm)}(R,\ta) &=& I_k^{(\pm)}\(\mathcal{R}_0^{(s,\pm)}\) \;,\quad (s>0) \;, \\
\label{eq:IkfunIksneg}
I_k^{(s',\pm)}(R,\ta) &=& I_k^{(\pm)}\(\mathcal{R}_0^{(s',\mp)}\) \;,\quad (s'=-s<0) \;,
\eeqa
which are equivalent to the evolution equations (\ref{eq:simplBurgers}), deduced at classical level.\\
Using the scaling property of the CFT charges , (\ref{eq:simplBurgers}) can be further simplified to
\beqa 
\label{eq:BurgersnoR}
\partial_\ta I_k^{(s,\pm)} &=& -\frac{2k \left(R-2\ta\,(s-1)\,I_s^{(s,\pm)}\right)I_s^{(s,\pm)}I_k^{(s,\mp)}}{2\ta \left(R-2\ta \left(s^2-1\right)I_s^{(s,+)}\right)I_s^{(s,-)}+R \left(R+2\ta I_s^{(s,+)}\right)} \;,\quad (s>0) \\
\partial_\ta I_k^{(s',\pm)} &=& -\frac{2 k\,I_k^{(s',\pm)}\,I_{s}^{(s',\pm)}}{R + 2\ta\,(s+1)\,I_s^{(s',\pm)}} \;,\quad (s'=-s<0) \;.
\eeqa
Notice that, setting $k=s=1$ in (\ref{eq:BurgersnoR}) we get
\beq 
\partial_\ta I_1^{(\pm)} = \frac{-2\,I_1^{(+)}I_1^{(-)}}{R+2\ta\(I_1^{(+)}+I_1^{(-)}\)} \;,
\eeq
which matches with the $\TbT$ result quoted in \cite{Jiang:2019hxb}.
\vspace{0.5cm}\\
\fbox{\parbox{5.0cm}{\bf
Perturbations with $\Bs<0$\,:}}
\vspace{0.2cm}\\
\noindent
For simplicity, let us discuss first the cases with $\Bs<0$. Since the generic $k$-th charges of a CFT scale as\footnote{Separately, the NLIEs in (\ref{eq:DDVCFT}), with generic  parameters $\beta$ and $\alpha_0^{(\pm)}$  can be also associated to the  quantum KdV theory, as extensively discussed in \cite{Bazhanov:1994ft,Bazhanov:1996dr}.  The coefficients $a_k$ for $k=3,5$ can be recovered from \cite{Bazhanov:1994ft}.} $R^{-k}$ according to (\ref{eq:DDVcharges}) and the corresponding deformed charges $I_k^{(s',\pm)}(R,\ta)$ fulfil the same equation with $R \rightarrow \mathcal{R}_0^{(s',\pm)}$, then
\beq
\label{eq:EsCFTdef}
I_{k}^{(s',\pm)}(R,\ta) = I_{k'}^{(s',\mp)}(R,\ta) = 
\frac{R^k\,I_{k'}^{(\mp)}(R)}{\(\mathcal{R}_0^{(s',\pm)}\)^{k}}
 = \frac{2\pi\,a_{k'}^{(\mp)}}{\Bigl(\mathcal{R}_0^{(s',\pm)}\Bigr)^k}  \;,
\eeq
where $a_k^{(\pm)}=a_{k'}^{(\mp)}$, $k'=-k,\; s'=\Bs$, and $\mathcal{R}_0^{(s',\pm)}$ is defined in (\ref{eq:R00}).\\

\begin{figure}
    \centering
    \includegraphics[scale=0.55]{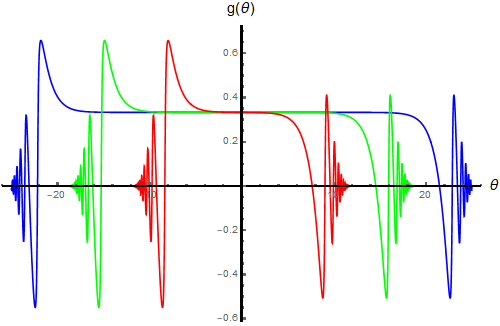}
    \caption{The formation of the plateau in $g(\theta) = -\frac{1}{2 \pi}\cI m
    \left( \log\left(1+e^{-f_\nu(\theta + i 0^+)} \right)\right)$ as $m \rightarrow 0$ at $\beta^2=1/2$, $\alpha_0=1/6$. For this choice of the parameters, the height of the plateau of $g(\theta)$ is $1/3$.}
    \label{fig:plateau}
\end{figure}

In order to find the solution to (\ref{eq:EsCFTdef}) for generic $k$, one must first solve (\ref{eq:EsCFTdef}) for $k=s$ $(k'=s')$. In this case, the solution can be reconstructed perturbatively as
\beq
\label{eq:solEs}
I_{s}^{(s',\pm)}(R,\ta) =I_{s'}^{(s',\mp)}(R,\ta) = \sum_{j=0}^{\infty} \frac{(-1)^j}{j+1}\, \binom{j(1+s)+(s-1)}{j}\,(2\ta)^j\,\frac{\(2\pi\,a_{s}^{(\pm)}\)^{j+1}}{R^{j(1+s)+s}} \;.
\eeq
This expression can be resummed as
\begin{itemize}
    \item $\Bs=-1$ : \beq
    I_1^{(-1,\pm)}(R,\ta)  = \frac{R}{4\ta}\( -1+\sqrt{1+8\ta\,\frac{2\pi\,a_1^{(\pm)}}{R^2}} \) \;.
    \eeq
    Both the classical (see the $\Bs=-1$ case in (\ref{eq:cIdens})) and quantum results suggest that the leading perturbing operator corresponds to the Lorentz breaking operator typically appearing in effective field theories for discrete lattice models \cite{Lukyanov:1997wq}. 
    \item $\Bs=-2$ : 
    \beq 
    I_2^{(-2,\pm)}(R,\ta) = \frac{4R}{6\ta}\,\sinh{\( \frac{1}{3}\,\arcsinh{\( \frac{3\sqrt{3}}{2}\( 2\ta\,\frac{2\pi\,a_2^{(\pm)}}{R^3} \)^{1/2} \)} \)}\,.
    \eeq
\end{itemize}
For generic spin $\Bs<0$  the result can be written in terms of a single  generalised hypergeometric function:
\beq
I_{s}^{(s',\pm)}(R,\ta)=I_{s'}^{(s',\mp)}(R,\ta) = 
\frac{s\,R}{2\ta\,(1+s)} \( -1 +\,F_{s}\(-2\ta\,\frac{(1+s)^{1+s}}{s^s} \frac{2\pi\,a_{s}^{(\pm)}}{R^{1+s}} \) \) \;,
\label{eq:solEs2}
\eeq
where $F_s(x)$ is defined in (\ref{eq:hypergeo2}).
The total momentum and energy are
\beq
E(R)= I_1^{(s',+)}(R,\ta) +I_1^{(s',-)}(R,\ta)\,,\quad P(R)= I_1^{(s',+)}(R,\ta) -I_1^{(s',-)}(R,\ta)\;,
\eeq
where $I_1^{(s',\pm)}(R,\ta)$ are obtained by solving (\ref{eq:EsCFTdef}) with $k'=-k=-1$ using (\ref{eq:solEs2}).\\
Finally, notice that even spin charges do not, in general, correspond to local conserved currents in the sine-Gordon model. They can occasionally emerge from the  set of non-local charges, at specific rational values of  $\beta^2$. Our results concerning the exact quantum spectrum can be smoothly deformed in $\Bs$, therefore  they formally also describe  deformations of the sine-Gordon model by  non-local currents \cite{Bernard:1990ys}. Moreover, there are many integrable systems with extended symmetries where even spin charges appear. The sign of the corresponding  eigenvalues depends on the internal flavor of the specific soliton configuration considered. Since the flow equations (\ref{eq:sol})--(\ref{eq:energyBu}), should properly describe the evolution of the spectrum  driven by analogous deformations in a very wide class of systems, perturbations by  currents with  $\Bs$ even, may lead to interesting quantum gravity toy models where the effective sign of the perturbing parameter $\ta$ depends on the specific state under consideration. 
\vspace{0.5cm}\\
\fbox{\parbox{5.0cm}{\bf
Perturbations with $\Bs>0$\,:}}
\vspace{0.2cm}
\\
\noindent
In the  case $\Bs>0$, the left- and right-mover sectors are now coupled and the solution of the generalised Burgers equations become equivalent to the set of equations
\beq
I_k^{(s,\pm)}(R,\ta) = \frac{2\pi\,a_{k}^{(\pm)}}{\( R+2\ta\,I_s^{(s,\mp)}(R,\ta) \)^k} \;. 
\label{eq:spositive}
\eeq
While, for models with  $\mathcal{P}_k^{(s)}=I_k^{(s,+)}(R,\ta)-I_k^{(s,-)}(R,\ta)=0$  the left- and right-mover sectors are completely symmetric and the relations (\ref{eq:spositive}) formally reduce to the equations (\ref{eq:EsCFTdef}), for $\mathcal{P}_k^{(s)} \ne 0$ the solution to (\ref{eq:spositive}) for $k=s$ are
\beqa
I_{s}^{(s,\pm)}(R,\ta) &=& \frac{2\pi\,a_{s}^{(\pm)}}{R^s}\Biggl[1+2\pi\,a_{s}^{(\mp)}\sum_{k=1}^{\infty}\sum_{l=0}^{k-1}\frac{1}{l+1} \, 
\left(\begin{array}{c} (k-l)s+l-1 \\ l
\end{array}\right) \, \left(\begin{array}{c} (l+1)s+k-l-1 \\ k-l
\end{array}\right) \notag \\
&\times& \left(2\pi\,a_{s}^{(\pm)}\right)^l \left(2\pi\,a_{s}^{(\mp)}\right)^{k-l-1}\frac{(-1)^k(2\ta)^k}{R^{(s+1)k}}\Biggr]\,.
\label{eq:Isum}
\eeqa
We were not able to find a general compact expression of (\ref{eq:Isum}), except for the already known ($\Bs=1$) $\TbT$-related result
\beqa
\mathcal{E}_1^{(1)}(R,\ta) &=& \frac{R}{2\ta}\( -1 + \sqrt{1+\frac{4\ta}{R}\(\frac{2\pi\,\bigl(a_1^{(+)} + a_1^{(-)}\bigr)}{R}\)+\frac{4\ta^2}{R^2}\(\frac{2\pi\,\bigl(a_1^{(+)} - a_1^{(-)}\bigr)}{R}\)^2} \) \;, \notag \\
\mathcal{P}_1^{(1)}(R) &=& \frac{2\pi\,\bigl(a_1^{(+)} - a_1^{(-)}\bigr)}{R} \;.
\eeqa
\vspace{0.5cm}\\
\fbox{\parbox{5.0cm}{\bf
Perturbations with $\Bs\rightarrow 0$\,:}}
\vspace{0.2cm}\\
\noindent
Let us first perform the $\Bs \rightarrow 0$ limit in the massive case. In this limit $\gamma_s \rightarrow 0$ but, rescaling $\tau$, we can recast the driving term in the form
\beq 
\nu(R,\alpha_0\,|\,\theta)=i 2\pi \alpha_0 -i\, m\rr \sinh(\theta) + i  \tau \mathcal{G}\; \cosh(\theta)\,,
\eeq
where
\beq
\mathcal{G} =  -\frac{1}{2} \left( \int_{\mathcal{C}_1} \frac{d\theta}{2 \pi i}    \log\left( 1 + e^{-f_{\nu}(\theta)} \right)-\int_{\mathcal{C}_2} \frac{d\theta}{2 \pi i}   \log\left( 1 + e^{f_{\nu}(\theta)} \right)\right) \;.
\label{eq:ints0}
\eeq
Next, in the limit $m \rightarrow 0$, the leading contribution  to  $\mathcal{G}$ in (\ref{eq:EECFT}) is coming from a large  plateau  of the integrand ($g(\theta)$ in figure \ref{fig:plateau}) of  height
\beq
-i f_\nu(\theta)  \sim 4 \pi \alpha_0 \beta^2\,,
\eeq
and width growing as  $\sim 2 \log(m R/2)$: 
\beq
\mathcal{G}  \sim    \log\left ( \frac{m R}{2} \right)   2\alpha_0  \beta^2\,.
\eeq
The emergence of the plateau can be deduced analytically from the NLIE (see for example \cite{ZamolodchikovTBA} for a related  discussion in the TBA context). 
Therefore, at fixed finite $R$,  after  a further rescaling 
$\tau \rightarrow \tau \pi / \log(mR/2)$, as $m$ tends  to zero, we obtain for the driving terms in the NLIEs:
\beq 
\nu^{(\pm)}=\nu^{(\pm)}(\mathcal{R}_0^{(\pm)},\alpha_0\,|\,\theta)=i 2\pi \alpha_0 \mp i\, \frac{\hat{m}}{2}\,e^{\pm \theta} \mathcal{R}_0^{(\pm)} \,,
\label{eq:s0nu}
\eeq
with
\beq
\mathcal{R}_0^{(\pm)} = R  \mp  \ta Q_0\;,\quad  Q_0 =2\pi \alpha_0 \beta^2 \, .
\label{eq:R01}
\eeq
Thus,  perturbing the theory with the  phase factor (\ref{eq:CDDs}) with $\Bs=0$ is equivalent, in an appropriate scaling limit, to a constant shift of the volume $R$. At fixed normalisation 
of the deforming parameter $\tau$, the shift turns out to be directly proportional to the topological charge $Q_0$. We see, from  (\ref{eq:s0nu}) that the left- and right-mover sectors remain decoupled also at $\tau \ne 0$. However, starting directly from the CFT limit, one can argue that there exist less trivial ways to couple the two sectors. The most general variant involves four different  coupling constants $\vec \tau =\{ \tau^{(a|b)} \}$ with $a,b \in \{+,- \}$:
\beq
\mathcal{R}_0^{(\pm)}(\vec \tau) = R  \mp  \left(\ta^{(\pm|+)} Q_0^{(+)}+\ta^{(\pm|-)} Q_0^{(-)}\right) \;,\quad Q_0^{(\pm)} =2 \pi \alpha_0^{(\pm)} \beta^2\,.
\label{eq:s0four}
\eeq
Notice that in (\ref{eq:s0four}) we allowed the possibility  for two different topological charges $Q^{(\pm)}_0$, associated  to the $U(1)_R \times U(1)_L$ symmetry of the $c=1$ free (compactified) boson model, corresponding to the CFT limit of the sine-Gordon theory.

\subsection{Further deformations involving the topological charge}
\label{eq:furhterD}
A further, natural extension of the  quantum models studied in  section \ref{sec:generalisedBurgers}, corresponds to scattering phase factors of the form
\beq
\delta^{(\tilde \Bs,\Bs)}( \theta, \theta') = \tau \, \gamma_{\tilde \Bs} \gamma_{\Bs} \sinh(\tilde \Bs\,\theta - \Bs\,\theta' ) \;.
\label{eq:CDDsJT}
\eeq
Deforming the kernel according to (\ref{eq:modkernel}) and  using (\ref{eq:CDDsJT}) instead of (\ref{eq:CDDs}), the driving term of (\ref{eq:sGDDV}) becomes
\beq
\label{eq:modsGDDV3}
\nu=\nu(R,\alpha_0\,|\,\theta) -i \tau \tilde \Bs\,\gamma_{\tilde s} \Bigl( e^{\tilde \Bs\,\theta} I^{(\tilde \Bs,\Bs,-)}_\Bs -  e^{-\tilde \Bs\,\theta} I^{(\tilde \Bs,\Bs,+)}_\Bs \Bigr) \;,
\eeq
where $I_{\Bk}^{(\tilde \Bs,\Bs,\pm)}$ are defined through (\ref{eq:EECFT}) and (\ref{eq:DDVcharges}) with the deformed driving term (\ref{eq:modsGDDV3}). The cases with $\tilde \Bs=\pm 1$, have been extensively discussed in the previous sections, they all correspond  to gravity-like theories, where the effect of the perturbation can be re-absorbed into a redefinition of the volume $\rr$ plus a shift in the rapidity $\theta$, according to (\ref{eq:modsGDDV2}).  From equation (\ref{eq:modsGDDV3}), we see that  for generic values of $\tilde \Bs \ne \pm 1$ it is not no longer possible to re-absorb the perturbation in the same fashion.  However, another interesting example is recovered by considering  the scaling limit $\{ \tilde \Bs,\gamma_{\tilde \Bs}, m \}  \rightarrow \{0,0,0\}$, such that
$\tilde \Bs\,\gamma_{\tilde \Bs}$ remains finite. In the following we shall set 
$\tilde \Bs\,\gamma_{\tilde \Bs} = 2\pi$. This situation corresponds to the standard massless limit where, in the undeformed $(\ta=0)$ NLIE, the right- $(+)$ and the left-mover $(-)$ sectors are completely decoupled, while a residual interaction between the two sectors is still present at $\tau \ne 0$. In fact, the deformed versions of $f_{\nu^{(\pm)}}(\theta)$ fulfil (\ref{eq:DDVCFT}) with
\beq 
\label{eq:modCFTDDV}
\nu^{(\pm)} = \nu^{(\pm)}\bigl(\rr,\alpha_0^{(\pm)}-\alpha^{(\Bs,\pm)}\,|\,\theta\bigr) \;,\quad \alpha^{(\Bs,\pm)} = \Bigl( \ta^{(\pm)}  I^{(\Bs,-)}_\Bs(\rr,\vec \ta ) -  \ta^{(\mp)}  I^{(\Bs,+)}_\Bs(\rr,\vec \ta) \Bigr) \;,
\eeq
where  the charges  $I^{(\Bs,\pm)}_\Bk$ are defined through (\ref{eq:EECFT}) with  driving term (\ref{eq:modCFTDDV}), and $\{ \tau^{(-)} , \tau^{(+)} \} = \vec \tau$ are two coupling  parameters defined as
\beq 
\tau^{(\pm)} = \tau e^{\pm\sigma}\,,\quad 
\pm\sigma = \lim_{\substack{\theta\to \pm\infty \\ \tilde \Bs\to 0^+}} \tilde \Bs\,\theta \;.
\label{eq:scaling}
\eeq
Therefore, the contributions of the nontrivial interaction can be formally re-absorbed in a redefinition of the vacuum  parameters $\alpha_0^{(\pm)}$. In turn, this affects the value of the effective central charges as
\beq
c^{(\Bs,\pm)} = c_0^{(\pm)} + 24\beta^2  (\alpha_0^{(\pm)})^2 - 24 \beta^2 \left((\alpha_0^{(\pm)})^2 - \alpha^{(\Bs,\pm)} \right)^2 \;,\quad c_{0}^{(\pm)}= c-24\beta^2(\alpha_0^{(\pm)})^2 \;.
\label{eq:cred}
\eeq
Considering the formal identification $Q_0^{(\pm)} =2\pi \alpha_0^{(\pm)} \beta^2$ made in (\ref{eq:R01}), the redefinition (\ref{eq:modCFTDDV}) corresponds to the following dressing of the topological charges
\beq
Q^{(\Bs,\pm)}(R,\vec \tau)= Q_0^{(\pm)} -  \frac{\kappa}{4 \pi} \Bigl( \ta^{(-)}  I^{(\Bs,-)}_\Bs(\rr,\vec \ta ) -  \ta^{(+)}  I^{(\Bs,+)}_\Bs(\rr,\vec \ta) \Bigr)  \;,
\label{eq:Qdef}
\eeq
with 
\beq
\kappa = 8\pi^2\beta^2\,. 
\eeq
Let us focus on the $\Bs=1$ case. The equations for the spectrum are
\beq
I^{(\pm)}_1(R,\vec \ta) = \frac{2 \pi}{\rr}\left( n^{(\pm)} - \fract{c^{(\pm)}}{24}\right) \;,\quad (I^{(\pm)}_1=I^{(1,\pm)}_1 \;,\; c^{(\pm)} = c^{(1,\pm)}) \;,
\label{eq:JTT}
\eeq
which can be solved exactly for $I^{(\pm)}_1(\rr,\vec \ta)$ at any values of the parameters $\tau^{(\pm)}$. However, since the general analytic expressions for $I^{(\pm)}_1$ are very cumbersome, we will restrict the discussion to specific scaling limits:
\begin{itemize}
\item $\tau^{(+)}=\tau^{(-)}=\tau$ :
\begin{gather}
I^{(+)}_1(\rr,\ta) - I^{(-)}_1(\rr,\ta) = P(\rr) = \frac{2 \pi}{\rr} \left(h^{(+)}  -h^{(-)} \right) \;, \label{eq:JTfirst1}\\
I^{(+)}_1(\rr,\ta) = I_1^{(+)}(\rr) + \tau \frac{Q_0^{(\pm)}}{\rr} P(\rr) + \tau^2 \frac{\kappa}{8 \pi \rr} P^2(\rr) \;,
\label{eq:JTfirst2}
\end{gather}
with $h^{(\pm)}= h_0^{(\pm)}+  n^{(\pm)}$ and (\ref{eq:Qdef}) becomes
\beq
Q^{(\pm)}(R,\tau)= Q_0^{(\pm)} + \tau \frac{\kappa}{4 \pi}\,P(R) \;.
\label{eq:R02}
\eeq
Surprisingly, with the replacement
\beq
I^{(\pm)}_1(R,\ta) = \frac{2 \pi}{\rr}\left( h^{(\pm)}(\tau) - \frac{c}{24}\right) \;,
\label{eq:JTT0}
\eeq
equations (\ref{eq:JTfirst1}) and (\ref{eq:JTfirst2}) lead to  exact expressions for the conformal dimensions $h^{(\pm)}(\tau)$, which precisely match the form of the (left) conformal dimension in the $\JbT$ model, as recently shown in \cite{Guica:2019vnb}:
\beqa
h^{(\pm)}(\tau) = h^{(\pm)} + \tau \frac{Q_0^{(\pm)}}{2 \pi}  P +  \tau^2 \frac{\kappa}{ 16 \pi^2} P^2\,.
\eeqa
As we will shortly see,  this is  the first instance among many that link the phase factor (\ref{eq:CDDsJT}) with $\Bs=1$, in the scaling limit (\ref{eq:scaling})  to the quantum $\JbT$ model. 

    \item $\tau^{(+)}=0$ , $\tau^{(-)} =\ta$ : the two sectors $(\pm)$ are completely decoupled
    \beqa
    \label{eq:JTsymeq}
    I^{(\pm)}_1(\rr,\ta) &=&
    I^{(\pm)}_1(\rr) \pm \tau \frac{Q_0^{(\pm)}}{\rr} I^{(\pm)}_1(\rr,\ta) + \tau^2 \frac{\kappa}{8 \pi \rr} \(I^{(\pm)}_1(\rr,\ta)\)^2 \;, \\
    Q^{(\pm)}(R,\ta) &=& Q_0^{(\pm)} \pm\ta\frac{\kappa}{4\pi}\,I_1^{(\pm)}(R,\ta) \;,
    \eeqa
    and the solutions to (\ref{eq:JTsymeq}) are
    \beq
    \label{eq:JTsymsol}
    I^{(\pm)}_1(\rr,\ta) = \frac{4\pi}{\kappa \ta^2}\(\rr\mp\ta Q_0^{(\pm)}-\sqrt{\left(\rr\mp\ta Q_0^{(\pm)}\right)^2-\ta^2 \kappa \(h^{(\pm)}-\frac{c}{24}\)}\) \;.
    \eeq
    We notice that (\ref{eq:JTsymeq}) can be rewritten as
    \beq 
    I_1^{(\pm)}(\rr,\ta) = \frac{2\pi\(h^{(\pm)}-\frac{c}{24}\)}{ \rr-\ta\(\pm Q_0^{(\pm)} +\ta\frac{\kappa}{8 \pi}I_1^{(\pm)}(\rr,\ta)\)} \;,
    \eeq
    which suggests that, in this limit, the perturbation has a dual geometric description, since it can be interpreted as a redefinition of the length $\rr$. The dual deformation of the NLIE corresponds to a deformed version of $f_{\nu^{(\pm)}}(\theta)$ which fulfils (\ref{eq:DDVCFT}) with 
    \beq 
    \label{eq:modCFTDDV2}
    \nu^{(\pm)}=\nu^{(\pm)}\(\mathcal{\rr}^{(\pm)}_0,\alpha_0\;\vline\;\theta\) \;,\quad \mathcal{\rr}^{(\pm)}_0 = \rr - \ta\(\pm Q_0^{(\pm)} +\ta\frac{\kappa}{8 \pi}I_1^{(\pm)}(\rr,\ta)\) \;.
    \eeq
   Expressions (\ref{eq:JTsymsol}) are trivially solutions of two decoupled Burgers-like equations
    \beq  
    \partial_\ta I_1^{(\pm)}(\rr,\ta) \pm Q^{(\pm)}(R,\ta)\,\partial_R I_1^{(\pm)}(\rr,\ta) = 0 \;.
    \eeq
    Therefore, $I_1^{(\pm)}(R,\ta)$ fulfils a Burgers-type equation analogous to (\ref{eq:Burgers}), where $Q^{(\pm)}(R,\ta)$ play the role of velocities.
\end{itemize}
\subsection{The quantum $\JbT$ model}
\label{sec:TJ}
Formula (\ref{eq:JTsymsol}) strongly resembles the expression for  the energy of the right-movers of  the  $\JbT$-deformed CFT derived in \cite{Bzowski:2018pcy,Chakraborty:2018vja}. To obtain the $\JbT$ model, one must treat the $(\pm)$ sectors in a non-symmetric way. For $\Bs=1$, a possible asymmetric generalisation of (\ref{eq:modCFTDDV}) with four free parameters $\vec{\tau}= \{\tau^{(a|b)} \}$ with  $a,b \in \{+,-\}$  is
\begin{gather}
\nu^{(\pm)} = \nu^{(\pm)}\bigl(\rr,\alpha_0^{(\pm)}-\alpha^{(\pm)}(\vec \tau)\,|\,\theta\bigr) \;, \notag \\
\alpha^{(\pm)}(\vec \tau) = \Bigl( \tau^{(\pm|-)}  I^{(-)}_s(\rr,\vec \ta) -  \tau^{(\mp|+)}  I^{(+)}_s(\rr,\vec \ta) \Bigr) \;.
\label{eq:nunonsym}
\end{gather}
In (\ref{eq:nunonsym}) we also included the possibility to have two different initial values of the twist parameters $\alpha_0^{(\pm)}$ in the two sectors.
From (\ref{eq:nunonsym}), the $\JbT$ model is recovered with the choice
\beq
\tau^{(\pm|-)} = 0\;,\quad \tau^{(\mp|+)}=\tau\;,\quad \alpha_0^{(+)}=\alpha_0^{(-)} = \alpha_0 \;.
\eeq
Correspondingly, relations (\ref{eq:nunonsym}) become
\beq 
\nu^{(\pm)} = \nu_{\JbT}^{(\pm)}\(R,\alpha_0-\alpha_{\JbT}^{(\pm)}\;\vline\;\theta\) \;,\quad \alpha_{\JbT}^{(\pm)} = -\ta\,I_{\JbT}^{(+)}(R,\ta) \;.
\eeq
Alternatively, the $(+)$ sector can be equivalently described with a redefinition of the length $R$ as
\beq  
\label{eq:DDVJTp}
\nu^{(+)} = \nu^{(+)}\( \mathcal{\rr}_{\JbT},\alpha_0\;\vline\;\theta \) \;,\quad \mathcal{\rr}_{\JbT} = \rr - \ta\( Q_0 + \ta\frac{\kappa}{8\pi}\,I_{\JbT}^{(+)}(\rr,\ta) \) \;.
\eeq
The right-moving solution in (\ref{eq:JTsymsol}) and the topological charge become
\begin{gather}
\label{eq:JTIpIm}
I^{(+)}_1(\rr,\ta) = I^{(+)}_{\JbT}(\rr,\ta) = \frac{4\pi}{\kappa\ta^2}\(\rr-\ta Q_0-\sqrt{\left(\rr-\ta Q_0\right)^2-\ta^2 \kappa\(h^{(+)}-\frac{c}{24}\)}\) \;,\quad \\
I_{\JbT}^{(+)}(\rr,\ta) - I_{\JbT}^{(-)}(\rr,\ta) = P(R) = \frac{2\pi}{\rr}\(h^{(+)}-h^{(-)}\) \;, \\
\label{eq:JTQpdef}
Q^{(+)}(R,\ta) = Q_{\JbT}^{(+)}(R,\ta) = Q_0 + \ta\,\frac{\kappa}{4\pi}\,I_{\JbT}^{(+)}(R,\ta) \;.
\end{gather} 
Therefore, $Q_0 = 2\pi \alpha_0\beta^2$ and $\kappa= 8\pi^2\beta^2$ have been again consistently identified with the topological charge and the chiral anomaly, respectively. The results described here are in full agreement with \cite{Bzowski:2018pcy} and the classical results presented in section \ref{sec:TTsHC}.
\subsection{A simple example involving a pair of scattering phase factors}
As already discussed, in principle one may introduce several scattering phase factors to deform the NLIEs. In this section we will consider a particular combination of $\Bs\to 0$ and $\tilde\Bs \to 0$ scattering factors which allows us to match with the classical results (\ref{eq:s0symIpm})--(\ref{eq:s0symQpm}). We consider a double perturbation made of a length redefinition (\ref{eq:s0four}) with $\ta^{(\pm|+)}=\mp\ta$ and $\ta^{(\pm|-)}=\pm\ta$, together with a shift of the twist parameter (\ref{eq:nunonsym}) with $\ta^{(\mp|+)}= +\ta $ and $\ta^{(\pm|-)} = +\ta$. The corresponding deformed driving term is then
\beq 
\nu^{(\pm)}=\nu^{(\pm)}\( \mathcal{R}_0^{(\pm)},\alpha_0^{(\pm)}-\alpha^{(\pm)}(R,\ta)\;\vline\;\theta \) \;,
\eeq
with
\begin{gather}  
\alpha^{(\pm)}(R,\ta) = -\ta\,P(R) \;, \notag \\
\mathcal{R}_0^{(\pm)} = R+\ta\( Q^{(+)}(R,\ta) - Q^{(-)}(R,\ta) \) = R+\ta\(Q_0^{(+)} - Q_0^{(-)}\) \;,
\end{gather}
where in the last equality we used the fact that $Q^{(\pm)}(R,\ta) = 2\pi\,\beta^2\(\alpha_0^{(\pm)} - \alpha^{(\pm)}(R,\ta)\)$. Since the central charges are affected by the deformation as
\beq 
c^{(\pm)}(R,\ta) = c + 24\beta^2\bigl(\alpha_0^{(\pm)}\bigr)^2 -24\beta^2\(\alpha_0^{(\pm)} - \alpha^{(\pm)}(R,\ta) \)^2 \;,
\eeq
one finds
\begin{gather}
\label{eq:s0IpIm}
I_1^{(\pm)}(R,\ta) = \frac{2\pi}{\mathcal{R}_0^{(\pm)}}\( n^{(\pm)} - \frac{c^{(\pm)}(R,\ta)}{24} \) = \frac{R\,I_1^{(\pm)}(R)+\ta\,Q_0^{(\pm)}P(R)+\frac{\kappa}{8\pi}\,\ta^2\,P^2(R)}{R+\ta\(Q_0^{(+)} - Q_0^{(-)}\)} \;, \\
I_1^{(+)}(R,\ta) - I_1^{(-)}(R,\ta) = P(R) \;, \\
\label{eq:s0QpQmdef}
Q^{(\pm)}(R,\ta) = Q_0^{(\pm)} + \ta\frac{\kappa}{4\pi}\,P(R) \;,
\end{gather}
which match exactly with the results (\ref{eq:s0symIpm})--(\ref{eq:s0symQpm}) obtained at the classical level for the corresponding densities.
\section{Conclusions}
In this paper, we have introduced and studied novel types of systems obtained by a local space-time deformation induced by  higher-spin conserved charges.
We have argued that these geometrical maps correspond
to classical and quantum field theories perturbed by irrelevant Lorentz-breaking composite fields. The possibility of non-Lorentz invariant generalisations of the $\TbT$ operator was first briefly discussed in  \cite{Smirnov:2016lqw} and 
the $\JbT$ and the $\TbJ$ models studied in \cite{Guica:2017lia, Bzowski:2018pcy, Apolo:2018qpq, Aharony:2018ics, Chakraborty:2018vja, Nakayama:2018ujt, Araujo:2018rho} are concrete realisations   of these ideas, though with some important difference in the actual details. Another concise discussion on non-Lorentz invariant perturbations has appeared more recently in \cite{Cardy:2018jho} were it was underlined the possibility to replace the $\TbT$ operator with composite fields built out of a pair of conserved currents, a somehow obvious generalisation of the earlier comments \cite{Smirnov:2016lqw}  and results \cite{Guica:2017lia, Bzowski:2018pcy, Apolo:2018qpq, Aharony:2018ics, Chakraborty:2018vja, Nakayama:2018ujt, Araujo:2018rho} which however also encompasses the systems effectively studied in the current paper. Here we would like to stress again that the spin $s$ in the scattering phase factor (\ref{eq:modkernel}) can also assume non-integer values, therefore the extension to non-local conserved charges \cite{Bazhanov:1996dr} appears to be straightforward, as also underlined in \cite{Smirnov:2016lqw}.

Many open problems deserve further attention. First of all, the identification between the classical and the quantum results is, in our opinion, very convincing, but certainly more work is needed to fully confirm the match. In particular, a Lagrangian description for the higher-spin deformed models would lead the way to more stringent tests. Furthermore, it is natural to try to identify within the AdS$_3$/CFT$_2$ setup the corresponding (bulk) JJ and Yang-Baxter type  deformations, as it was successfully done for their close relatives studied \cite{Guica:2017lia, Bzowski:2018pcy, Apolo:2018qpq, Aharony:2018ics, Chakraborty:2018vja, Nakayama:2018ujt, Araujo:2018rho}. It would be important to extend the gravity setup adopted in \cite{Dubovsky:2017cnj, Dubovsky:2018bmo}  to accommodate also these novel geometric deformations. 

While the $\TbT$ perturbation is compatible with supersymmetry \cite{Baggio:2018gct, Dei:2018mfl, Baggio:2018rpv,Chang:2018dge,Nakayama:2018ujt},  higher spin Hamiltonians should, at least partially,  lift the degeneracy of the states. Therefore, in general, supersymmetry should be explicitly broken by the more general perturbations discussed here.  However, it would be interesting  to find specific examples were a residual  supersymmetry survives.  Finally, it would be also nice to study the effect of these perturbations on 2D Yang-Mills \cite{Conti:2018jho, Santilli:2018xux} and, following the suggestion of \cite{Jiang:2019tcq}, to study the expectation values on curved spacetimes, of the composite operators discussed here. 
Finally, while this paper was in its  final writing stage the  works \cite{Giveon:2019fgr,LeFloch:2019rut} appeared. The  latter articles contain many interesting results on the $\JbT$ model and the combined $\JbT$/$\TbT$ perturbations. There are only minor overlaps with the current paper. As already mentioned, our methods are easily generalisable to encompass such a two-coupling extension of the models described here, both at classical and quantum levels.

\medskip
\noindent{\bf Acknowledgments --}
We are especially grateful to Riccardo Borsato, Patrick Dorey, Ferdinando Gliozzi, Yunfeng Jiang, Zohar Komargodski,  Bruno Le Floch, Sergei Lukyanov, Marc Mezei, Domenico Orlando, Alessandro Sfondrini, Kentaroh Yoshida and Sasha Zamolodchikov for useful discussions.
This project was partially supported by the INFN project SFT, the EU network GATIS+, NSF Award  PHY-1620628, and by the FCT Project PTDC/MAT-PUR/30234/2017 ``Irregular connections on algebraic curves and Quantum Field Theory".
RT gratefully acknowledges support from the Simons Center for Geometry and Physics, 
Stony Brook University at which some of the research for this paper was performed.

\appendix
\section{Deformed classical solutions}
\label{sec:deflogsol}
Following \cite{Conti:2018tca}, we shall now describe how to derive explicitly a deformed solution $\phi^{(\Bs)}(\mathbf z,\ta)$ associated to the change of variables (\ref{eq:JacobianTTs}) starting from the undeformed solution $\phi(\mathbf w)$. First of all we compute the relation between the sets of coordinates $\mathbf w$ and $\mathbf z$ by integrating, for $\Bs \ge 0$, (\ref{eq:generalJacobianHC2}) for $\mathbf z$
\beq
\label{eq:system}
\begin{cases}
\frac{\partial z^{(s)}(\mathbf w)}{\partial w} = 1 + 2\ta\,\bar\Th_{s-1}(\mathbf w)\,, \\
\frac{\partial z^{(s)}(\mathbf w)}{\partial\bar w} = 2\ta\,\bar\T_{s+1}(\mathbf w)\,,
\end{cases} \;\quad
\begin{cases}
\frac{\partial\bar z^{(s)}(\mathbf w)}{\partial w} = 2\ta\,\T_{s+1}(\mathbf w)\,, \\
\frac{\partial\bar z^{(s)}(\mathbf w)}{\partial\bar w} = 1 + 2\ta\,\Th_{s-1}(\mathbf w)\,,
\end{cases} 
\eeq
where we have denoted $\mathbf z = \mathbf z^{(s)}$ to distinguish between the different perturbations. The components of the higher charges $\T_{s+1}(\mathbf w)$, $\Th_{s-1}(\mathbf w)$ along with their complex conjugates are implicitly evaluated on the specific field configuration $\phi(\mathbf w)$. Inverting the relation $\mathbf z^{(s)}(\mathbf w)$, we find the deformed solution as
\beq 
\label{eq:solrel}
\phi^{(s)}\bigl(\mathbf z,\ta\bigr) = \phi\Bigl(\mathbf w\bigl(\mathbf z^{(s)}\bigr)\Bigr) \;.
\eeq
Similarly, we can apply the same strategy to derive a deformed solution associated to the change of variables (\ref{eq:generalJacobianHC1}), for $\Bs \leq 0$. In this case we should, instead, integrate the system
\beq
\label{eq:systemsneg}
\begin{cases}
\frac{\partial z^{(s')}(\mathbf w)}{\partial w} = 1 + 2\ta\,\T_{s+1}(\mathbf w)\,, \\
\frac{\partial z^{(s')}(\mathbf w)}{\partial\bar w} = 2\ta\,\Th_{s-1}(\mathbf w)\,,
\end{cases} \;\quad
\begin{cases}
\frac{\partial\bar z^{(s')}(\mathbf w)}{\partial w} = 2\ta\,\bar\Th_{s-1}(\mathbf w) \,,\\
\frac{\partial\bar z^{(s')}(\mathbf w)}{\partial\bar w} = 1 + 2\ta\,\bar\T_{s+1}(\mathbf w)\,,
\end{cases} \;
\eeq
and repeat the same procedure described above.
\subsection{Explicit Lorentz breaking in a simple example}
\begin{figure}[t]
  \centering
  \hspace{-0.5cm}
  \subfloat[$\Bs=1$]{\includegraphics[scale=0.20]{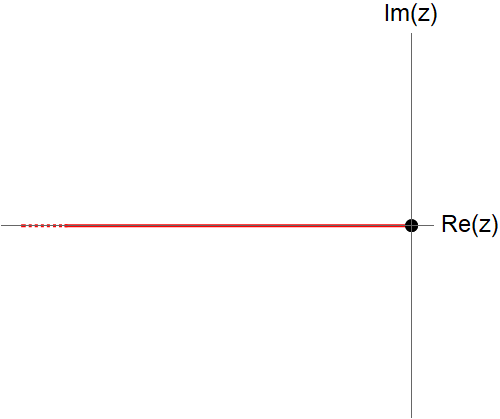}
  \label{fig:logsolplotp1}}
  \hspace{0.5cm}
  \subfloat[$\Bs=-1$]{\includegraphics[scale=0.20]{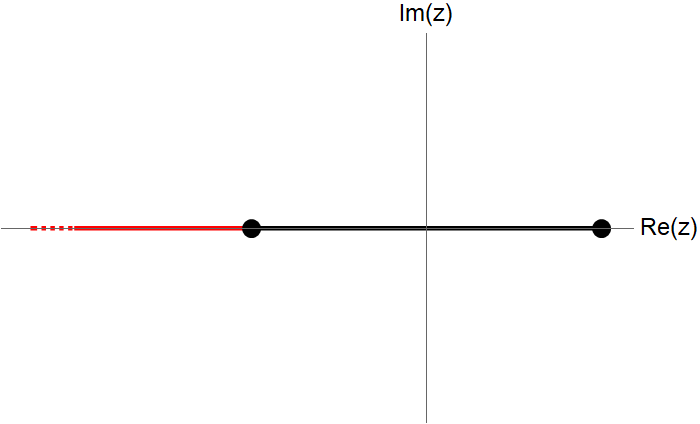}
  \label{fig:logsolplotm1}}
  \hspace{0.5cm}
  \subfloat[$\Bs=-3$]{\includegraphics[scale=0.20]{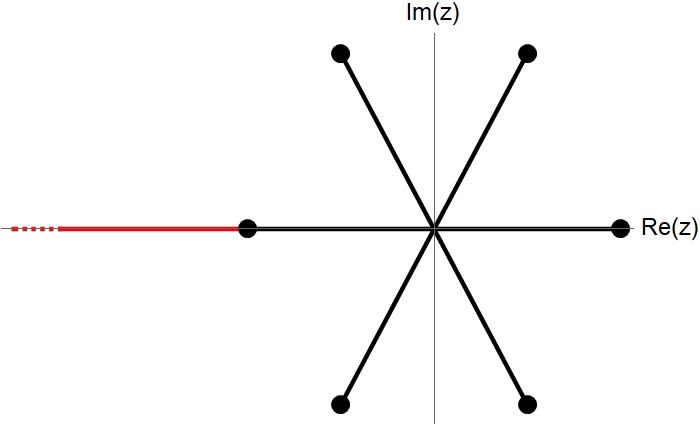}
  \label{fig:logsolplotm2}}
  \vspace{0.1cm}
 \caption{Analytic structure of the deformed solution (\ref{eq:deflogsol1}) in the complex plane of $z$ for different values of $\Bs$, with $\xi=\bar{\xi}=0$.  The black lines correspond to the square root branch cuts connecting the singularities (\ref{eq:branchpts}), while the red lines correspond to the logarithmic cuts.}
    \label{fig:logsolplot}
\end{figure}

\begin{figure}[t]
  \centering
  \subfloat[]{\includegraphics[scale=0.40]{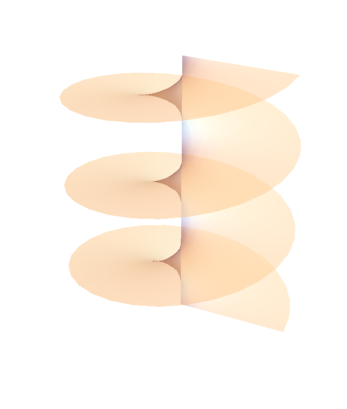}
  \label{fig:Riemannp1}}
  \hspace{0.5cm}
  \subfloat[]{\includegraphics[scale=0.40]{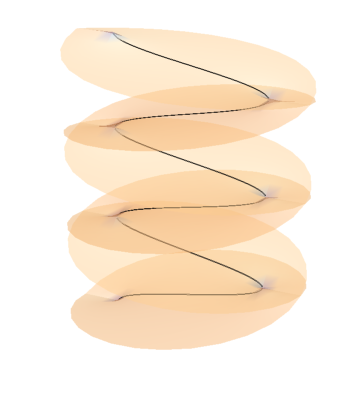}
  \label{fig:Riemannm1}}
  \vspace{0.1cm}
 \caption{Riemann surface of the deformed solution $\phi^{(\Bs)}(\mathbf z,\ta)$ , $(\xi=\bar\xi=0)$, in the complex plane of $z$ for $\Bs=1$ (a), and $\Bs=-1$ (b).}
    \label{fig:Riemann}
\end{figure}

The purpose of this section is to highlight the difference between the $\TbT$ perturbation and those corresponding to spins $\Bs \ne 1$, through
the study of a simple example. Since perturbations with higher spins should break explicitly Lorentz symmetry, it is convenient to start  from a solution of the Laplace equation which is particularly symmetric under rotations i.e. the  ``spiral staircase" solution of \cite{Gutshabash:2017qxl}:
\beq
\label{eq:logsol}
\phi(\mathbf w) = 
d\,\log{\(\frac{w+\xi}{\bar w+\bar\xi}\)} \;,\quad( \xi,\bar\xi \in \mathbb{C} \;,\quad d \in \mathbb{R}) \;,
\eeq
where $w$ and $\bar w$ are complex conjugated variables related to the cartesian coordinates $\mathbf y = (y^1,y^2)$ through (\ref{eq:EuclLC}). Using
\beq
\partial_{\bar w} \frac{1}{w+\xi} = -\pi\,\delta(w+\xi) \;,\quad \partial_w \frac{1}{\bar w+\bar\xi} = -\pi\,\delta(\bar w+\bar\xi) \;,
\eeq
one can show that (\ref{eq:logsol}) is indeed solution to the undeformed EoMs: $\partial_w\partial_{\bar w}\phi=0$.\\
For sake of brevity, we will only consider  perturbations induced by the set of charges (\ref{eq:NG1formcomp2}), the deformations associated to (\ref{eq:NG1formcomp1}) can be obtained from the results described here  through a simple redefinition of the coupling $\ta$.\\
Integrating (\ref{eq:system}) using the set of charges (\ref{eq:NG1formcomp2}) evaluated on the solution (\ref{eq:logsol}) we obtain
\beqa
\label{eq:systemsol1log}
z^{(s)}(\mathbf w) = w + \ta\,\frac{\Gamma\bigl(\frac{s+1}{2}\bigr)^2}{s}\frac{d^2}{(\bar w + \bar \xi)^{s}} \;, \quad
\bar z^{(s)}(\mathbf w) = \bar w + \ta\,\frac{\Gamma\bigl(\frac{s+1}{2}\bigr)^2}{s}\frac{d^2}{(w + \xi)^{s}} \;. 
\eeqa
The latter equations cannot explicitly be inverted, for generic spin $s$, as $\mathbf w\(\mathbf z^{(s)}\)$.
However, for $s=1$, {\it i.e.} the $\TbT$ perturbation, (\ref{eq:systemsol1log}) can be written as 
\beqa 
z + \xi = (w + \xi)\Biggl[ 1 + \ta \frac{d^2}{(w + \xi)(\bar w + \bar \xi)}\Biggr] \;, \quad
\bar z + \bar \xi &=& (\bar w + \bar \xi)\Biggl[ 1 + \ta \frac{d^2}{(w + \xi)(\bar w + \bar \xi)}\Biggr] \;,
\eeqa
with $\mathbf z=\mathbf z^{(1)}(\mathbf w)$, from which 
\beq
\phi(\mathbf w) = d\,\log{\(\frac{w+\xi}{\bar w+\bar \xi}\)} = d\,\log{\(\frac{z+\xi}{\bar z+\bar \xi}\)} = \phi(\mathbf z,\ta) \;.
\eeq
Therefore, the solution (\ref{eq:logsol}) is a fixed point of the $\TbT$ flow. 
To clearly see how  the $\Bs \ne 1$ perturbations affect the solution (\ref{eq:logsol}), we shall restrict  to  the $\Bs<0$ perturbations, where the deformed solutions can be found explicitly. Integrating (\ref{eq:systemsneg}) using the set of charges (\ref{eq:NG1formcomp2}), again evaluated on (\ref{eq:logsol}), we find
\beqa
\label{eq:systemsol1lognegs}
z^{(s')}(\mathbf w) = w + \ta\,\frac{\Gamma\bigl(\frac{s+1}{2}\bigr)^2}{s}\frac{d^2}{(w + \xi)^{s}} \;,\quad
\bar z^{(s')}(\mathbf w) = \bar w + \ta\,\frac{\Gamma\bigl(\frac{s+1}{2}\bigr)^2}{s}\frac{d^2}{(\bar w + \bar\xi)^{s}} \;.
\eeqa
Comparing (\ref{eq:systemsol1log}) with (\ref{eq:systemsol1lognegs}) we see that the difference between the $\Bs>0$ and $\Bs<0$ perturbations lies in the substitution $w + \xi \longleftrightarrow \bar w + \bar\xi$ in the term proportional to $\ta$, which implies that, for $\Bs<0$, there is no mixing between holomorphic and anti-holomorphic components.\\
Since holomorphic and anti-holomorphic parts are completely decoupled, we can integrate explicitly (\ref{eq:systemsol1lognegs}) for $w$ and $\bar w$. The result is
\beqa
\label{eq:inversionsystemsol1lognegs}
w\bigl(\mathbf z^{(s')}\bigr) &=& z^{(s')} + \frac{s}{1+s}\[ -1 + F_s\( \(\frac{1+s}{s}\)^{1+s} \Gamma\biggl(\frac{1+s}{2}\biggr)^2\frac{\ta\,d^2}{\bigl(z^{(s')} + \xi\bigr)^{1+s}} \) \]\bigl(z^{(s')} + \xi\bigr) \;, \notag \\
\bar w\bigl(\mathbf z^{(s')}\bigr) &=& \bar z^{(s')} + \frac{s}{1+s}\[ -1 + F_s\( \(\frac{1+s}{s}\)^{1+s} \Gamma\biggl(\frac{1+s}{2}\biggr)^2\frac{\ta\,d^2}{\bigl(\bar z^{(s')} + \bar\xi\bigr)^{1+s}} \) \]\bigl(\bar z^{(s')} + \bar\xi\bigr) \;, \notag \\
\eeqa
Finally, the deformed solutions are recovered by  plugging (\ref{eq:inversionsystemsol1lognegs}) into (\ref{eq:logsol})
\beq
\label{eq:deflogsol1}
\phi^{(s')}(\mathbf z,\ta) = d\,\log{\(\frac{z+\xi}{\bar z+\bar \xi}\)} + d\,\log{\(\frac{1+s\,F_s\( \(\frac{1+s}{s}\)^{1+s} \Gamma\bigl(\frac{1+s}{2}\bigr)^2\frac{\ta\,d^2}{(z + \xi)^{1+s}} \)}{1+s\,F_s\( \(\frac{1+s}{s}\)^{1+s} \Gamma\bigl(\frac{1+s}{2}\bigr)^2\frac{\ta\,d^2}{(\bar z + \bar \xi)^{1+s}} \)}\)} \;.
\eeq
Let us discuss in detail the case $\Bs=-1$ of (\ref{eq:deflogsol1}),
\beq
\label{eq:deflogsol2s-1}
\phi^{(-1)}(\mathbf z,\ta) =d\,\log{\(\frac{z+\xi}{\bar z+\bar \xi}\)} + d\,\log{\(\frac{1+\sqrt{1-\frac{4\ta\,d^2}{(z+\xi)^2}}}{1+\sqrt{1-\frac{4\ta\,d^2}{(\bar z+\bar \xi)^2}}}\)} \;.
\eeq
We observe that, as soon as the perturbation is switched on, a pair of square root branch points appears at $z = \pm 2d\sqrt{\ta}-\xi$. Considering for simplicity $\xi,\bar \xi \in \mathbb{R}$, they form a branch cut on the real axis of the complex plane of $z$
\beq  
\mathcal{C} = \( -2d\sqrt \ta - \xi ; +2d\sqrt \ta - \xi \) \;,
\eeq
{\it i.e.} the black line in Figure \ref{fig:logsolplotm1}.
Instead, the logarithmic singularity of the undeformed solution (\ref{eq:logsol}) at $w=-\xi$ cancels out with the singularities coming from the additional term in (\ref{eq:deflogsol2s-1}). Therefore, the logarithmic cut of (\ref{eq:logsol}) which runs, in our convention, along the real axis from $w=- \infty$ to $w=-\xi$, now connects $z=- \infty$  on the first sheet to $z=- \infty$  on the secondary branches reached by passing through $\mathcal{C}$ (see the red line in Figure \ref{fig:logsolplotm1}). This implies that the behaviour of (\ref{eq:deflogsol2s-1}) at $z=\infty$ is different according to the choice of the branch. On the first sheet one has
\beq  
\phi^{(-1)}(\mathbf z,\ta) \underset{\underset{(1\text{-th sheet})}{\mathbf z \to \infty}}{\sim} d\,\log{\(\frac{z+\xi}{\bar z+\bar \xi}\)} + \mathcal{O}(z^0) \;,
\eeq
while, on the second sheet, flipping the $+$ sign in front of the square roots in (\ref{eq:deflogsol2s-1}) into a $-$ sign one finds
\beq  
\phi^{(-1)}(\mathbf z,\ta) \underset{\underset{(2\text{-nd sheet})}{\mathbf z \to \infty}}{\sim} -d\,\log{\(\frac{z+\xi}{\bar z+\bar \xi}\)} + \mathcal{O}(z^0)\;.
\eeq
In Figure \ref{fig:Riemann}, is represented the Riemann surface of the solution (\ref{eq:deflogsol2s-1}) (Figure \ref{fig:Riemannm1}) together with that of the bare solution (Figure \ref{fig:Riemannp1}), which coincides with the $\TbT$ deformed solution. Notice that the analytic structure of (\ref{eq:deflogsol2s-1}) can be read out from the implicit map (\ref{eq:systemsol1lognegs}). In fact, for $\Bs=-1$, equation (\ref{eq:systemsol1lognegs}) reduces to the Zhukovsky transformation
\beqa
\label{eq:systemsollognegs}
z + \xi = (w + \xi) + \ta\,\frac{d}{(w + \xi)} \;, \quad
\bar z + \bar \xi = (\bar w + \bar \xi) + \ta\,\frac{d}{(\bar w + \bar \xi)} \;,
\eeqa
from which we see that $z=\infty$ on the first sheet is mapped into $w=\infty$, while $z=\infty$ on the second sheet is mapped into $w=-\xi$. Moreover (\ref{eq:systemsollognegs}) captures the large $z$ behaviour of the solution. In fact, 
\beqa
d\,\log{\(\frac{w + \xi}{\bar w + \bar \xi}\)} \underset{w \to \infty}{\sim} d\,\log{\(\frac{z + \xi}{\bar z + \bar \xi}\)} \;, \quad
d\,\log{\(\frac{w + \xi}{\bar w + \bar \xi}\)} \underset{w \to -\xi}{\sim} -d\,\log{\(\frac{z + \xi}{\bar z + \bar \xi}\)} \;.
\eeqa
Let us  now consider the generic solution (\ref{eq:deflogsol1}). The hypergeometric functions appearing in (\ref{eq:deflogsol1}) are of the form $\,_{p+1}F_p(a_1,\dots,a_{p+1};b_1,\dots,b_p;x)$, with coefficients $\lbrace a_i \rbrace_{i=1}^{p+1},\;\lbrace b_j \rbrace_{j=1}^{p}\in\mathbb{Q}$. Generally, these hypergeometric functions have branch points at $x=\infty$ and $x=1$, which in our case are mapped into $z=0$ and
\beq 
\label{eq:branchpts}
x_n = \frac{1+s}{s}\(d\,\Gamma\(\frac{1+s}{2}\)\)^{\frac{2}{1+s}}\ta^{\frac{1}{1+s}} e^{\frac{2\pi i}{1+s}n}-\xi \;,\quad (n=0,\dots,s) \;,
\eeq
respectively.  The branch points (\ref{eq:branchpts}) are all of square root type.
In conclusion, roughly speaking,  starting from a rotational-symmetric solution, the perturbations with  $\Bs<0$ have explicitly broken the original $U(1)$ symmetry down to a discrete $Z_{2s}$.
\section{Burgers equations}
\label{Appendix:burgers}
We want to prove that the general Burgers equations (\ref{eq:BurgersIpm}), which we report here for convenience
\beq
\partial_\ta I_{\Bk}^{(\Bs,\pm)}(R,\tau) + R' \partial_R  I_{\Bk}^{(\Bs,\pm)}(R,\tau)  = \pm  \Bk \, \theta'_0  I_{\Bk}^{(\Bs,\pm)}(R,\tau) \;,
\eeq
with $R'$ and $\theta_0'$ defined through (\ref{eq:sol}), reduce to
\beqa
\label{eq:BurgersIkredspos}
\partial_\ta I_k^{(s,\pm)}(R,\ta) &=& 2 I_s^{(s,\mp)}(R,\ta)\,\partial_R I_k^{(s,\pm)}(R,\ta) \;,\quad (\mathbf{s}=s>0) \;, \\
\label{eq:BurgersIkredsneg}
\partial_\ta I_{k'}^{(s',\pm)}(R,\ta) &=& 2 I_{s'}^{(s',\pm)}(R,\ta)\,\partial_R I_{k'}^{(s',\pm)}(R,\ta) \;,\quad (\mathbf{s}=s'=-s<0) \;,
\eeqa
in the CFT limit.\\ 
Let us begin with the $\mathbf{s}<0$ case. From the implicit relation (\ref{eq:IkfunIksneg})
\beq 
I_{k'}^{(s',\pm)}(R,\ta) = I_{k'}^{(s',\pm)}\( \mathcal{R}_0^{(s',\mp)} \) \;,
\eeq
using (\ref{eq:R00}), we have 
\beq  
\label{eq:app21}
I_{k'}^{(s',\pm)}(R,\ta) = \frac{2\pi a_{k'}}{\(R+2\ta I_{s'}^{(s',\pm)}(R,\ta)\)^{k}} \;.
\eeq
Differentiating (\ref{eq:app21}) w.r.t. $R$ for $k'=s'$ $(k=s)$, we first express $I_{s'}^{(s',\pm)}(R,\ta)$ as a function of $\partial_R I_{s'}^{(s',\pm)}(R,\ta)$ as
\beq
\label{eq:app22}
I_{s'}^{(s',\pm)}(R,\ta) = -\frac{R\,\partial_R I_{s'}^{(s',\pm)}(R,\ta)}{2\ta s(1+s)\,\partial_R I_{s'}^{(s',\pm)}(R,\ta)} \;.
\eeq
Then, differentiating again (\ref{eq:app21}) w.r.t. $R$ for generic $k'$ $(k)$ and using (\ref{eq:app22}), we express $I_{k'}^{(s',\pm)}(R,\ta)$ as a function of 
$\partial_R I_{s'}^{(s',\pm)}(R,\ta)$ and $\partial_R I_{k'}^{(s',\pm)}(R,\ta)$ as follows
\beq  
\label{eq:app23}
I_{k'}^{(s',\pm)}(R,\ta) = -\frac{s}{k}\frac{R\,\partial_R I_{k'}^{(s',\pm)}(R,\ta)}{2\ta s(1+s)\,\partial_R I_{s'}^{(s',\pm)}(R,\ta)} \;.
\eeq 
Finally, using (\ref{eq:app23}), it is a matter of simple algebraic manipulation to show that
\beq  
-R'\partial_R I_{k'}^{(s',\pm)}(R,\ta) + k'\theta_0' I_{k'}^{(s',\pm)}(R,\ta) = 2 I_{s'}^{(s',\pm)}(R,\ta)\,\partial_R I_{k'}^{(s',\pm)}(R,\ta)\;,
\eeq
which proves that (\ref{eq:BurgersIpm}) for $\Bk=k'$ and $\Bs=s'$ reduce to (\ref{eq:BurgersIkredsneg}).\\
Considering the $\mathbf{s}>0$ case, from the implicit relation (\ref{eq:IkfunIkspos})
\beq 
I_{k}^{(s,\pm)}(R,\ta) = I_{k}^{(s,\pm)}\( \mathcal{R}_0^{(s,\pm)} \) \;,
\eeq 
using (\ref{eq:R00}), we have
\beq  
\label{eq:app24}
I_{k}^{(s,\pm)}(R,\ta) = \frac{2\pi a_{k}}{\(R+2\ta I_{s}^{(s,\mp)}(R,\ta)\)^{k}} \;.
\eeq
Repeating the same procedure as in the $\Bs<0$ case, we express $I_{k}^{(s,\pm)}(R,\ta)$ as a function of $\partial_R I_{k}^{(s,\pm)}$ and $\partial_R I_{s}^{(s,\pm)}$ as follows
\beq
\label{eq:app25}
I_{k}^{(s,\pm)} = -\frac{s}{k}\frac{R\,\partial_R I_k^{(s,\pm)} \left( 2\ta s\,\partial_R I_s^{(s,\pm)}-2 \ta  \partial_R I_k^{(s,\mp)}+s\right)}{s^2 \left(1+2 \tau\, \partial_R I_s^{(s,+)}+2 \tau\, \partial_R I_s^{(s,-)}\right)+ 4 \left(s^2-1\right) \ta^2\,\partial_R I_s^{(s,+)}\,\partial_R I_s^{(s,-)}}\;,
\eeq
Again, using (\ref{eq:app25}), one can show that the following relation holds
\beq  
-R'\partial_R I_{k}^{(s,\pm)} + k\theta_0' I_{k}^{(s,\pm)} = 2 I_{s}^{(s,\mp)}\,\partial_R I_{k}^{(s,\pm)}\;,
\eeq
which proves that (\ref{eq:BurgersIpm}) for $\Bk=k$ and $\Bs=s$ reduce to (\ref{eq:BurgersIkredspos}).\\

\bibliography{Biblio3}

\end{document}